\documentclass[%
reprint, 
amsmath,amssymb,
aps,
prx,
]{revtex4-2}

\usepackage{graphicx}
\usepackage{epsfig} 
\usepackage{dcolumn}
\usepackage{bm}
\usepackage{epstopdf}
\usepackage{multirow}
\usepackage{amsmath}
\usepackage{booktabs}
\usepackage{color}
\usepackage[dvipsnames]{xcolor}
\usepackage{gensymb}
\usepackage{kantlipsum}  
\usepackage[colorlinks=true, allcolors=blue]{hyperref}
\usepackage{ulem}
\usepackage{braket}
\usepackage{physics}
\usepackage{url}
\usepackage[utf8]{inputenc}
\usepackage[T1]{fontenc}
\usepackage{mathptmx}
\usepackage{lineno}
\usepackage[frozencache,cachedir=.]{minted} 

\definecolor{burntorange}{rgb}{0.8, 0.33, 0.0}
\definecolor{darkviolet}{RGB}{148,0,211}

\begin{document}

\title{Automated Graph-Based Detection of Quantum Control Schemes: \\ Application to Molecular Laser Cooling}

\author{Anna~Dawid$^1$}
\author{Niccolò~Bigagli$^2$}
\author{Daniel~W.~Savin$^3$}
\author{Sebastian~Will$^2$}
\affiliation{%
$^1$Center for Computational Quantum Physics, Flatiron Institute, 162 Fifth Avenue, New York, NY 10010, USA
}
\affiliation{%
$^2$Department of Physics, Columbia University, New York, New York 10027, USA
}
\affiliation{%
$^3$Columbia Astrophysics Laboratory, Columbia University, New York, New York 10027, USA
}
\date{\today}

\begin{abstract}
One of the demanding frontiers in ultracold quantum science is identifying laser cooling schemes for complex atoms and molecules out of their vast spectra of internal states. Motivated by the prospect of expanding the set of available ultracold molecules for applications in fundamental physics, chemistry, astrochemistry, and quantum simulation, we propose and demonstrate an automated graph-based search approach for viable laser cooling schemes. The method is time efficient, reproduces the results of previous manual searches, and reveals a plethora of new potential laser cooling schemes. We discover laser cooling schemes for YO, C$_2$, OH$^+$, CN, and CO$_2$, including surprising schemes that start from highly excited states or do not rely on a strong main transition. A central insight of this work is that the reinterpretation of quantum states and transitions between them as a graph can dramatically enhance the ability to identify new quantum control schemes for complex quantum systems. As such, this approach will also apply to complex atoms and, in fact, any complex many-body quantum system with a discrete spectrum of internal states. 
\end{abstract}

\maketitle

\section{Introduction}
Ultracold atomic and molecular systems are rich playgrounds for modern efforts to understand quantum many-body physics and its emergent phenomena \cite{Bloch2008RMP, Lamata18AdvPhys, Tomza19RevModPhys, Kjaergaard20AnnuRev, Monroe21RevModPhys, baranov2012condensed} as well as platforms for quantum computing \cite{jaksch2000fast, demille2002quantum, morgado2021quantum} and quantum sensing \cite{ludlow2015optical, bongs2019taking}. 
In order to use atoms and molecules for quantum applications, control of internal quantum states and their state of motion is a critical prerequisite. In particular, the quantum control of molecules poses a significant experimental challenge due to their rich internal state structure~\cite{Langen2023ultracoldreview}.
Ultracold molecules hold great promise for exploring the role of quantum effects in chemistry \cite{Ospelkaus10, Ni10, Miranda11, TomzaPRL15, Klein17, Puri17, Jongh20, Liu2021molreview}, which may lead to a better understanding of chemical reactions and their control via external fields \cite{Krems2008, Lemeshko13, Park2023resonance, Chen2023resonance}. Moreover, ultracold molecules enable ultraprecise measurements of fundamental constants \cite{mitra2022quantum}, including the most accurate constraints of the electron electric dipole moment to date \cite{acme2018improved, Roussy2023}. Cold molecules are also needed for laboratory studies of astrochemically important reactions~\cite{Herbst2008, Herbst2014}. 

To pursue these applications, 
laser cooling of molecules~\cite{Fitch2021chapter} has been successfully demonstrated over recent years using molecules with nearly diagonal Franck-Condon (FC) factors such as SrF \cite{Shuman09PRL, ShumanNature10, barry2014magneto}, CaF \cite{zhelyazkova2014laser, hemmerling2016laser, anderegg2017radio}, YbF \cite{lim2018laser}, BaH \cite{mcnally2020optical}, CaOH \cite{baum20201d}, CaOCH$_3$ \cite{mitra2020direct}, YO \cite{hummon2013YO, yeo2015rotational}, SrOH \cite{kozyryev2017sisyphus}, and YbOH \cite{augenbraun2020laser}. Another fruitful approach has been the assembly of ultracold molecules from samples of ultracold alkali atoms, leading to ultracold Cs$_2$ \cite{Danzl08Science}, KRb \cite{Ni08, deMarco2019}, RbCs \cite{TakekoshiPRL14, Molony14}, NaK \cite{Park15}, NaRb \cite{Guo2016}, and NaCs \cite{Cairncross2021NaCs, Stevenson23NaCs, Bigagli2023NaCs}. The dominant factor driving the selection of these molecules was the expectation that they would be amenable to cooling. As a result, the available ultracold molecules come from a narrow range of species with applications mostly in fundamental and many-body quantum physics.

\begin{figure}[t]
    \centering
    \includegraphics[width=\columnwidth]{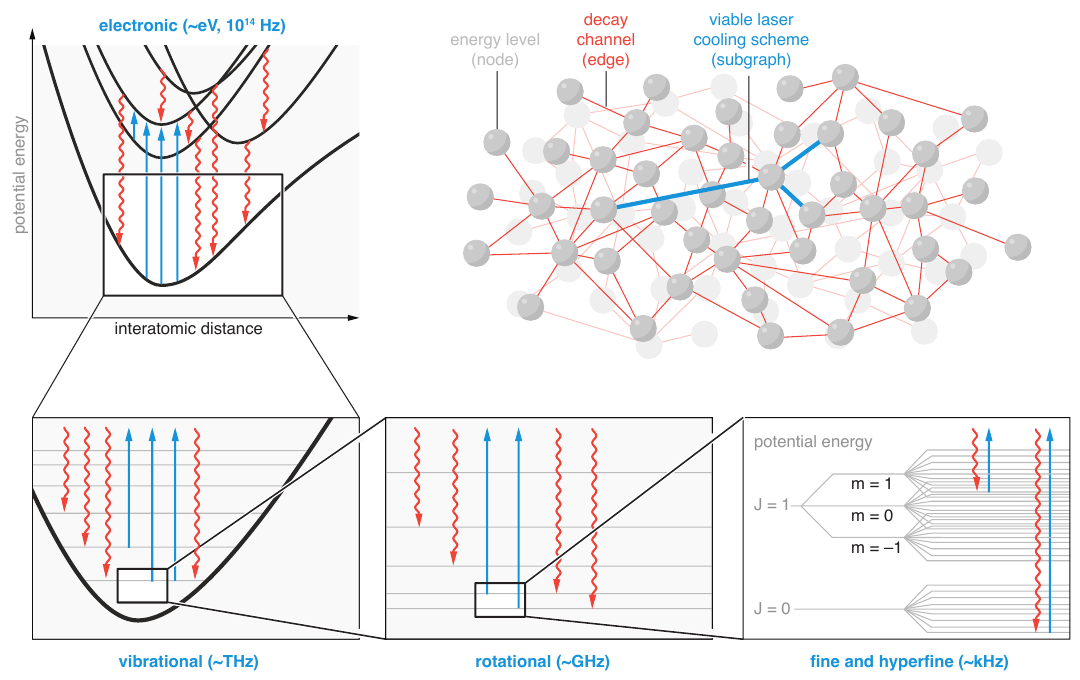}
    \caption{The search for laser cooling schemes needs to address electronic, vibrational, rotational, fine, and hyperfine states spanning multiple energy scales. The transformation to a graph-based representation of energy levels and decay channels enables the use of highly parallelizable search algorithms to automatically detect viable laser cooling schemes.}
    \label{fig:intro}
\end{figure}

In order to unlock a wider potential for ultracold molecules and, in particular, to understand chemical reactions at a quantum-state-resolved level, the field will benefit from access to a greater diversity of ultracold molecular species. These include species that are relevant for chemistry, biology, industrial applications, etc. \cite{Softley2023twenties}. A promising step in this direction was the recent realization that molecules with significant off-diagonal FC factors can also be amenable to laser cooling, as discussed in Refs.~\cite{Bigagli22carbon, Bigagli2023OH}. While near-diagonal FC factors make it more probable that a laser cooling scheme exists, these works show that focusing on them introduces a restricting bias that narrowed the search range for molecular laser cooling. This motivates the development of a bias-free method that enables a broad search for any molecule for which sufficient spectroscopic data are available. The task of identifying laser cooling schemes is, however, highly non-trivial as it requires combing through millions of internal quantum states across many orders of magnitude in energy and billions of decay channels between them. Manual searches have resulted in the identification of laser cooling schemes for molecules at a rate on the order of one per year.

With the goal of avoiding limiting assumptions and biases, we have developed an automated approach to search for laser cooling schemes. Typically, the complex spectra of molecules are given as a list of electronic, vibrational, rotational, fine, and hyperfine energy levels and transitions between them. Our method introduces a change of perspective, reinterpreting molecular spectra in the form of a graph (Fig.~\ref{fig:intro}). The graph representation allows us to harness the power of highly efficient graph processing tools, mining the graphs to identify subgraphs representing viable laser cooling schemes. To represent a technically viable laser cooling scheme, a subgraph needs to meet a set of carefully chosen conditions determined by the required level of closure, cooling time, and the wavelengths of available lasers. For our search, we leverage spectroscopic data of various molecules that are available in large databases, such as ExoMol~\cite{Tennyson2016Exomol}, HITRAN~\cite{Gordon2022HITRAN}, Splatalogue~\cite{Splatalogue23}, and more \cite{NISTdatabase, CDMS23, Liu2020database, Bernath2020MoLLIST}. An automated search can dramatically speed up the identification of laser cooling schemes for atoms and molecules, which can then be confirmed in an experiment. More generally, the graph approach can help to identify quantum control schemes in any complex quantum system that has discrete energy levels and transitions between them.

In this work, we describe our graph-based approach and demonstrate its application. We discuss the methodology in Sec.~\ref{sec:method}. In Sec.~\ref{sec:results}, we give examples for five different molecules, YO, C$_2$, OH$^{+}$, CN, and CO$_2$, where the automated search has identified new laser cooling schemes. We especially describe cooling schemes that can be viewed as surprising and counterintuitive and that showcase the power of automation and avoiding human bias. In Sec.~\ref{sec:conclusion}, we conclude with a discussion of potential extensions and improvements of our method. Finally, our goal is to make the graph-based approach available and accessible to everyone. In App.~\ref{app:schemes}, we present a list of selected cooling schemes that we have identified; and in App.~\ref{app:NH-tutorial}, we present a step-by-step tutorial describing the graph platform, query language, and search algorithm.

\section{Method}\label{sec:method}
\subsection{Introduction to graphs and graph processing}\label{ssec:intro_to_graphs}

Graphs originated in mathematics~\cite{Trudeau1994book} from the insight that there are problems that are simplified by a change of representation, where one focuses on connections between objects rather than on the objects themselves. Today, graphs are widely considered a useful and informative way of modeling and analyzing data \cite{barabasi2002linked}. The objects that make up a graph are called nodes or vertices, and the connections between them are called relationships, links, or edges. For the remainder of the paper, we call graph elements \textit{nodes} and \textit{relationships} to align with the documentation of the graph platform used, which is Neo4j~\cite{Neo4j}. Additionally, each node and relationship can have \textit{properties}, also called attributes, which can contain a variety of data types, from numbers and strings to spatial and temporal data. For example, a node representing an internal molecular quantum state can have properties, such as energy and lifetime. Relationships in directed graphs have not only properties but are also characterized by the \textit{direction}. In our work, the relationship represents a spontaneous decay channel from one energy level to another, and its property is the energy difference between the connected energy levels. In addition, each element of a graph can be assigned a \textit{label}, which marks its belonging to a group. For example, a node representing an energy level can belong to a group of metastable states.

Instead of just representing a problem in the form of a graph, we also want to process the graph to reveal additional information (e.g., find a grouping of nodes in a transportation network responsible for congestion). This is the domain of \textit{graph algorithms}~\cite{Needham2019book}. They can be roughly categorized into local and global. Global graph processing is usually interested in global patterns and structures, and its output may be an enriched graph or some aggregate value, such as a score. 
For example, one may be interested in the toxicity of a molecule represented as a graph~\cite{Miao2023toxicity} or symmetries in atomic spectra~\cite{Wellnitz2023network}. In this work, we are interested in locating subgraphs within a graph, that is, graphs within a larger graph. This approach belongs to local graph processing and is usually connected to \textit{querying a graph}, which can be understood as sending questions to a graph and obtaining a response. 
In our case, the subgraphs represent laser cooling schemes and, therefore, have to meet a series of conditions. In the next section, we present a primer on laser cooling to understand the origin of those conditions.

\subsection{Primer on laser cooling}\label{ssec:laser_cooling}

Doppler cooling \cite{Metcalf1999book} entails repeated photon absorption and spontaneous emission events that aim to decelerate and, therefore, cool atoms or molecules. What makes it successful is an exploitation of the Doppler shift so that the absorption event is velocity-dependent. Particles moving toward the laser beam preferentially absorb its photons, while for particles moving away from the laser, absorption is suppressed. Absorption of a photon induces a transition of the ground state to an excited internal state of the particle. Additionally, the particle experiences a kick from the transfer of one photon momentum, decelerating it in the direction of the laser beam. Then, the particle spontaneously emits a photon, which, due to its random direction, results in an average net momentum transfer of zero. In this way, we can decelerate particles whose internal states are well approximated by a closed two-level system up to the Doppler cooling limit, given by the random momentum kicks generated by the spontaneous emission~\cite{castin1989limit}. The set of states involved in the repeated absorption-emission events is known as a \textit{laser cooling} or \textit{cycling} scheme, and the whole process as \textit{photon cycling}.

This picture becomes more complex when we consider systems that go beyond the two-level approximation. In this case, we need to account for many possible decay channels from the excited state to various lower-energy states. Every such additional decay channel requires an additional laser to pump back the particles to the excited state. Otherwise, the particle leaks into a dark state, that is, the particle is not amenable to further laser cooling as the dark state cannot absorb another photon. The presence of multiple decay channels from an excited state and the need for additional repump lasers constitute the primary challenge in cooling molecules and multi-electron atoms \cite{FITCH2021157}.

Molecular laser cooling schemes that are used today comprise up to 12 lasers \cite{Vilas2022CaOH}, which address up to 12 decay channels. However, there may be thousands of decay channels from a single excited molecular state. \textit{A viable laser cooling scheme needs to be sufficiently closed so that the cooling to a desired temperature is completed before all molecules leak into a dark state and get lost from the sample}. In particular, we compare here two estimated quantities: the time $t_{10\%}$ after which 10\% of the molecules remain in bright states of the laser cooling scheme (that is, until 90\% of the molecules are lost) and the time $t_{\rm cool}$ needed to cool down a molecule to the Doppler limit. For a viable laser cooling scheme, it is reasonable to require the condition $t_{\rm cool} < t_{10\%}$. If this condition can be met, we call the laser cooling scheme sufficiently closed. We note that at the end of the cooling process, the population of the molecular sample will be distributed among all states addressed by the cycling scheme. By selectively turning off one of the laser beams, the sample can be optically pumped into a specific state if desired.  

The closure of a laser cooling scheme depends on the number of available repump lasers. 
Fundamentally, it depends on the probability that a molecule follows a particular decay channel $i$, which is expressed as a branching ratio, $\mathrm{BR}_{i}$. The branching ratio depends on the Einstein coefficients, $A_{j}$, of allowed decay paths out of the excited state \cite{Rosa2004EPJD} as
\begin{equation}\label{eq:br}
    \mathrm{BR}_{i} = \frac{A_i}{\sum_j A_j}\,,
\end{equation}
where $j$ runs over all decay paths starting at the same excited state. The lifetime of the excited state is
\begin{equation}\label{eq:lifetime}
    \tau = \frac{1}{\sum_j A_j}\,.
\end{equation}
The closure of a cooling scheme is simply given by
\begin{equation}\label{eq:closure}
    p = \sum_i \mathrm{BR}_{i}\,,
\end{equation}
where, importantly, $i$ runs only over driven transitions (the main cooling transition and the addressed repump transitions); these are the decay paths that are addressed with the lasers, which repump the molecule back to the desired excited state after a spontaneous emission event. If all decay channels are driven by lasers, then $p=1$, and the laser cooling scheme is perfectly closed. In practice, well-closed laser cooling schemes have $p > 0.9999$. After $n$ photon scatterings, the fraction of molecules in bright states, i.e., those addressed by the lasers, is given by $p^n$. Therefore, the number of scatterings that retain 10\% of the molecules in a bright state is
\begin{equation}\label{eq:n10}
    n_{10\%} = \frac{\ln{0.1}}{\ln{p}}\,,
\end{equation}
and the corresponding time is \cite{Tarbutt2013NJP, Fitch2021chapter}
\begin{equation}\label{eq:t10}
    t_{10\%} = n_{10\%} \times R^{-1}\,,
\end{equation}
where
\begin{equation}\label{eq:R}
    R^{-1} = \tau\left[ (G+1) + \frac{2}{3}\Gamma \pi h c \sum_i \frac{1}{I_i \lambda_i^3}\right] \,.
\end{equation}
Here, $G$ is the number of driven transitions, $\Gamma \equiv 1/\tau$, $h$ is Planck’s constant, $c$ is the speed of light, $I_i$ is the intensity of the laser addressing the $i$-th transition, and $\lambda_i$ is the wavelength of the $i$-th addressed transition. In this expression, the term multiplying the lifetime represents the probability of occupying the excited state. The first part highlights the equal distribution of population among all addressed states at the steady state, while the second part is a correction taking into account the saturation intensities. In the following, we assume a constant intensity $I_i$ of $10^3$ mW$\,$cm$^{-2}$ for all transitions as an experimentally reasonable number. This value was picked to simplify calculations. However, it may not be experimentally feasible for all transitions (e.g. for UV lasers). If any molecular properties or technical constraints require a different condition, it can be adjusted in the algorithm as needed.

To estimate $t_{\rm cool}$, we calculate the approximate number of scatterings for molecules to reach a standstill, $n_{\rm cool}$. To this end, we divide the initial momentum of the molecule by that of an average photon that participates in the cooling process \cite{Rosa2004EPJD}. The initial momentum follows from the Maxwell-Boltzmann distribution, 
\begin{equation}\label{eq:MaxwellBoltzmann}
    p_{\rm init} = m \expval{v}_{\rm init} = \sqrt{3 k_{\rm B} T_{\rm init} m}\,,
\end{equation}
where $k_\mathrm{B}$ is the Boltzmann constant, $T_{\rm init}$ is the initial temperature of the molecular gas, $m$ is the molecular mass, and $\expval{ \cdot }$ denotes averaging over the Boltzmann distribution. Importantly, $T_{\rm init}$ of the gas also determines which internal molecular states are thermally occupied and, therefore, can become starting points for the cooling procedure. We explain how we estimate the required $T_{\rm init}$ of a cooling scheme in Sec.~\ref{ssec:limitations}. The average photon momentum is \cite{Bigagli22carbon}
\begin{equation}\label{eq:pmean}
   p_{\rm mean} = \sum_i \frac{h}{\lambda_i} \frac{\mathrm{BR}_i}{\sum_j \mathrm{BR}_j} = \frac{h}{p} \sum_i \frac{\mathrm{BR}_i}{\lambda_i} \,,
\end{equation}
where $i$ runs over all driven transitions. This allows the calculation of the number of scattering processes
\begin{equation}\label{eq:ncool}
    n_{\rm cool} = \frac{p_{\rm init}}{p_{\rm mean}} = \frac{p}{h} \sqrt{3 k_{\rm B} T_{\rm init} m} \left( \sum_i \frac{\mathrm{BR}_i}{\lambda_i} \right)^{-1} \,.
\end{equation}
Then, following Eq.~\eqref{eq:t10}, we compute the cooling time as
\begin{equation}\label{eq:tcool}
    t_{\rm cool} = n_{\rm cool} \times R^{-1}\,.
\end{equation}

To ground these derivations in practical considerations, it is useful to give some order of magnitude estimates of these quantities. Typically, one can expect $n_{\rm cool}\sim 10^3$-$10^4$, $t_{\rm cool}\sim$ 1-100 ms, and $R\sim 0.1$-10 MHz.

\begin{figure*}[t]
    \centering
    \includegraphics[width=\textwidth]{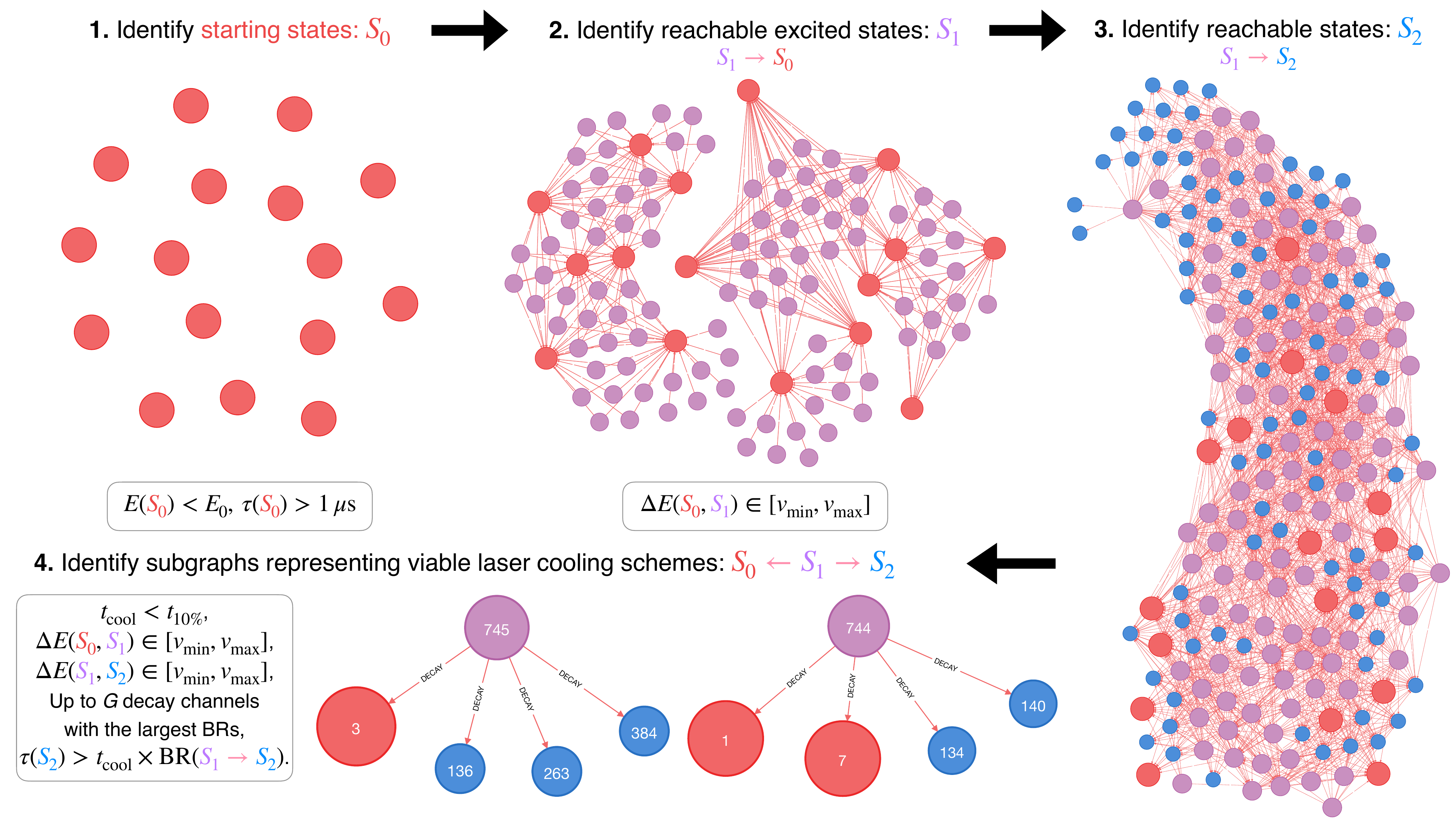}
    \caption{Illustration of the steps of the graph-based search for laser cooling schemes: 1. The search starts with identifying potential starting states, $S_0$ (red nodes), i.e., all states with an energy lower than the preset $E_0$ with lifetimes longer than 1 $\mu$s. 2. Then it identifies all excited states $S_1$ (purple nodes) that can be reached from $S_1$ using laser frequencies from a preset range. 3. The search identifies all states $S_2$ (blue nodes) to which any of the $S_1$ states can decay. 4. Finally, within the graph of $S_0$, $S_1$, and $S_2$ states, subgraphs representing laser cooling schemes are identified. Here, we use as an example the ${}^{14}$NH molecule and the following parameters: $G=4$, $\lambda_{\rm min} = 330\,\,$nm, $\lambda_{\rm max} = 1000\,\,$nm, and $T_0 = 500\,\,$K. To reproduce those results, follow the tutorial in App.~\ref{app:NH-tutorial}.}
    \label{fig:graph_algo}
\end{figure*}

\subsection{Graph search algorithm to identify laser cooling schemes}\label{ssec:graph_algo}

A manual search for a laser cooling scheme is generally a very hard task. Imagine locating a connected system of up to a dozen states that allow nearly closed photon cycling within a complex spectrum of electronic, vibrational, rotational, and hyperfine molecular states with a plethora of decay paths between them. For example, CO$_2$ has over 3.5 million states with over 2.5 billion decay paths between them \cite{Yurchenko2020CO2}.

Instead, we take advantage of powerful and highly parallelizable graph algorithms to search for schemes automatically. This becomes possible by representing the molecular spectrum and laser cooling schemes within it in the form of a graph. For example, a molecule in one state decaying to another state can be represented as a two-node directed graph, with nodes representing the electronic, vibrational, rotational, fine, and hyperfine energy levels and the relationship representing the direction of the decay process. In this way, we obtain a graph that simultaneously represents both the energy levels of a molecule and the relationships between the levels.

Next, we search for a subgraph corresponding to a viable laser cooling scheme. The graph search algorithm is presented schematically in Fig.~\ref{fig:graph_algo}. We start by identifying all possible \textit{starting states}, $S_0$ (first panel of Fig.~\ref{fig:graph_algo}). A ground state is always a viable starting state; but interestingly, a higher initial temperature of the molecular gas may unlock different laser cooling schemes starting from $S_0$ levels that are excited states. Therefore, the available $S_0$ levels are determined by the preset maximum initial temperature of the molecular gas, $T_0$. Assuming that we want at least 10\% occupancy of the excited state compared to the ground state, it follows from the Boltzmann distribution that $E_{S_0} < E_0 = -\ln (0.1) k_{\rm B} T_0$. Here, $E_{S_0}$ is the energy of a potential starting state relative to the ground state, and $E_0$ is the maximum energy allowed for a starting state in $S_0$ at the preset starting temperature $T_0$. Moreover, note that here $S_0$ does not need to be the starting point for the main cooling transition. We only require that the lifetime $\tau_{S_{0}}$ of $S_0$ be longer than $\sim 1~\mu$s, in order to allow enough time to pump from this state to an excited state.

The second step is to identify all higher energy states to which $S_0$ states can be driven by accessible lasers. We call those states \textit{reachable excited states} and mark them as $S_1$. The major conditions here are that (1) an electric dipole allowed transition channel must exist between $S_1$ and $S_0$, which ensures the possibility of efficient laser driving between $S_0$ and $S_1$ and (2) the energy difference between $S_0$ and $S_1$ has to be within the range of accessible laser frequencies, $\Delta E (S_0, S_1) \in [ \nu_{\rm min}, \nu_{\rm max} ]$, where $\nu$ is the laser frequency with a corresponding wavelength, $\lambda = c/\nu$, where $c$ is the speed of light. This is schematically shown in the second panel of Fig.~\ref{fig:graph_algo}.

The third step is to identify all lower energy states to which the $S_1$ states can spontaneously decay. We call those states \textit{reachable states} and mark them $S_2$. These states constitute all possible decay paths of the molecule excited to any of the $S_1$ states. In particular, note that $S_0 \subset S_2$ \footnote{To keep the notation simple, we chose to use $S_0$, $S_1$, and $S_2$ interchangeably to symbolize individual states belonging to the set of starting states, reachable excited states, and reachable states and to symbolize the sets themselves.}.

\begin{figure}[t]
    \centering
    \includegraphics[width=\columnwidth]{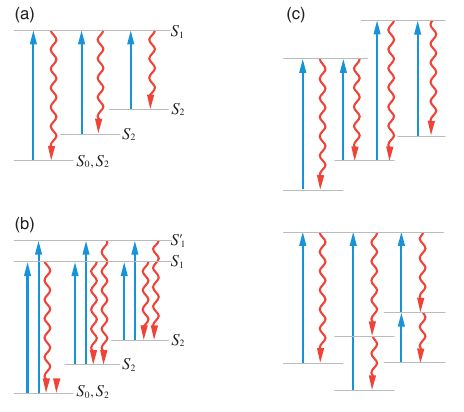}
    \caption{Representative classes of laser cooling schemes. Our graph-based tool searches for schemes of the type as in (a) and (b). (a) A simple laser cooling scheme based on one excited state. (b) A scheme with two excited states that decay to the same $S_2$ states. Including an extra excited state $S_1$ is of interest, as it usually leads to a shorter cooling time $t_{\rm cool}$. (c) Examples of complex laser cooling schemes that a graph-based approach could search for if a computationally cheap estimate of the cooling time were available.}
    \label{fig:schemes}
\end{figure}

At this stage, we have identified all possible states $\in \{ S_0, S_1, S_2 \}$ that could participate in a laser cooling scheme (third panel of Fig.~\ref{fig:graph_algo}). Now, among them, we need to identify \textit{viable} laser cooling schemes. Here, we focus on schemes of the type as in Fig.~\ref{fig:schemes}(a) that are based on a single excited state $S_1$. In App.~\ref{app:algo_extension}, we also describe the search for schemes as in Fig.~\ref{fig:schemes}(b), based on two excited states $S_1$ and $S_1'$ that decay to the same states $S_2$. As explained in Sec.~\ref{ssec:laser_cooling}, the central condition is that the cooling time is shorter than the time after which only 10\% of molecules remain in bright states, $t_{\rm cool} < t_{10\%}$, from Eqs.~\eqref{eq:t10} and~\eqref{eq:tcool}. Taking advantage of highly parallel graph processing, we fix the number of lasers, $G$, and identify for each state $S_1$ the $G$ decay channels with the largest $\textrm{BR}$s that can be driven with accessible preset laser wavelengths, $\lambda \in [\lambda_{\rm min}, \lambda_{\rm max}]$. For all schemes found, we compute $t_{\rm cool}$ and $t_{10\%}$ and save those that meet the condition $t_{\rm cool} < t_{10\%}$ (or $n_{\rm cool} < n_{10\%}$). To find all relevant laser cooling schemes, we repeat the search for $G = 1, 2, \ldots, G_{\rm max}$. This step is schematically shown in the fourth panel of Fig.~\ref{fig:graph_algo}.

A viable laser cooling scheme, in addition to $t_{\rm cool} < t_{10\%}$, also requires that the lifetimes of the reachable states, $S_2$, are long enough to pump the molecules back to the reachable excited states $S_1$ before $S_2$ decays to other lower energy unmarked states. Hence, we could require $\tau (S_2) > t_{\rm cool}$. However, this is quite restricting. In particular, it rules out $S_2$ states that are short-lived but have an extremely small $\mathrm{BR}$ from the upper level $S_1$. Therefore, we relax this condition to 
\begin{equation}\label{eq:S2lifetime_condition}
    \tau (S_2) > t_{\rm cool} \times \mathrm{BR}(S_1 \rightarrow S_2)\,,
\end{equation}
where $t_{\rm cool} \times \mathrm{BR}(S_1 \rightarrow S_2)$ is an estimated average time that a molecule spends in state $S_2$ during the cooling process. Note that $S_0 \subset S_2$, and therefore this ensures that also the lifetime of $S_0$ meets the condition in Eq.~\eqref{eq:S2lifetime_condition}.

\begin{table}[t]
\centering
\vspace{-0.5cm}
\caption{Parameters of the graph algorithm that need to be preset by the user}
\label{tab:algo-params}
\resizebox{\columnwidth}{!}{%
\begin{tabular}{l|l}
\textbf{Parameter}                                                      & \textbf{Unit} \\ \hline
Number of lasers: $G$                                                    & unitless      \\
Range of laser wavelengths: $\lambda_{\rm min}$ and $\lambda_{\rm max}$ & nm            \\
Maximal initial temperature of the molecular gas: $T_0$                  & K             \\
Molecular mass: $m$                                                          & u            
\end{tabular}%
}
\end{table}

The graph search algorithm described above requires the user to preset some parameters, as listed in Tab.~\ref{tab:algo-params}. The first is the number of lasers $G$, i.e., the maximum number of driven transitions in a cooling scheme. Other parameters to set are the minimum and maximum laser wavelengths available for cooling, $\lambda_{\rm min}$ and $\lambda_{\rm max}$. The choice of parameters is up to the user, but setups that we consider technologically feasible nowadays use up to twelve repumping lasers with wavelengths ranging from 450 nm to 3.5 $\mu$m. Given the rapid speed of technological developments, perhaps a laser wavelength range of 300 nm - 5 $\mu$m and the simultaneous use of more than a dozen lasers should be realistic in the relatively near future.

The user also needs to preset the maximum initial temperature of the molecular gas, $T_0$. This parameter restricts the states that are considered starting states, $S_0$, due to the condition of $E_{S_0} < E_0 = - \ln(0.1) k_{\mathrm{B}} T_0$. We chose $T_0 = 500\,$K, which therefore sets $E_0 \approx 800\,\mathrm{cm}^{-1}$. Counterintuitively, sometimes it is beneficial to start with a hotter molecular gas because it enables access to different efficient laser cooling schemes that make up for the higher temperature. It can also enable the use of more readily available laser wavelengths. Finally, the mass of the molecule $m$ is needed to approximate the initial momentum of the molecule in Eq.~\eqref{eq:ncool}.

\subsection{Limitations of the method}\label{ssec:limitations}

While the graph-based search can quickly find laser cooling schemes of molecules, it also has some limitations and avenues for further development. Sometimes a deeper understanding of the molecular structure may help to uncover even more laser cooling schemes than the automated search alone.
For example, when deciding on a viable laser cooling scheme, we compare $t_{10\%}$ and $\tau (S_2)$ to $t_{\rm cool}$. The quantity $t_{\rm cool}$ depends, in part, on the initial temperature of the molecular gas, $t_{\rm cool} \propto \sqrt{T_{\rm init}}$. We approximate the required initial temperature of a cooling scheme as
\begin{equation}\label{eq:Tinit}
    T_{\rm init} = \max{[4\,\mathrm{K}, -E_{S_{0}}/\ln(0.1) k_{\mathrm{B}}]}\,, 
\end{equation}
where $E_{S_{0}}$ is the energy of the starting state. In other words, for low-energy starting states, we assume 4 K to be the lowest achievable temperature using buffer-gas cooling \cite{hutzler2012buffer}. For excited starting states, we assume a temperature where the state has an occupancy of 10\% compared to the ground state, based on the Boltzmann distribution at that temperature. So far, we have assumed that kinetic and internal energies are in equilibrium. However, this assumption does not account for the details of the molecular gas preparation. In particular, it may happen that starting the laser cooling scheme from a highly excited state does not require a high initial kinetic temperature. This is the case, e.g., in the Swan and Duck cooling scheme for the C$_2$ molecule \cite{Bigagli22carbon}, where the triplet and singlet energy manifolds have a negligible coupling. Therefore, an initial condition can be created in which the molecular gas is prepared in the highly excited triplet ground state at low kinetic temperature without decaying to much lower-lying singlet states. Moreover, the assumption of 10\% occupancy can be easily relaxed if one has access to a high-flux molecular source. To account for these two special cases, it is useful to output the results also assuming $T_{\rm init} = 4\,\,$K (instead of equal to $\max{[4\,\mathrm{K}, -E_{S_{0}}/\ln(0.1) k_{\mathrm{B}}]}$), which can recover additional laser cooling schemes that need to be carefully analyzed by a researcher to determine if they are viable. Such special cases can also be included in the automated search but go beyond its current implementation.

Moreover, in this work, we focus on searches for two types of laser cooling schemes with a \textit{single} excited state or \textit{two} excited states that decay to the same reachable states. We present these two types of schemes in Fig.~\ref{fig:schemes}(a) and (b), respectively. In these simple scenarios, our search already identifies a plethora of new laser cooling schemes. In principle, more complex cooling schemes are possible, for example, involving multiple excited states that do not decay to the same reachable states~\cite{Vilas2022CaOH, Bigagli2023OH} or multi-level decay channels, as illustrated in Fig.~\ref{fig:schemes}(c). Including more complex laser cooling schemes will be a meaningful extension. It requires a computationally simple approach for calculating cooling times. Nowadays, such calculations are performed, for example, using Monte Carlo simulations, which would render the graph search prohibitively slow. Hence, finding ways to efficiently estimate the cooling time for complex laser cooling schemes represents one of the current theoretical frontiers for the field. 

In general, we are conservative in our estimates of cooling parameters. Therefore, the number of schemes detected via our automated search is a lower bound on the number of all possible schemes. Every addressed limitation described above will only increase the number of viable cooling schemes. 

However, in some cases, the graph search may yield results that are too optimistic. For example, potential off-resonant excitation of nearby allowed transitions driven by repumping lasers may constitute loss channels that are unaccounted for. An analysis of such effects could be incorporated into the search by taking into account the proximity of nearby states. Moreover, we assume that the lifetime of the starting state $\tau(S_0)$ needs to be larger than 1 $\mu$s to allow for effective pumping into the excited state. More precisely, it should be larger than the duration of electric dipole transitions between $S_0$ and $S_1$.

Finally, we note that the quality of the graph search is limited by the quality and completeness of the spectroscopic data available. The results obtained are reliable only to the extent that the input spectroscopic data are accurate. The example of YO discussed in Sec.~\ref{ssec:YO} illustrates, though, that there are spectroscopic data sets sufficiently accurate to allow the practical use of our tool. Joint theoretical and experimental efforts are expected to continuously improve open-access databases of molecular lines and transitions and, as a result, increase the detection power of our search method.

\subsection{ExoMol database}\label{ssec:Exomol}
For this work, our main data source is the ExoMol database \cite{Tennyson2016Exomol, Exomol23}, which uses experimental data on molecular energy levels and transitions combined with first-principles theoretical calculations. From these data, we extract information on the quantum states, such as their energy and various quantum numbers, and transitions between the states, including transition frequencies and Einstein $A$ coefficients, from which we calculate branching ratios (BR) of possible decay paths via Eq.~\eqref{eq:br}. The branching ratios and decay rates fully characterize the spontaneous emission pathways and the available molecule-light interactions. As a result, a separate evaluation of the FC factors is not necessary. Here, we have analyzed data on ${}^{89}$Y${}^{14}$O~\cite{Yurchenko2023YO}, ${}^{12}$C$_2$~\cite{Yurchenko2018C2, mckemmish2020update}, ${}^{16}$O${}^{1}$H$^+$~\cite{Hodges2017OH, Bernath2020MoLLIST}, ${}^{12}$C${}^{14}$N~\cite{Brooke2014CN, Syme2020CN, Syme2021CN}, and ${}^{12}$C${}^{16}$O$_2$~\cite{Yurchenko2020CO2}.

It should be noted that, in addition to ExoMol, there are multiple other databases on molecular spectroscopic data, such as the HITRAN2020 molecular spectroscopic database \cite{Gordon2022HITRAN}, Splatalogue \cite{Splatalogue23}, the National Institute of Standards and Technology Diatomic Spectral Database \cite{NISTdatabase}, the Cologne Database for Molecular Spectroscopy \cite{CDMS23}, and the Diatomic Molecular Spectroscopy Database \cite{Liu2020database}.

\subsection{Numerical implementation: Neo4j, Cypher, scaling, and hardware details}\label{ssec:Neo4j_Cypher_scaling}
We implement our graph search using the Neo4j graph database \cite{Neo4j}, which contains an integrated, highly optimized library with graph algorithms, called the Graph Data Science Library \cite{Neo4jGDSLibrary}. The framework can handle graph problems with up to a trillion relationships and nodes \cite{Neo4jtrillion}. The graph procedures can be executed via the Cypher query language \cite{Cypher}.

An important aim of our work is to make the graph-based algorithm accessible to any researcher interested in finding laser cooling schemes for an arbitrary atom or molecule, neutral or charged, given the information on the energy levels and transition rates between them. To this end, we provide two versions of the algorithm: a simplified one that can be run in the cloud and plays a mostly educational role and a full-scale one that needs to be run with a local installation of the Neo4j framework.

The simplified version of the algorithm is described in App.~\ref{app:NH-tutorial}. We provide a step-by-step tutorial on how to build a Neo4j database instance for ${}^{14}$NH from data available in ExoMol and how to search for viable laser cooling schemes using Cypher. The instance is built in the cloud, so no software installation is needed. The code provided in the Appendix is sufficient to understand all concepts of the graph search algorithm, but we also recommend taking advantage of detailed general tutorials provided by Neo4j~\cite{Neo4j}.

To tackle larger molecules, such as CO$_2$ with more than 3.5 million states and more than 2.5 billion decay paths between them, additional numerical tricks are needed to load the data and comb through them efficiently. These tricks are presented in detail in the tutorial and code available on GitHub \cite{OurMolRepo}. Running this code requires a local installation of the Neo4j framework.

The graph search algorithm that we propose in this work can easily find viable laser cooling schemes for any system with up to 2.5 billion decay channels. Taking CO$_2$ as an example, after downloading and preparing the data files, the data import into the database instance takes 28 mins, the preparation of the database instance for the search takes roughly 30 mins, and the search itself takes less than 1 min (run on a Linux workstation with Intel(R) Xeon(R) Gold 6128 CPU \@ 3.40 GHz and 196 GB RAM). Each of these steps is described in more detail in App.~\ref{app:NH-tutorial}.
Although the number of energy levels has a negligible impact on the search's run time, the algorithm scales poorly with the number of decay channels. For example, tackling NH$_3$ with more than 5 million energy levels and almost 17 billion decay channels is beyond the scope of the current implementation, as it increases the run time to weeks. However, this is not a fundamental limitation; rather, it is dependent on the efficiency of implementation and the availability of computational resources.

\section{Results: detected laser cooling schemes}\label{sec:results}
\subsection{YO: Benchmarking an experimentally laser-cooled molecule}\label{ssec:YO}

We start with YO as a testing ground for the automated detection algorithm. This molecule has already been successfully laser cooled in experiments, including the demonstration of laser cooling in a three-dimensional magneto-optical trap and of sub-Doppler cooling \cite{hummon2013YO, Collopy2015YO,yeo2015rotational,collopy20183d,ding2020sub}. Therefore, it provides an ideal testing ground for the practical applicability of our approach. The experimentally demonstrated cycling schemes for YO use as $S_0$ the $|v=0,\, N = 1\rangle$ state in the ground electronic state $X^2\Sigma^+$, and as $S_1$ the $|v=0,\, J=1/2\rangle$ and $|v=1,\, J=1/2\rangle$ states in the electronically excited $A^2\Pi_{1/2}$ manifold. Repumping transitions are added from the $v=1,2$ states in the $X^2\Sigma^+$ potential to $S_1$. The main leakage in this cycling scheme arises from decays to the $A^{\prime 2}\Delta_{3/2}$ electronic state at a $10^{-4}$ level \cite{collopy20183d}. ExoMol provides all its data using the $J$ rotational quantum number associated with Hund's case (a), while the ground and excited electronic states of YO follow different Hund's cases, (b) and (a), respectively, with case (b) described by the quantum number $N$. For YO, the relation $J\rightarrow N-1/2$ for $f$ parity and  $J\rightarrow N+1/2$ for $e$ parity  \cite{bernard1979fourier} allows for the translation between the cases. Once this translation is taken into account, together with the microwave mixing used to maintain rotational closure, the laser cooling scheme identified in \cite{Collopy2015YO} is observed in the form of two cycling schemes with, respectively, $A^2\Pi_{1/2}$ $|v = 0\rangle$ or $|v = 1\rangle$ as the $S_1$ state, the same $S_0$ state, and many overlapping $S_2$ states. Both have a predicted cooling time from 4 K of below 1.5~ms. Interestingly, our automated scheme also identifies the unwanted decay to $A^{\prime 2}\Delta_{3/2}$ as a leakage that needs to be fixed to improve the cycling scheme and correctly calculates its branching ratio from the state $A^2\Pi_{1/2}$ $|v=0\rangle$ as larger than that to the $X^2\Sigma^+$ $|v=3\rangle$ state, as shown in \cite{Collopy2015YO}.

Our automated search demonstrates that YO is an ideal molecule for laser cooling. In addition to the cycling scheme discussed above, we find over 1700 different cooling schemes based on about 140 different excited states that, unlike the other molecules discussed in this paper, have low cooling times ($\leq10$ ms from 4 K) and thus lend themselves to Doppler cooling. These schemes all use lasers with easily accessible frequencies between 450 and 1000 nm, except that a couple of the weakest repumping transitions would require going up to 5,500 nm. If we limit ourselves to schemes with $\leq 10$ repumping lasers, we find cycling schemes that make use of both states in the $A^2\Pi$ and the $B^2\Sigma$ electronic manifolds as excited states for cycling and can cool starting from $J$ states as high as $15/2$. In particular, we find a scheme based on a $B^2\Sigma$ excited state that requires only three lasers and achieves a closure of $p = 0.9992$ and $t_{\rm cool} = 615\,\mu$s. The newly found schemes employing states from the $B^2 \Sigma$ manifold as $S_1$ states are faster and have more accessible laser wavelengths than those using $S_1$ states from the $A^2 \Pi_{1/2}$ manifold, which are used in the experiment. Moreover, we identify several viable laser cooling schemes based on two $S_1$ states of the type as in Fig.~\ref{fig:schemes}(b), using the algorithm extension described in App.~\ref{app:algo_extension}. The fastest double-$S_1$ scheme is an extension of the scheme discussed above, based on a $B^2\Sigma$ excited state. Its second $S_1$ state is the lowest-energy state from another electronic manifold, $A^2 \Pi$, that decays to the same states $S_2$. Most excitingly, it offers a speedup to the single-$S_1$ cooling scheme at the expense of a larger number of lasers and worse closure. We present both schemes in more detail in App.~\ref{app-ss:YO}.

\subsection{\texorpdfstring{C\textsubscript{2}}{C2} and schemes based on high rotational states}

After YO, we consider the carbon dimer as a testing ground for our automated detection algorithm. In previous work, we manually identified laser cycling schemes for C$_2$ based on data from the ExoMol database \cite{Bigagli22carbon}.

Although it does not feature near-diagonal FC factors, C$_2$ has a favorable internal structure that supports laser cooling schemes, as discussed in Ref.~\cite{Bigagli22carbon}. In brief, C$_2$ is a tightly bound and light molecule with completely decoupled singlet and triplet manifolds. For the most abundant isotope, $^{12}$C$_2$, there is no nuclear spin and hence no hyperfine structure. Its ground electronic state, $X ^1\Sigma^+_g$, is radiatively connected to the excited state $A ^1\Pi_u$ via the Phillips band. At the same time, the metastable $a^3\Pi_u$ in the triplet manifold is radiatively connected to the excited $d^3\Pi_g$ and $b^3\Sigma_g^-$ states, generating the Swan and Ballik-Ramsay bands, respectively. Furthermore, $d^3\Pi_g \leftrightarrow c^3\Sigma_u^+$ transitions are observed to form the so-called Duck band. Our manual search identified a cycling scheme capable of Doppler cooling using 9 lasers in the Swan and Duck bands and two additional narrow-line cycling schemes using 8 and 6 lasers in the Phillips and Ballik-Ramsay bands, respectively. These schemes make use of low-lying vibrational and rotational quantum numbers, $v$ and $J$, in the respective electronic potentials.  

\begin{figure}[t]
    \centering
    \includegraphics[width=0.65\columnwidth]{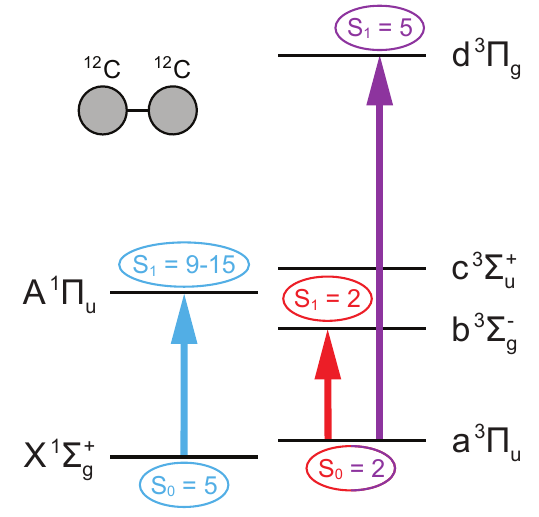}
    \caption{Electronic level diagram for C$_2$. The arrows represent possible first excitations, where $S_0$ is the number of initial states identified by the automated scheme and $S_1$ is the number of possible excited states for each of these initial states.}
    \label{fig:C2new}
\end{figure}

By applying our automated scheme, we find $\sim 60$ possible first excitations (from $S_0$ to $S_1$) that lead to cycling schemes with up to 15 lasers. These do not all represent independent cycling schemes but rather correspond to different choices for the first excitation within similar schemes. Figure~\ref{fig:C2new} shows a pictorial representation of the findings of the automated search. In the $X ^1\Sigma^+_g$ electronic ground state, there are five states that can serve as the initial state $S_0$, all in the $v=0$ vibrational state and with $J = 0, 2, 4, 6, 8$. For each of these, there exist between 9 and 15 possible choices for the first excitation. Among these schemes, we also retrieve the Phillips scheme that we found in the manual search~\cite{Bigagli22carbon}. The $a^3\Pi_u$ manifold has two possible $S_0$ states, $|v=0;J=1,\Omega=0,1\rangle$, which both couple to the same five states in the $d^3\Pi_{g}$ manifold or the same two states of the $b^3\Sigma_g^-$ manifold. Among these schemes are the Ballik-Ramsay and Swan-Duck schemes found in the manual search.

The results of the automated approach for C$_2$ are encouraging. All cycling schemes identified in the manual search of Ref.~\cite{Bigagli22carbon} have been retrieved by our algorithm. Moreover, while we confirm that the previous Swan-Duck scheme is the one that provides the fastest Doppler cooling, we see that it belongs to a set of two equivalent cycling schemes that use identical $S_2$ states and address either $d^3\Pi_{g}$ $|0;0,0\rangle$ or $d^3\Pi_{g}$ $|1;0,0\rangle$ as the excited state. This means that the laser wavelengths needed are the same relative to each other, with a shift of about 43 nm between them. The second most promising scheme identified is neither the Ballik-Ramsay nor the Phillips schemes from Ref.~\cite{Bigagli22carbon}. It is a five-laser scheme starting in $X ^1\Sigma_{g}^+$ $|0;2,0\rangle$ that uses $A^1\Pi_{u}$ $|2;1,1\rangle$ as the excited state and with $X ^1\Sigma_{g}^+$, $|v=0,1,2,3,4;2;0\rangle$ as the other lower cycling states (see Fig.~\ref{fig:C2new}). This scheme is notable as it uses both rotationally and vibrationally excited levels. We present it in more detail in App.~\ref{app-ss:C2}. The required cooling time from $T_\mathrm{init} = 4$ K of $t_\mathrm{{4K}} = 0.3$ s may not allow for direct cooling, but it could be an option for narrow-line cooling to a lower Doppler temperature. Moreover, the $J=2$ scheme belongs to a series of schemes relying on even higher rotational numbers, which we also present in App.~\ref{app-ss:C2}. The identification of new schemes, especially with large $J$ quantum numbers, showcases the power of the automated approach. 

\subsection{\texorpdfstring{OH$^+$}{OH+} and a scheme starting from an excited state}

The prospect of laser cooling the hydroxyl cation is enticing from an astrophysical and quantum information standpoint. The OH$^+$ levels and transitions are available in the ExoMol database, and it was also previously studied in a manual search \cite{Bigagli2023OH}.  Here, we use it as a third benchmark.

OH$^+$ has a relatively simple level structure, similar to C$_2$ in some respects. Its lowest-lying electronic states have triplet or singlet character, with a relatively weak (relative to vibrational spacings) spin-orbit coupling constant of 75 cm$^{-1}$ \cite{merer1975ultraviolet}. ExoMol tabulates states and transitions between the ground state, X$^3\Sigma^-$, and the excited $A^3\Pi$ state. Losses to other electronic states are minimal due to selection rules~\cite{Bigagli2023OH}.  Finally, since ion traps have storage times that are much longer than those of trapping techniques for neutral atoms or molecules, we can explore cooling times that are longer than a few milliseconds. Via a manual approach, we identified a laser cycling scheme that could cool OH$^+$ from cryogenic temperatures using 5 to 6 lasers and from room temperature using 8 lasers~\cite{Bigagli2023OH}. 

With our automatic search, we find 132 possible first excitations using up to 15 lasers. These can be broken down into 39 different schemes with unique excited states $S_1$ and different choices of initial states $S_0$ from the same pool of states. In particular, we retrieve the scheme we identified in Ref.~\cite{Bigagli2023OH}, which uses the states $X^3\Sigma^-$ $|v=0,N=0,J=1\rangle$ and $A^3\Pi$ $|0,1,0\rangle$ as $S_0$ and $S_1$, respectively. Interestingly, while it is among the quicker cooling schemes, it is neither the quickest nor the one with the smallest number of required lasers. The most promising one, with a cooling time of 37 ms from 21.5 K and using only 3 lasers, turns out to be a scheme using $X^3\Sigma^-$ $|0,1,1\rangle$ as $S_0$ and $A^3\Pi$ $|0,1,0\rangle$ as $S_1$. The wavelengths needed by this scheme are within the range of 357-450 nm, which are easily accessible with current technology. We present this scheme in detail in App.~\ref{app-ss:OH}. Finally, this scheme uses an excited $J$ quantum number as its launching point, further showing how our automated scheme can find promising laser cycling schemes that may be overlooked in a manual search. It also corroborates our claim that it can be worthwhile to start with a system at a higher temperature in order to access higher-energy states that may enable cooling schemes whose speed and efficiency can compensate for the higher initial temperature.

\subsection{CN and state selectivity}

\begin{figure}[t]
    \centering
    \includegraphics[width=\columnwidth]{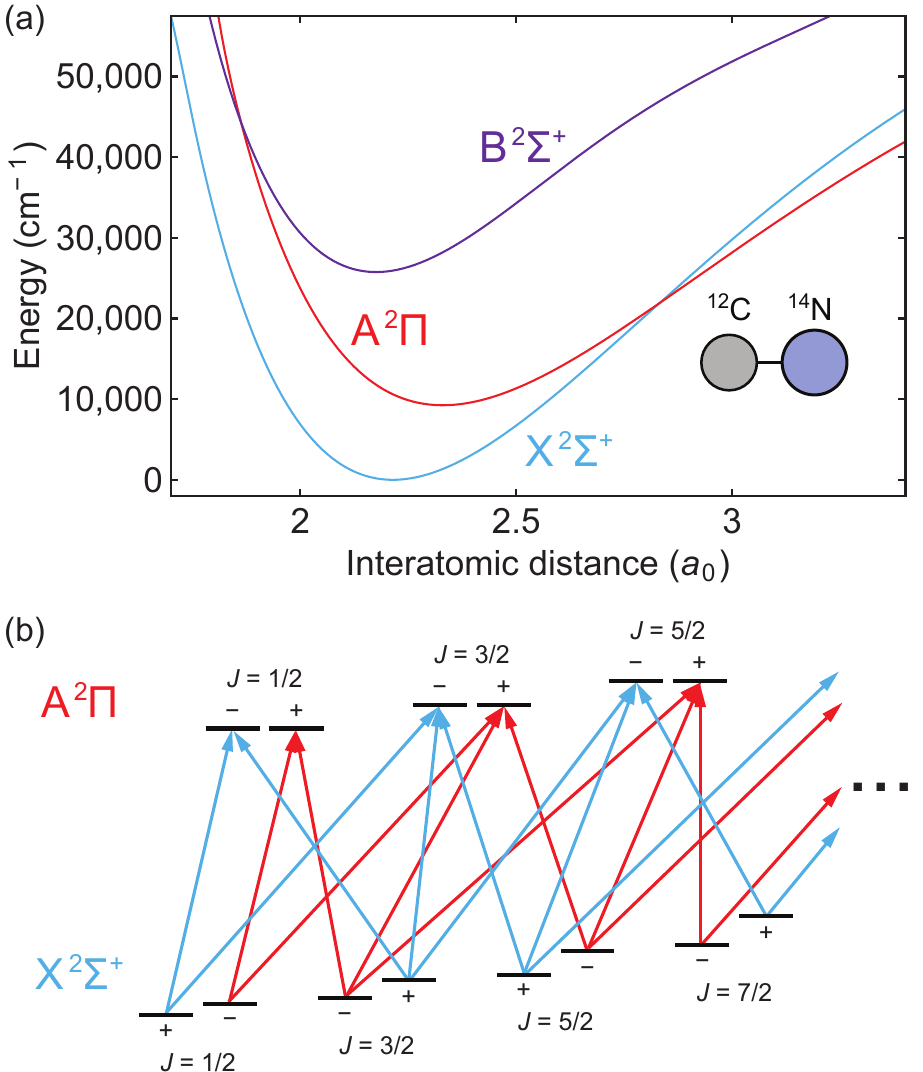}
    \caption{(a) Potential energy curves for the low-lying electronic states of CN \cite{Brooke2014CN}. (b) $X^2\Sigma^+ \leftrightarrow  A^2\Pi$ cycling schemes for CN. Each $J \leftrightarrow J, J\pm1$ cycling scheme is divided into two separate families depending on the $\pm$ symmetry of the initial state. Each state in the $A^2\Pi$ electronic manifold can have $v=0$ or $1$. States in the $X^2\Sigma^+$ electronic manifold can have $v = 0, \ldots, N$, where $N$ is arbitrary. The diagram is not to scale, and states with energy differences < 1 GHz are shown as degenerate. }
    \label{fig:CN}
\end{figure}

The fourth molecule to which we apply our automated approach is CN. The relevant potential energy curves are shown in Fig.~\ref{fig:CN}(a). The potential for laser cooling of CN has been previously explored in Ref.~\cite{zhang2018theoretical}. However, that analysis only accounted for electronic and vibrational states. It did not consider the rotational or hyperfine structure of the molecule, which is necessary to assess the viability of a laser cooling scheme. The authors proposed a vibrational cycling scheme between the $X^2\Sigma^+$ electronic manifold and the $B^2\Sigma^+$ manifold, which involves four vibrational levels in the ground-state electronic manifold ($v=0,1,2,3$) and two vibrational levels in the excited electronic manifold ($v=0,1$). Indeed, our graph-based method identifies similar cycling schemes with $v=0$ or $v=1$ as excited states.

In our automated search, we have accounted for additional spectroscopic details of CN that were not included in the study of Ref.~\cite{zhang2018theoretical}. The ExoMol database provides information on the rotational structure,  enabling a more complete understanding of possible cycling schemes. The main feature is that the $v=0 \leftrightarrow v = 0,1,2, \ldots, n$ scheme is really comprised of two classes of cycling schemes, depending on the original parity ($+/-$) of the $S_0$ state used. Given the $+\leftrightarrow-$ parity selection rules, if $S_0$ has $+(-)$ parity, then so will all $S_2$ states, while $S_1$ will have $-(+)$ parity. This creates two isolated cycling systems that can be exploited to achieve state preparation in a state of a specific parity. The two systems are separated by over 200 GHz in frequency, so different laser systems would be needed to address each. Focusing on the rotational structure, we discover that the cycling scheme mentioned above uses transitions $J_{S_1} = 1/2 \leftrightarrow J_{S_0,S_2} = 1/2, 3/2$. Finally, our algorithm identifies a fast cooling time for this scheme of 1~ms from 4 K.

Our study further highlights how the predominantly $X^2\Sigma^+ \leftrightarrow  B^2\Sigma^+$ scheme proposed in \cite{zhang2018theoretical} is the only one using $S_1$ states from the $B^2\Sigma^+$ manifold. Our search additionally reveals CN cycling schemes based on $X^2\Sigma^+ \leftrightarrow  A^2\Pi$ transitions. These schemes require longer cooling times from cryogenic temperatures (of the order of 100s or 1000s of milliseconds, compared to the cooling times of the order of 1~ms for the previous schemes), but are interesting due to their rotational structure. While these schemes exhibit the same $\pm$ parity and vibrational structure described above, different schemes are made up of a ladder of rotational states, with transitions $J_{S_0,S_2} = J_{S_1},  J_{S_1} \pm1 $. This may enable laser cooling of different rotational states for state-selective applications. These findings again highlight that the automated detection approach can find cycling schemes with higher quantum numbers, which tend to be overlooked by manual searches. Figure \ref{fig:CN}(b) schematically summarizes the findings of these $X^2\Sigma^+ \leftrightarrow  A^2\Pi$ cycling schemes. We present them in more detail in App.~\ref{app-ss:CN}.          

\subsection{\texorpdfstring{CO$_2$}{CO2} and three billion decay channels}

Finally, we go beyond diatomic molecules and apply our method to the more complex CO$_2$ molecule. Going from a diatomic to a triatomic molecule significantly increases the number of degrees of freedom and, hence, the number of quantum states and transitions that need to be taken into account. In particular, we analyze over 3.5 million states connected with over 2.5 billion decay channels.

One surprising find is that our automated scheme identifies from one to three orders of magnitude fewer cycling schemes (only four) than for the diatomic molecules that we discussed above. This is likely a consequence of the significantly increased number of decay channels from a given energy level. One of the schemes identified is quite promising, as it employs six lasers to achieve closure, starting from the absolute ground state as $S_0$. We present it in detail in App.~\ref{app-ss:CO2}. Although this scheme is theoretically viable due to the full closure of 1, it will not be able to cool CO$_2$ through a Doppler scheme on experimentally relevant times scales; due to its low scattering rates the cooling time would be about 6 minutes. Nevertheless, the presence of a closed cycling scheme may lend itself for narrow-line cooling techniques. 

The fact that we identified only a few slow cooling schemes for CO$_2$ helps illustrating that further improvements to our algorithm may reveal additional cooling schemes. The main limiting factor to our approach is that many possible $S_2$ states we identify have very short lifetimes and, therefore, repumping lasers cannot efficiently close certain cycling schemes. However, it is possible that these states are only intermediate decay channels to other metastable states that would allow repumping. We currently do not consider these \textit{indirect} repumping schemes, which we present schematically in Fig.~\ref{fig:schemes}(c), but an extension of our algorithm to this end may be extremely fruitful.

Furthermore, our analysis overestimates the cooling time of the schemes that it identifies, as already discussed in Sec.~\ref{ssec:limitations}. Relaxing some of our constraints may reveal schemes with experimentally relevant cooling time. The first assumption that leads to longer cooling times is the use of the mean initial velocity in Eq.~\eqref{eq:MaxwellBoltzmann} to calculate the momentum from which molecules are to be slowed down. In an actual experiment, that would mean that half of the Maxwell-Boltzmann distribution would be slowed to a halt. However, this is a very strict assumption that is not typically realized in molecular cooling experiments. In fact, it is quite common that only molecules in the low-velocity tail of the Maxwell-Boltzmann distribution are cooled, sufficient to create sizeable cold samples. Relaxing this condition can lead to a cooling time shorter by up to two orders of magnitude. A second assumption is that the occupancy of the starting state $S_0$ relative to the absolute ground state is $10\%$. This can easily be relaxed to $1\%$ or lower, allowing the use of lower starting temperatures and, therefore, lower initial momenta. This improvement scales as $\sqrt{T_{\rm init}}$ (e.g., Eq.~\ref{eq:ncool}). Relaxing the above two conditions is only advisable for molecules that can be produced in large quantities and with high-flux sources. For a molecule such as CO$_2$, which is easily obtained as a product of combustion, such relaxed conditions are likely appropriate, although a precise quantitative assessment is beyond the scope of this work. Improvements in the initial cooling in the form of better cryogenics would also shorten the cooling time. Finally, the use of coherent cooling schemes, such as bichromatic \cite{corder2015laser} or polychromatic \cite{wenz2020large} cooling, would not be limited by low scattering rates, thus bypassing the limitations on Doppler cooling that arise from low scattering rates.
    
\section{Conclusion}\label{sec:conclusion}

In this work, we have described and demonstrated a novel method to automatically identify laser cooling schemes for molecules. A key element of our approach is the change in representation, expressing the list of internal states and transition rates in the form of a graph. In our initial tests of the approach, we have identified thousands of laser cooling schemes for YO, C$_2$, OH$^{+}$, CN, and CO$_2$, all of which are molecules with non-diagonal FC factors. Many of the laser cooling schemes identified would have been difficult to find manually, not only due to the large number of energy levels and decay channels but also due to human biases, such as focusing on schemes that start from the absolute ground state, reliance on diagonal FC factors, or omission of higher rotational and vibrational states. In particular:

\begin{itemize}
    \item For every molecule that we analyzed in this work, we have recovered the known schemes and identified unknown ones.
    \item For the previously unknown schemes, higher rotational and vibrational branches were involved, enabling the production of ultracold molecules with a high rotational quantum number without additional state transfers. In the case of CN, they also allow for parity selection.
    \item For YO, we identify over 1700 cooling schemes based on about 140 different excited states with accessible laser wavelengths and fast cooling times.
    \item For OH$^{+}$, the automated search identified a laser cooling scheme that is faster and requires fewer lasers than the one found manually~\cite{Bigagli2023OH}. This scheme does not start from the absolute ground state, illustrating our claim that it can be worth starting the laser cooling process at a higher initial temperature to gain access to efficient cycling schemes at higher-energy internal states,
    \item For CO$_2$, we identified a scheme that may be difficult to use for Doppler cooling due to slow cooling time but may lend itself to narrow-line cooling.
\end{itemize}

We highlight that among the newly identified schemes, some start from highly excited states. This opens the prospect for laser cooling experiments that may not require a cryogenic buffer gas source, which could be a significant simplification over current approaches. We also note that our method can be directly applied to other molecules as well as to multi-electron atoms. Of course, the search quality ultimately relies on the completeness and accuracy of the spectral databases used; improved available spectroscopic data may lead to additional findings. Moreover, the search returns only schemes that meet predetermined conditions, including the range of possible laser wavelengths. As a broader range of laser wavelengths becomes available in laboratories, also additional laser cooling scheme may become viable.

We focused here on relatively simple laser cooling schemes involving only one excited state or two excited states decaying to the same lower-energy states. Despite this restriction, we have been able to identify many cooling schemes. We also showed in the example of YO that the automatically identified scheme involving two excited states $S_1$ \cite{hummon2013YO, Bigagli2023OH} offers a speedup to the fastest scheme with only one excited state. In future work, the automated search can be expanded to more exotic schemes, e.g., ones with multiple reachable excited states $S_1$ decaying to different lower-energy states or with multi-level decays. The main challenge that needs to be solved before incorporating such complex schemes into the graph search is of a theoretical nature. Currently, an efficient parallel search is not possible, as we lack a closed-form estimate of the cooling time in such cases. For example, adding Monte Carlo simulations of the cooling time on top of the search would make it computationally prohibitive. The search method can be further improved by taking into account losses caused by potential off-resonant excitation of nearby allowed transitions driven by repumping lasers and by tying a required lifetime of the starting state (instead of a rigid 1 $\mu$s) to the achievable coupling strength to the excited state.

We stress that -- due to limits in the accuracy or completeness of available spectroscopic data -- ultimately the feasibility of a cooling scheme can only be confirmed in an experiment for manual and automated searches alike. With the method discussed here, we identify cooling scheme candidates far more rapidly and extensively than in a manual search, which then makes it easier for experimentalists to quickly survey molecules of interest and confirm a candidate.

Finally, we point out that a variety of problems in quantum physics may benefit from a reformulation as a search on an appropriate graph. Focusing on molecules, the graph representation can be generally helpful to quickly identify allowed transitions, without the need for detailed knowledge of the sometimes complex selection rules. For example, this may help identifying new quantum control schemes and state pairs that can serve as molecular qubits \cite{Yelin2006molqubit}. Also, it may be possible to identify state preparation and pumping schemes for precision measurements using molecules, such as the measurement of the electron-to-proton mass ratio \cite{Zelevinsky2008massratio}. A graph-based search can also simplify the identification of transitions that are useful for stimulated Raman adiabatic passage (STIRAP) \cite{Vitanov2017stirap, Panda2016STIRAP} or other coherent state transfer techniques. Finally, the utility of reformulating quantum control problems in the form of a graph may well go beyond molecules. The basic concepts developed in this work - the graph formulation and the search algorithms operating on the graph - apply to any complex quantum system with a discrete excitation spectrum.

\begin{acknowledgments}
We thank Mirosław Łękowski-Dawid for directing us to the Neo4j database and Cypher framework. We also thank Géraud Krawezik and the Scientific Computing Center of the Flatiron Institute for their help in setting up the graph platform. We thank Sergey Yurchenko for his help with the ExoMol database. We also thank Lucy Reading-Ikkanda for the graphical design of Figures~1 and~3. We thank Peter F. Bernath for kindly providing the potential energy curves for CN. N.B., D.W.S., and S.W.~work acknowledge support through a Columbia Research Initiative in Science and Engineering (RISE) award. D.W.S. was additionally supported, in part, by the NASA Astrophysics Research and Analysis program under 80NSSC19K0969 and the NSF Division of Astronomical Sciences Astronomy and Astrophysics Grants program under AST-2002661. S.W. acknowledges additional support from the Alfred P.~Sloan Foundation.
The Flatiron Institute is a division of the Simons Foundation.
\end{acknowledgments}

\bibliography{literature}

\begin{thebibliography}{104}%
\makeatletter
\providecommand \@ifxundefined [1]{%
 \@ifx{#1\undefined}
}%
\providecommand \@ifnum [1]{%
 \ifnum #1\expandafter \@firstoftwo
 \else \expandafter \@secondoftwo
 \fi
}%
\providecommand \@ifx [1]{%
 \ifx #1\expandafter \@firstoftwo
 \else \expandafter \@secondoftwo
 \fi
}%
\providecommand \natexlab [1]{#1}%
\providecommand \enquote  [1]{``#1''}%
\providecommand \bibnamefont  [1]{#1}%
\providecommand \bibfnamefont [1]{#1}%
\providecommand \citenamefont [1]{#1}%
\providecommand \href@noop [0]{\@secondoftwo}%
\providecommand \href [0]{\begingroup \@sanitize@url \@href}%
\providecommand \@href[1]{\@@startlink{#1}\@@href}%
\providecommand \@@href[1]{\endgroup#1\@@endlink}%
\providecommand \@sanitize@url [0]{\catcode `\\12\catcode `\$12\catcode
  `\&12\catcode `\#12\catcode `\^12\catcode `\_12\catcode `\%12\relax}%
\providecommand \@@startlink[1]{}%
\providecommand \@@endlink[0]{}%
\providecommand \url  [0]{\begingroup\@sanitize@url \@url }%
\providecommand \@url [1]{\endgroup\@href {#1}{\urlprefix }}%
\providecommand \urlprefix  [0]{URL }%
\providecommand \Eprint [0]{\href }%
\providecommand \doibase [0]{https://doi.org/}%
\providecommand \selectlanguage [0]{\@gobble}%
\providecommand \bibinfo  [0]{\@secondoftwo}%
\providecommand \bibfield  [0]{\@secondoftwo}%
\providecommand \translation [1]{[#1]}%
\providecommand \BibitemOpen [0]{}%
\providecommand \bibitemStop [0]{}%
\providecommand \bibitemNoStop [0]{.\EOS\space}%
\providecommand \EOS [0]{\spacefactor3000\relax}%
\providecommand \BibitemShut  [1]{\csname bibitem#1\endcsname}%
\let\auto@bib@innerbib\@empty
\bibitem [{\citenamefont {Bloch}\ \emph {et~al.}(2008)\citenamefont {Bloch},
  \citenamefont {Dalibard},\ and\ \citenamefont {Zwerger}}]{Bloch2008RMP}%
  \BibitemOpen
  \bibfield  {author} {\bibinfo {author} {\bibfnamefont {I.}~\bibnamefont
  {Bloch}}, \bibinfo {author} {\bibfnamefont {J.}~\bibnamefont {Dalibard}},\
  and\ \bibinfo {author} {\bibfnamefont {W.}~\bibnamefont {Zwerger}},\
  }\bibfield  {title} {\bibinfo {title} {Many-body physics with ultracold
  gases},\ }\href {https://doi.org/10.1103/RevModPhys.80.885} {\bibfield
  {journal} {\bibinfo  {journal} {Rev. Mod. Phys.}\ }\textbf {\bibinfo {volume}
  {80}},\ \bibinfo {pages} {885} (\bibinfo {year} {2008})}\BibitemShut
  {NoStop}%
\bibitem [{\citenamefont {Lamata}\ \emph {et~al.}(2018)\citenamefont {Lamata},
  \citenamefont {Parra-Rodriguez}, \citenamefont {Sanz},\ and\ \citenamefont
  {Solano}}]{Lamata18AdvPhys}%
  \BibitemOpen
  \bibfield  {author} {\bibinfo {author} {\bibfnamefont {L.}~\bibnamefont
  {Lamata}}, \bibinfo {author} {\bibfnamefont {A.}~\bibnamefont
  {Parra-Rodriguez}}, \bibinfo {author} {\bibfnamefont {M.}~\bibnamefont
  {Sanz}},\ and\ \bibinfo {author} {\bibfnamefont {E.}~\bibnamefont {Solano}},\
  }\bibfield  {title} {\bibinfo {title} {Digital-analog quantum simulations
  with superconducting circuits},\ }\href
  {https://doi.org/10.1080/23746149.2018.1457981} {\bibfield  {journal}
  {\bibinfo  {journal} {Adv. Phys.: X}\ }\textbf {\bibinfo {volume} {3}},\
  \bibinfo {pages} {1457981} (\bibinfo {year} {2018})}\BibitemShut {NoStop}%
\bibitem [{\citenamefont {Tomza}\ \emph {et~al.}(2019)\citenamefont {Tomza},
  \citenamefont {Jachymski}, \citenamefont {Gerritsma}, \citenamefont
  {Negretti}, \citenamefont {Calarco}, \citenamefont {Idziaszek},\ and\
  \citenamefont {Julienne}}]{Tomza19RevModPhys}%
  \BibitemOpen
  \bibfield  {author} {\bibinfo {author} {\bibfnamefont {M.}~\bibnamefont
  {Tomza}}, \bibinfo {author} {\bibfnamefont {K.}~\bibnamefont {Jachymski}},
  \bibinfo {author} {\bibfnamefont {R.}~\bibnamefont {Gerritsma}}, \bibinfo
  {author} {\bibfnamefont {A.}~\bibnamefont {Negretti}}, \bibinfo {author}
  {\bibfnamefont {T.}~\bibnamefont {Calarco}}, \bibinfo {author} {\bibfnamefont
  {Z.}~\bibnamefont {Idziaszek}},\ and\ \bibinfo {author} {\bibfnamefont
  {P.~S.}\ \bibnamefont {Julienne}},\ }\bibfield  {title} {\bibinfo {title}
  {Cold hybrid ion-atom systems},\ }\href
  {https://doi.org/10.1103/RevModPhys.91.035001} {\bibfield  {journal}
  {\bibinfo  {journal} {Rev. Mod. Phys.}\ }\textbf {\bibinfo {volume} {91}},\
  \bibinfo {pages} {035001} (\bibinfo {year} {2019})}\BibitemShut {NoStop}%
\bibitem [{\citenamefont {Kjaergaard}\ \emph {et~al.}(2020)\citenamefont
  {Kjaergaard}, \citenamefont {Schwartz}, \citenamefont {Braum\"{u}ller},
  \citenamefont {Krantz}, \citenamefont {Wang}, \citenamefont {Gustavsson},\
  and\ \citenamefont {Oliver}}]{Kjaergaard20AnnuRev}%
  \BibitemOpen
  \bibfield  {author} {\bibinfo {author} {\bibfnamefont {M.}~\bibnamefont
  {Kjaergaard}}, \bibinfo {author} {\bibfnamefont {M.~E.}\ \bibnamefont
  {Schwartz}}, \bibinfo {author} {\bibfnamefont {J.}~\bibnamefont
  {Braum\"{u}ller}}, \bibinfo {author} {\bibfnamefont {P.}~\bibnamefont
  {Krantz}}, \bibinfo {author} {\bibfnamefont {J.~I.-J.}\ \bibnamefont {Wang}},
  \bibinfo {author} {\bibfnamefont {S.}~\bibnamefont {Gustavsson}},\ and\
  \bibinfo {author} {\bibfnamefont {W.~D.}\ \bibnamefont {Oliver}},\ }\bibfield
   {title} {\bibinfo {title} {Superconducting qubits: Current state of play},\
  }\href {https://doi.org/10.1146/annurev-conmatphys-031119-050605} {\bibfield
  {journal} {\bibinfo  {journal} {Annu. Rev. Condens. Matter Phys.}\ }\textbf
  {\bibinfo {volume} {11}},\ \bibinfo {pages} {369} (\bibinfo {year}
  {2020})}\BibitemShut {NoStop}%
\bibitem [{\citenamefont {Monroe}\ \emph {et~al.}(2021)\citenamefont {Monroe},
  \citenamefont {Campbell}, \citenamefont {Duan}, \citenamefont {Gong},
  \citenamefont {Gorshkov}, \citenamefont {Hess}, \citenamefont {Islam},
  \citenamefont {Kim}, \citenamefont {Linke}, \citenamefont {Pagano},
  \citenamefont {Richerme}, \citenamefont {Senko},\ and\ \citenamefont
  {Yao}}]{Monroe21RevModPhys}%
  \BibitemOpen
  \bibfield  {author} {\bibinfo {author} {\bibfnamefont {C.}~\bibnamefont
  {Monroe}}, \bibinfo {author} {\bibfnamefont {W.~C.}\ \bibnamefont
  {Campbell}}, \bibinfo {author} {\bibfnamefont {L.-M.}\ \bibnamefont {Duan}},
  \bibinfo {author} {\bibfnamefont {Z.-X.}\ \bibnamefont {Gong}}, \bibinfo
  {author} {\bibfnamefont {A.~V.}\ \bibnamefont {Gorshkov}}, \bibinfo {author}
  {\bibfnamefont {P.~W.}\ \bibnamefont {Hess}}, \bibinfo {author}
  {\bibfnamefont {R.}~\bibnamefont {Islam}}, \bibinfo {author} {\bibfnamefont
  {K.}~\bibnamefont {Kim}}, \bibinfo {author} {\bibfnamefont {N.~M.}\
  \bibnamefont {Linke}}, \bibinfo {author} {\bibfnamefont {G.}~\bibnamefont
  {Pagano}}, \bibinfo {author} {\bibfnamefont {P.}~\bibnamefont {Richerme}},
  \bibinfo {author} {\bibfnamefont {C.}~\bibnamefont {Senko}},\ and\ \bibinfo
  {author} {\bibfnamefont {N.~Y.}\ \bibnamefont {Yao}},\ }\bibfield  {title}
  {\bibinfo {title} {Programmable quantum simulations of spin systems with
  trapped ions},\ }\href {https://doi.org/10.1103/RevModPhys.93.025001}
  {\bibfield  {journal} {\bibinfo  {journal} {Rev. Mod. Phys.}\ }\textbf
  {\bibinfo {volume} {93}},\ \bibinfo {pages} {025001} (\bibinfo {year}
  {2021})}\BibitemShut {NoStop}%
\bibitem [{\citenamefont {Baranov}\ \emph {et~al.}(2012)\citenamefont
  {Baranov}, \citenamefont {Dalmonte}, \citenamefont {Pupillo},\ and\
  \citenamefont {Zoller}}]{baranov2012condensed}%
  \BibitemOpen
  \bibfield  {author} {\bibinfo {author} {\bibfnamefont {M.~A.}\ \bibnamefont
  {Baranov}}, \bibinfo {author} {\bibfnamefont {M.}~\bibnamefont {Dalmonte}},
  \bibinfo {author} {\bibfnamefont {G.}~\bibnamefont {Pupillo}},\ and\ \bibinfo
  {author} {\bibfnamefont {P.}~\bibnamefont {Zoller}},\ }\bibfield  {title}
  {\bibinfo {title} {Condensed matter theory of dipolar quantum gases},\ }\href
  {https://doi.org/10.1021/cr2003568} {\bibfield  {journal} {\bibinfo
  {journal} {Chem. Rev.}\ }\textbf {\bibinfo {volume} {112}},\ \bibinfo {pages}
  {5012} (\bibinfo {year} {2012})}\BibitemShut {NoStop}%
\bibitem [{\citenamefont {Jaksch}\ \emph {et~al.}(2000)\citenamefont {Jaksch},
  \citenamefont {Cirac}, \citenamefont {Zoller}, \citenamefont {Rolston},
  \citenamefont {C{\^o}t{\'e}},\ and\ \citenamefont {Lukin}}]{jaksch2000fast}%
  \BibitemOpen
  \bibfield  {author} {\bibinfo {author} {\bibfnamefont {D.}~\bibnamefont
  {Jaksch}}, \bibinfo {author} {\bibfnamefont {J.~I.}\ \bibnamefont {Cirac}},
  \bibinfo {author} {\bibfnamefont {P.}~\bibnamefont {Zoller}}, \bibinfo
  {author} {\bibfnamefont {S.~L.}\ \bibnamefont {Rolston}}, \bibinfo {author}
  {\bibfnamefont {R.}~\bibnamefont {C{\^o}t{\'e}}},\ and\ \bibinfo {author}
  {\bibfnamefont {M.~D.}\ \bibnamefont {Lukin}},\ }\bibfield  {title} {\bibinfo
  {title} {Fast quantum gates for neutral atoms},\ }\href
  {https://doi.org/10.1103/PhysRevLett.85.2208} {\bibfield  {journal} {\bibinfo
   {journal} {Phys. Rev. Lett.}\ }\textbf {\bibinfo {volume} {85}},\ \bibinfo
  {pages} {2208} (\bibinfo {year} {2000})}\BibitemShut {NoStop}%
\bibitem [{\citenamefont {DeMille}(2002)}]{demille2002quantum}%
  \BibitemOpen
  \bibfield  {author} {\bibinfo {author} {\bibfnamefont {D.}~\bibnamefont
  {DeMille}},\ }\bibfield  {title} {\bibinfo {title} {Quantum computation with
  trapped polar molecules},\ }\href
  {https://doi.org/10.1103/PhysRevLett.88.067901} {\bibfield  {journal}
  {\bibinfo  {journal} {Phys. Rev. Lett.}\ }\textbf {\bibinfo {volume} {88}},\
  \bibinfo {pages} {067901} (\bibinfo {year} {2002})}\BibitemShut {NoStop}%
\bibitem [{\citenamefont {Morgado}\ and\ \citenamefont
  {Whitlock}(2021)}]{morgado2021quantum}%
  \BibitemOpen
  \bibfield  {author} {\bibinfo {author} {\bibfnamefont {M.}~\bibnamefont
  {Morgado}}\ and\ \bibinfo {author} {\bibfnamefont {S.}~\bibnamefont
  {Whitlock}},\ }\bibfield  {title} {\bibinfo {title} {Quantum simulation and
  computing with {Rydberg}-interacting qubits},\ }\href
  {https://doi.org/10.1116/5.0036562} {\bibfield  {journal} {\bibinfo
  {journal} {AVS Quantum Sci.}\ }\textbf {\bibinfo {volume} {3}},\ \bibinfo
  {pages} {023501} (\bibinfo {year} {2021})}\BibitemShut {NoStop}%
\bibitem [{\citenamefont {Ludlow}\ \emph {et~al.}(2015)\citenamefont {Ludlow},
  \citenamefont {Boyd}, \citenamefont {Ye}, \citenamefont {Peik},\ and\
  \citenamefont {Schmidt}}]{ludlow2015optical}%
  \BibitemOpen
  \bibfield  {author} {\bibinfo {author} {\bibfnamefont {A.~D.}\ \bibnamefont
  {Ludlow}}, \bibinfo {author} {\bibfnamefont {M.~M.}\ \bibnamefont {Boyd}},
  \bibinfo {author} {\bibfnamefont {J.}~\bibnamefont {Ye}}, \bibinfo {author}
  {\bibfnamefont {E.}~\bibnamefont {Peik}},\ and\ \bibinfo {author}
  {\bibfnamefont {P.~O.}\ \bibnamefont {Schmidt}},\ }\bibfield  {title}
  {\bibinfo {title} {Optical atomic clocks},\ }\href
  {https://doi.org/10.1103/RevModPhys.87.637} {\bibfield  {journal} {\bibinfo
  {journal} {Rev. Mod. Phys.}\ }\textbf {\bibinfo {volume} {87}},\ \bibinfo
  {pages} {637} (\bibinfo {year} {2015})}\BibitemShut {NoStop}%
\bibitem [{\citenamefont {Bongs}\ \emph {et~al.}(2019)\citenamefont {Bongs},
  \citenamefont {Holynski}, \citenamefont {Vovrosh}, \citenamefont {Bouyer},
  \citenamefont {Condon}, \citenamefont {Rasel}, \citenamefont {Schubert},
  \citenamefont {Schleich},\ and\ \citenamefont {Roura}}]{bongs2019taking}%
  \BibitemOpen
  \bibfield  {author} {\bibinfo {author} {\bibfnamefont {K.}~\bibnamefont
  {Bongs}}, \bibinfo {author} {\bibfnamefont {M.}~\bibnamefont {Holynski}},
  \bibinfo {author} {\bibfnamefont {J.}~\bibnamefont {Vovrosh}}, \bibinfo
  {author} {\bibfnamefont {P.}~\bibnamefont {Bouyer}}, \bibinfo {author}
  {\bibfnamefont {G.}~\bibnamefont {Condon}}, \bibinfo {author} {\bibfnamefont
  {E.}~\bibnamefont {Rasel}}, \bibinfo {author} {\bibfnamefont
  {C.}~\bibnamefont {Schubert}}, \bibinfo {author} {\bibfnamefont {W.~P.}\
  \bibnamefont {Schleich}},\ and\ \bibinfo {author} {\bibfnamefont
  {A.}~\bibnamefont {Roura}},\ }\bibfield  {title} {\bibinfo {title} {Taking
  atom interferometric quantum sensors from the laboratory to real-world
  applications},\ }\href {https://doi.org/10.1038/s42254-019-0117-4} {\bibfield
   {journal} {\bibinfo  {journal} {Nat. Rev. Phys.}\ }\textbf {\bibinfo
  {volume} {1}},\ \bibinfo {pages} {731} (\bibinfo {year} {2019})}\BibitemShut
  {NoStop}%
\bibitem [{\citenamefont {Langen}\ \emph {et~al.}(2024)\citenamefont {Langen},
  \citenamefont {Valtolina}, \citenamefont {Wang},\ and\ \citenamefont
  {Ye}}]{Langen2023ultracoldreview}%
  \BibitemOpen
  \bibfield  {author} {\bibinfo {author} {\bibfnamefont {T.}~\bibnamefont
  {Langen}}, \bibinfo {author} {\bibfnamefont {G.}~\bibnamefont {Valtolina}},
  \bibinfo {author} {\bibfnamefont {D.}~\bibnamefont {Wang}},\ and\ \bibinfo
  {author} {\bibfnamefont {J.}~\bibnamefont {Ye}},\ }\bibfield  {title}
  {\bibinfo {title} {Quantum state manipulation and cooling of ultracold
  molecules},\ }\href {https://doi.org/10.1038/s41567-024-02423-1} {\bibfield
  {journal} {\bibinfo  {journal} {Nat. Phys.}\ }\textbf {\bibinfo {volume}
  {20}},\ \bibinfo {pages} {702–712} (\bibinfo {year} {2024})}\BibitemShut
  {NoStop}%
\bibitem [{\citenamefont {Ospelkaus}\ \emph {et~al.}(2010)\citenamefont
  {Ospelkaus}, \citenamefont {Ni}, \citenamefont {Wang}, \citenamefont
  {de~Miranda}, \citenamefont {Neyenhuis}, \citenamefont {Quemener},
  \citenamefont {Julienne}, \citenamefont {Bohn}, \citenamefont {Jin},\ and\
  \citenamefont {Ye}}]{Ospelkaus10}%
  \BibitemOpen
  \bibfield  {author} {\bibinfo {author} {\bibfnamefont {S.}~\bibnamefont
  {Ospelkaus}}, \bibinfo {author} {\bibfnamefont {K.-K.}\ \bibnamefont {Ni}},
  \bibinfo {author} {\bibfnamefont {D.}~\bibnamefont {Wang}}, \bibinfo {author}
  {\bibfnamefont {M.~H.~G.}\ \bibnamefont {de~Miranda}}, \bibinfo {author}
  {\bibfnamefont {B.}~\bibnamefont {Neyenhuis}}, \bibinfo {author}
  {\bibfnamefont {G.}~\bibnamefont {Quemener}}, \bibinfo {author}
  {\bibfnamefont {P.~S.}\ \bibnamefont {Julienne}}, \bibinfo {author}
  {\bibfnamefont {J.~L.}\ \bibnamefont {Bohn}}, \bibinfo {author}
  {\bibfnamefont {D.~S.}\ \bibnamefont {Jin}},\ and\ \bibinfo {author}
  {\bibfnamefont {J.}~\bibnamefont {Ye}},\ }\bibfield  {title} {\bibinfo
  {title} {Quantum-state controlled chemical reactions of ultracold
  potassium-rubidium molecules},\ }\href
  {https://doi.org/10.1126/science.1184121} {\bibfield  {journal} {\bibinfo
  {journal} {Science}\ }\textbf {\bibinfo {volume} {327}},\ \bibinfo {pages}
  {853} (\bibinfo {year} {2010})}\BibitemShut {NoStop}%
\bibitem [{\citenamefont {Ni}\ \emph {et~al.}(2010)\citenamefont {Ni},
  \citenamefont {Ospelkaus}, \citenamefont {Wang}, \citenamefont {Quéméner},
  \citenamefont {Neyenhuis}, \citenamefont {de~Miranda}, \citenamefont {Bohn},
  \citenamefont {Ye},\ and\ \citenamefont {Jin}}]{Ni10}%
  \BibitemOpen
  \bibfield  {author} {\bibinfo {author} {\bibfnamefont {K.-K.}\ \bibnamefont
  {Ni}}, \bibinfo {author} {\bibfnamefont {S.}~\bibnamefont {Ospelkaus}},
  \bibinfo {author} {\bibfnamefont {D.}~\bibnamefont {Wang}}, \bibinfo {author}
  {\bibfnamefont {G.}~\bibnamefont {Quéméner}}, \bibinfo {author}
  {\bibfnamefont {B.}~\bibnamefont {Neyenhuis}}, \bibinfo {author}
  {\bibfnamefont {M.~H.~G.}\ \bibnamefont {de~Miranda}}, \bibinfo {author}
  {\bibfnamefont {J.~L.}\ \bibnamefont {Bohn}}, \bibinfo {author}
  {\bibfnamefont {J.}~\bibnamefont {Ye}},\ and\ \bibinfo {author}
  {\bibfnamefont {D.~S.}\ \bibnamefont {Jin}},\ }\bibfield  {title} {\bibinfo
  {title} {Dipolar collisions of polar molecules in the quantum regime},\
  }\href {https://doi.org/10.1038/nature08953} {\bibfield  {journal} {\bibinfo
  {journal} {Nature}\ }\textbf {\bibinfo {volume} {464}},\ \bibinfo {pages}
  {1324} (\bibinfo {year} {2010})}\BibitemShut {NoStop}%
\bibitem [{\citenamefont {de~Miranda}\ \emph {et~al.}(2011)\citenamefont
  {de~Miranda}, \citenamefont {Chotia}, \citenamefont {Neyenhuis},
  \citenamefont {Wang}, \citenamefont {Quéméner}, \citenamefont {Ospelkaus},
  \citenamefont {Bohn}, \citenamefont {Ye},\ and\ \citenamefont
  {Jin}}]{Miranda11}%
  \BibitemOpen
  \bibfield  {author} {\bibinfo {author} {\bibfnamefont {M.~H.~G.}\
  \bibnamefont {de~Miranda}}, \bibinfo {author} {\bibfnamefont
  {A.}~\bibnamefont {Chotia}}, \bibinfo {author} {\bibfnamefont
  {B.}~\bibnamefont {Neyenhuis}}, \bibinfo {author} {\bibfnamefont
  {D.}~\bibnamefont {Wang}}, \bibinfo {author} {\bibfnamefont {G.}~\bibnamefont
  {Quéméner}}, \bibinfo {author} {\bibfnamefont {S.}~\bibnamefont
  {Ospelkaus}}, \bibinfo {author} {\bibfnamefont {J.~L.}\ \bibnamefont {Bohn}},
  \bibinfo {author} {\bibfnamefont {J.}~\bibnamefont {Ye}},\ and\ \bibinfo
  {author} {\bibfnamefont {D.~S.}\ \bibnamefont {Jin}},\ }\bibfield  {title}
  {\bibinfo {title} {Controlling the quantum stereodynamics of ultracold
  bimolecular reactions},\ }\href {https://doi.org/10.1038/nphys1939}
  {\bibfield  {journal} {\bibinfo  {journal} {Nat. Phys.}\ }\textbf {\bibinfo
  {volume} {7}},\ \bibinfo {pages} {502} (\bibinfo {year} {2011})}\BibitemShut
  {NoStop}%
\bibitem [{\citenamefont {Tomza}(2015)}]{TomzaPRL15}%
  \BibitemOpen
  \bibfield  {author} {\bibinfo {author} {\bibfnamefont {M.}~\bibnamefont
  {Tomza}},\ }\bibfield  {title} {\bibinfo {title} {Energetics and control of
  ultracold isotope-exchange reactions between heteronuclear dimers in external
  fields},\ }\href {https://doi.org/10.1103/PhysRevLett.115.063201} {\bibfield
  {journal} {\bibinfo  {journal} {Phys. Rev. Lett.}\ }\textbf {\bibinfo
  {volume} {115}},\ \bibinfo {pages} {063201} (\bibinfo {year}
  {2015})}\BibitemShut {NoStop}%
\bibitem [{\citenamefont {Klein}\ \emph {et~al.}(2016)\citenamefont {Klein},
  \citenamefont {Shagam}, \citenamefont {Skomorowski}, \citenamefont
  {Żuchowski}, \citenamefont {Pawlak}, \citenamefont {Janssen}, \citenamefont
  {Moiseyev}, \citenamefont {van~de Meerakker}, \citenamefont {van~der Avoird},
  \citenamefont {Koch},\ and\ \citenamefont {Narevicius}}]{Klein17}%
  \BibitemOpen
  \bibfield  {author} {\bibinfo {author} {\bibfnamefont {A.}~\bibnamefont
  {Klein}}, \bibinfo {author} {\bibfnamefont {Y.}~\bibnamefont {Shagam}},
  \bibinfo {author} {\bibfnamefont {W.}~\bibnamefont {Skomorowski}}, \bibinfo
  {author} {\bibfnamefont {P.~S.}\ \bibnamefont {Żuchowski}}, \bibinfo
  {author} {\bibfnamefont {M.}~\bibnamefont {Pawlak}}, \bibinfo {author}
  {\bibfnamefont {L.~M.~C.}\ \bibnamefont {Janssen}}, \bibinfo {author}
  {\bibfnamefont {N.}~\bibnamefont {Moiseyev}}, \bibinfo {author}
  {\bibfnamefont {S.~Y.~T.}\ \bibnamefont {van~de Meerakker}}, \bibinfo
  {author} {\bibfnamefont {A.}~\bibnamefont {van~der Avoird}}, \bibinfo
  {author} {\bibfnamefont {C.~P.}\ \bibnamefont {Koch}},\ and\ \bibinfo
  {author} {\bibfnamefont {E.}~\bibnamefont {Narevicius}},\ }\bibfield  {title}
  {\bibinfo {title} {Directly probing anisotropy in atom–molecule collisions
  through quantum scattering resonances},\ }\href
  {https://doi.org/10.1038/nphys3904} {\bibfield  {journal} {\bibinfo
  {journal} {Nat. Phys.}\ }\textbf {\bibinfo {volume} {13}},\ \bibinfo {pages}
  {35–38} (\bibinfo {year} {2016})}\BibitemShut {NoStop}%
\bibitem [{\citenamefont {Puri}\ \emph {et~al.}(2017)\citenamefont {Puri},
  \citenamefont {Mills}, \citenamefont {Schneider}, \citenamefont {Simbotin},
  \citenamefont {Montgomery}, \citenamefont {C{\^o}t{\'e}}, \citenamefont
  {Suits},\ and\ \citenamefont {Hudson}}]{Puri17}%
  \BibitemOpen
  \bibfield  {author} {\bibinfo {author} {\bibfnamefont {P.}~\bibnamefont
  {Puri}}, \bibinfo {author} {\bibfnamefont {M.}~\bibnamefont {Mills}},
  \bibinfo {author} {\bibfnamefont {C.}~\bibnamefont {Schneider}}, \bibinfo
  {author} {\bibfnamefont {I.}~\bibnamefont {Simbotin}}, \bibinfo {author}
  {\bibfnamefont {J.~A.}\ \bibnamefont {Montgomery}}, \bibinfo {author}
  {\bibfnamefont {R.}~\bibnamefont {C{\^o}t{\'e}}}, \bibinfo {author}
  {\bibfnamefont {A.~G.}\ \bibnamefont {Suits}},\ and\ \bibinfo {author}
  {\bibfnamefont {E.~R.}\ \bibnamefont {Hudson}},\ }\bibfield  {title}
  {\bibinfo {title} {Synthesis of mixed hypermetallic oxide {BaOCa}$^+$ from
  laser-cooled reagents in an atom-ion hybrid trap},\ }\href
  {https://doi.org/10.1126/science.aan4701} {\bibfield  {journal} {\bibinfo
  {journal} {Science}\ }\textbf {\bibinfo {volume} {357}},\ \bibinfo {pages}
  {1370} (\bibinfo {year} {2017})}\BibitemShut {NoStop}%
\bibitem [{\citenamefont {de~Jongh}\ \emph {et~al.}(2020)\citenamefont
  {de~Jongh}, \citenamefont {Besemer}, \citenamefont {Shuai}, \citenamefont
  {Karman}, \citenamefont {van~der Avoird}, \citenamefont {Groenenboom},\ and\
  \citenamefont {van~de Meerakker}}]{Jongh20}%
  \BibitemOpen
  \bibfield  {author} {\bibinfo {author} {\bibfnamefont {T.}~\bibnamefont
  {de~Jongh}}, \bibinfo {author} {\bibfnamefont {M.}~\bibnamefont {Besemer}},
  \bibinfo {author} {\bibfnamefont {Q.}~\bibnamefont {Shuai}}, \bibinfo
  {author} {\bibfnamefont {T.}~\bibnamefont {Karman}}, \bibinfo {author}
  {\bibfnamefont {A.}~\bibnamefont {van~der Avoird}}, \bibinfo {author}
  {\bibfnamefont {G.~C.}\ \bibnamefont {Groenenboom}},\ and\ \bibinfo {author}
  {\bibfnamefont {S.~Y.~T.}\ \bibnamefont {van~de Meerakker}},\ }\bibfield
  {title} {\bibinfo {title} {Imaging the onset of the resonance regime in
  low-energy no-he collisions},\ }\href
  {https://doi.org/10.1126/science.aba3990} {\bibfield  {journal} {\bibinfo
  {journal} {Science}\ }\textbf {\bibinfo {volume} {368}},\ \bibinfo {pages}
  {626} (\bibinfo {year} {2020})}\BibitemShut {NoStop}%
\bibitem [{\citenamefont {Liu}\ and\ \citenamefont
  {Ni}(2021)}]{Liu2021molreview}%
  \BibitemOpen
  \bibfield  {author} {\bibinfo {author} {\bibfnamefont {Y.}~\bibnamefont
  {Liu}}\ and\ \bibinfo {author} {\bibfnamefont {K.-K.}\ \bibnamefont {Ni}},\
  }\bibfield  {title} {\bibinfo {title} {Bimolecular chemistry in the ultracold
  regime},\ }\href {https://doi.org/10.1146/annurev-physchem-090419-043244}
  {\bibfield  {journal} {\bibinfo  {journal} {Annu. Rev. Phys. Chem.}\ }\textbf
  {\bibinfo {volume} {73}},\ \bibinfo {pages} {1} (\bibinfo {year}
  {2021})}\BibitemShut {NoStop}%
\bibitem [{\citenamefont {Krems}(2008)}]{Krems2008}%
  \BibitemOpen
  \bibfield  {author} {\bibinfo {author} {\bibfnamefont {R.~V.}\ \bibnamefont
  {Krems}},\ }\bibfield  {title} {\bibinfo {title} {{Cold controlled
  chemistry}},\ }\href {https://doi.org/10.1039/b802322k} {\bibfield  {journal}
  {\bibinfo  {journal} {Phys. Chem. Chem. Phys.}\ }\textbf {\bibinfo {volume}
  {10}},\ \bibinfo {pages} {4079} (\bibinfo {year} {2008})}\BibitemShut
  {NoStop}%
\bibitem [{\citenamefont {Lemeshko}\ \emph {et~al.}(2013)\citenamefont
  {Lemeshko}, \citenamefont {Krems}, \citenamefont {Doyle},\ and\ \citenamefont
  {Kais}}]{Lemeshko13}%
  \BibitemOpen
  \bibfield  {author} {\bibinfo {author} {\bibfnamefont {M.}~\bibnamefont
  {Lemeshko}}, \bibinfo {author} {\bibfnamefont {R.~V.}\ \bibnamefont {Krems}},
  \bibinfo {author} {\bibfnamefont {J.~M.}\ \bibnamefont {Doyle}},\ and\
  \bibinfo {author} {\bibfnamefont {S.}~\bibnamefont {Kais}},\ }\bibfield
  {title} {\bibinfo {title} {Manipulation of molecules with electromagnetic
  fields},\ }\href {https://doi.org/10.1080/00268976.2013.813595} {\bibfield
  {journal} {\bibinfo  {journal} {Mol. Phys.}\ }\textbf {\bibinfo {volume}
  {111}},\ \bibinfo {pages} {1648} (\bibinfo {year} {2013})}\BibitemShut
  {NoStop}%
\bibitem [{\citenamefont {Park}\ \emph {et~al.}(2023)\citenamefont {Park},
  \citenamefont {Lu}, \citenamefont {Jamison}, \citenamefont {Tscherbul},\ and\
  \citenamefont {Ketterle}}]{Park2023resonance}%
  \BibitemOpen
  \bibfield  {author} {\bibinfo {author} {\bibfnamefont {J.~J.}\ \bibnamefont
  {Park}}, \bibinfo {author} {\bibfnamefont {Y.-K.}\ \bibnamefont {Lu}},
  \bibinfo {author} {\bibfnamefont {A.~O.}\ \bibnamefont {Jamison}}, \bibinfo
  {author} {\bibfnamefont {T.~V.}\ \bibnamefont {Tscherbul}},\ and\ \bibinfo
  {author} {\bibfnamefont {W.}~\bibnamefont {Ketterle}},\ }\bibfield  {title}
  {\bibinfo {title} {A feshbach resonance in collisions between triplet
  ground-state molecules},\ }\href {https://doi.org/10.1038/s41586-022-05635-8}
  {\bibfield  {journal} {\bibinfo  {journal} {Nature}\ }\textbf {\bibinfo
  {volume} {614}},\ \bibinfo {pages} {54} (\bibinfo {year} {2023})}\BibitemShut
  {NoStop}%
\bibitem [{\citenamefont {Chen}\ \emph {et~al.}(2023)\citenamefont {Chen},
  \citenamefont {Schindewolf}, \citenamefont {Eppelt}, \citenamefont {Bause},
  \citenamefont {Duda}, \citenamefont {Biswas}, \citenamefont {Karman},
  \citenamefont {Hilker}, \citenamefont {Bloch},\ and\ \citenamefont
  {Luo}}]{Chen2023resonance}%
  \BibitemOpen
  \bibfield  {author} {\bibinfo {author} {\bibfnamefont {X.-Y.}\ \bibnamefont
  {Chen}}, \bibinfo {author} {\bibfnamefont {A.}~\bibnamefont {Schindewolf}},
  \bibinfo {author} {\bibfnamefont {S.}~\bibnamefont {Eppelt}}, \bibinfo
  {author} {\bibfnamefont {R.}~\bibnamefont {Bause}}, \bibinfo {author}
  {\bibfnamefont {M.}~\bibnamefont {Duda}}, \bibinfo {author} {\bibfnamefont
  {S.}~\bibnamefont {Biswas}}, \bibinfo {author} {\bibfnamefont
  {T.}~\bibnamefont {Karman}}, \bibinfo {author} {\bibfnamefont
  {T.}~\bibnamefont {Hilker}}, \bibinfo {author} {\bibfnamefont
  {I.}~\bibnamefont {Bloch}},\ and\ \bibinfo {author} {\bibfnamefont {X.-Y.}\
  \bibnamefont {Luo}},\ }\bibfield  {title} {\bibinfo {title} {Field-linked
  resonances of polar molecules},\ }\href
  {https://doi.org/10.1038/s41586-022-05651-8} {\bibfield  {journal} {\bibinfo
  {journal} {Nature}\ }\textbf {\bibinfo {volume} {614}},\ \bibinfo {pages}
  {59} (\bibinfo {year} {2023})}\BibitemShut {NoStop}%
\bibitem [{\citenamefont {Mitra}\ \emph {et~al.}(2022)\citenamefont {Mitra},
  \citenamefont {Leung},\ and\ \citenamefont {Zelevinsky}}]{mitra2022quantum}%
  \BibitemOpen
  \bibfield  {author} {\bibinfo {author} {\bibfnamefont {D.}~\bibnamefont
  {Mitra}}, \bibinfo {author} {\bibfnamefont {K.~H.}\ \bibnamefont {Leung}},\
  and\ \bibinfo {author} {\bibfnamefont {T.}~\bibnamefont {Zelevinsky}},\
  }\bibfield  {title} {\bibinfo {title} {Quantum control of molecules for
  fundamental physics},\ }\href {https://doi.org/10.1103/PhysRevA.105.040101}
  {\bibfield  {journal} {\bibinfo  {journal} {Phys. Rev. A}\ }\textbf {\bibinfo
  {volume} {105}},\ \bibinfo {pages} {040101} (\bibinfo {year}
  {2022})}\BibitemShut {NoStop}%
\bibitem [{\citenamefont {{ACME Collaboration}}(2018)}]{acme2018improved}%
  \BibitemOpen
  \bibfield  {author} {\bibinfo {author} {\bibnamefont {{ACME
  Collaboration}}},\ }\bibfield  {title} {\bibinfo {title} {Improved limit on
  the electric dipole moment of the electron},\ }\href
  {https://doi.org/10.1038/s41586-018-0599-8} {\bibfield  {journal} {\bibinfo
  {journal} {Nature}\ }\textbf {\bibinfo {volume} {562}},\ \bibinfo {pages}
  {355} (\bibinfo {year} {2018})}\BibitemShut {NoStop}%
\bibitem [{\citenamefont {Roussy}\ \emph {et~al.}(2023)\citenamefont {Roussy},
  \citenamefont {Caldwell}, \citenamefont {Wright}, \citenamefont {Cairncross},
  \citenamefont {Shagam}, \citenamefont {Ng}, \citenamefont {Schlossberger},
  \citenamefont {Park}, \citenamefont {Wang}, \citenamefont {Ye},\ and\
  \citenamefont {Cornell}}]{Roussy2023}%
  \BibitemOpen
  \bibfield  {author} {\bibinfo {author} {\bibfnamefont {T.~S.}\ \bibnamefont
  {Roussy}}, \bibinfo {author} {\bibfnamefont {L.}~\bibnamefont {Caldwell}},
  \bibinfo {author} {\bibfnamefont {T.}~\bibnamefont {Wright}}, \bibinfo
  {author} {\bibfnamefont {W.~B.}\ \bibnamefont {Cairncross}}, \bibinfo
  {author} {\bibfnamefont {Y.}~\bibnamefont {Shagam}}, \bibinfo {author}
  {\bibfnamefont {K.~B.}\ \bibnamefont {Ng}}, \bibinfo {author} {\bibfnamefont
  {N.}~\bibnamefont {Schlossberger}}, \bibinfo {author} {\bibfnamefont {S.~Y.}\
  \bibnamefont {Park}}, \bibinfo {author} {\bibfnamefont {A.}~\bibnamefont
  {Wang}}, \bibinfo {author} {\bibfnamefont {J.}~\bibnamefont {Ye}},\ and\
  \bibinfo {author} {\bibfnamefont {E.~A.}\ \bibnamefont {Cornell}},\
  }\bibfield  {title} {\bibinfo {title} {An improved bound on the electron's
  electric dipole moment},\ }\href {https://doi.org/10.1126/science.adg4084}
  {\bibfield  {journal} {\bibinfo  {journal} {Science}\ }\textbf {\bibinfo
  {volume} {381}},\ \bibinfo {pages} {46} (\bibinfo {year} {2023})}\BibitemShut
  {NoStop}%
\bibitem [{\citenamefont {Herbst}\ and\ \citenamefont
  {Millar}(2008)}]{Herbst2008}%
  \BibitemOpen
  \bibfield  {author} {\bibinfo {author} {\bibfnamefont {E.}~\bibnamefont
  {Herbst}}\ and\ \bibinfo {author} {\bibfnamefont {T.~J.}\ \bibnamefont
  {Millar}},\ }\bibinfo {title} {The chemistry of cold interstellar cloud
  cores},\ in\ \href {https://doi.org/10.1142/9781848162105_0001} {\emph
  {\bibinfo {booktitle} {Low Temperatures and Cold Molecules}}}\ (\bibinfo
  {publisher} {Imperial College Press},\ \bibinfo {year} {2008})\
  Chap.~\bibinfo {chapter} {1}, pp.\ \bibinfo {pages} {1--54}\BibitemShut
  {NoStop}%
\bibitem [{\citenamefont {Herbst}(2014)}]{Herbst2014}%
  \BibitemOpen
  \bibfield  {author} {\bibinfo {author} {\bibfnamefont {E.}~\bibnamefont
  {Herbst}},\ }\bibfield  {title} {\bibinfo {title} {Three milieux for
  interstellar chemistry: gas, dust, and ice},\ }\href
  {https://doi.org/10.1039/c3cp54065k} {\bibfield  {journal} {\bibinfo
  {journal} {Phys. Chem. Chem. Phys.}\ }\textbf {\bibinfo {volume} {16}},\
  \bibinfo {pages} {3344} (\bibinfo {year} {2014})}\BibitemShut {NoStop}%
\bibitem [{\citenamefont {Fitch}\ and\ \citenamefont
  {Tarbutt}(2021{\natexlab{a}})}]{Fitch2021chapter}%
  \BibitemOpen
  \bibfield  {author} {\bibinfo {author} {\bibfnamefont {N.}~\bibnamefont
  {Fitch}}\ and\ \bibinfo {author} {\bibfnamefont {M.}~\bibnamefont
  {Tarbutt}},\ }\bibfield  {title} {\bibinfo {title} {Laser-cooled molecules},\
  }in\ \href {https://doi.org/10.1016/bs.aamop.2021.04.003} {\emph {\bibinfo
  {booktitle} {Advances In Atomic, Molecular, and Optical Physics}}}\ (\bibinfo
   {publisher} {Elsevier},\ \bibinfo {year} {2021})\ pp.\ \bibinfo {pages}
  {157--262}\BibitemShut {NoStop}%
\bibitem [{\citenamefont {Shuman}\ \emph {et~al.}(2009)\citenamefont {Shuman},
  \citenamefont {Barry}, \citenamefont {Glenn},\ and\ \citenamefont
  {DeMille}}]{Shuman09PRL}%
  \BibitemOpen
  \bibfield  {author} {\bibinfo {author} {\bibfnamefont {E.~S.}\ \bibnamefont
  {Shuman}}, \bibinfo {author} {\bibfnamefont {J.~F.}\ \bibnamefont {Barry}},
  \bibinfo {author} {\bibfnamefont {D.~R.}\ \bibnamefont {Glenn}},\ and\
  \bibinfo {author} {\bibfnamefont {D.}~\bibnamefont {DeMille}},\ }\bibfield
  {title} {\bibinfo {title} {Radiative force from optical cycling on a diatomic
  molecule},\ }\href {https://doi.org/10.1103/PhysRevLett.103.223001}
  {\bibfield  {journal} {\bibinfo  {journal} {Phys. Rev. Lett.}\ }\textbf
  {\bibinfo {volume} {103}},\ \bibinfo {pages} {223001} (\bibinfo {year}
  {2009})}\BibitemShut {NoStop}%
\bibitem [{\citenamefont {Shuman}\ \emph {et~al.}(2010)\citenamefont {Shuman},
  \citenamefont {Barry},\ and\ \citenamefont {DeMille}}]{ShumanNature10}%
  \BibitemOpen
  \bibfield  {author} {\bibinfo {author} {\bibfnamefont {E.}~\bibnamefont
  {Shuman}}, \bibinfo {author} {\bibfnamefont {J.}~\bibnamefont {Barry}},\ and\
  \bibinfo {author} {\bibfnamefont {D.}~\bibnamefont {DeMille}},\ }\bibfield
  {title} {\bibinfo {title} {Laser cooling of a diatomic molecule},\ }\href
  {https://doi.org/10.1038/nature09443} {\bibfield  {journal} {\bibinfo
  {journal} {Nature}\ }\textbf {\bibinfo {volume} {467}},\ \bibinfo {pages}
  {820} (\bibinfo {year} {2010})}\BibitemShut {NoStop}%
\bibitem [{\citenamefont {Barry}\ \emph {et~al.}(2014)\citenamefont {Barry},
  \citenamefont {McCarron}, \citenamefont {Norrgard}, \citenamefont
  {Steinecker},\ and\ \citenamefont {DeMille}}]{barry2014magneto}%
  \BibitemOpen
  \bibfield  {author} {\bibinfo {author} {\bibfnamefont {J.}~\bibnamefont
  {Barry}}, \bibinfo {author} {\bibfnamefont {D.}~\bibnamefont {McCarron}},
  \bibinfo {author} {\bibfnamefont {E.}~\bibnamefont {Norrgard}}, \bibinfo
  {author} {\bibfnamefont {M.}~\bibnamefont {Steinecker}},\ and\ \bibinfo
  {author} {\bibfnamefont {D.}~\bibnamefont {DeMille}},\ }\bibfield  {title}
  {\bibinfo {title} {Magneto-optical trapping of a diatomic molecule},\ }\href
  {https://doi.org/10.1038/nature13634} {\bibfield  {journal} {\bibinfo
  {journal} {Nature}\ }\textbf {\bibinfo {volume} {512}},\ \bibinfo {pages}
  {286} (\bibinfo {year} {2014})}\BibitemShut {NoStop}%
\bibitem [{\citenamefont {Zhelyazkova}\ \emph {et~al.}(2014)\citenamefont
  {Zhelyazkova}, \citenamefont {Cournol}, \citenamefont {Wall}, \citenamefont
  {Matsushima}, \citenamefont {Hudson}, \citenamefont {Hinds}, \citenamefont
  {Tarbutt},\ and\ \citenamefont {Sauer}}]{zhelyazkova2014laser}%
  \BibitemOpen
  \bibfield  {author} {\bibinfo {author} {\bibfnamefont {V.}~\bibnamefont
  {Zhelyazkova}}, \bibinfo {author} {\bibfnamefont {A.}~\bibnamefont
  {Cournol}}, \bibinfo {author} {\bibfnamefont {T.~E.}\ \bibnamefont {Wall}},
  \bibinfo {author} {\bibfnamefont {A.}~\bibnamefont {Matsushima}}, \bibinfo
  {author} {\bibfnamefont {J.~J.}\ \bibnamefont {Hudson}}, \bibinfo {author}
  {\bibfnamefont {E.~A.}\ \bibnamefont {Hinds}}, \bibinfo {author}
  {\bibfnamefont {M.~R.}\ \bibnamefont {Tarbutt}},\ and\ \bibinfo {author}
  {\bibfnamefont {B.~E.}\ \bibnamefont {Sauer}},\ }\bibfield  {title} {\bibinfo
  {title} {Laser cooling and slowing of {CaF} molecules},\ }\href
  {https://doi.org/10.1103/PhysRevA.89.053416} {\bibfield  {journal} {\bibinfo
  {journal} {Phys. Rev. A}\ }\textbf {\bibinfo {volume} {89}},\ \bibinfo
  {pages} {053416} (\bibinfo {year} {2014})}\BibitemShut {NoStop}%
\bibitem [{\citenamefont {Hemmerling}\ \emph {et~al.}(2016)\citenamefont
  {Hemmerling}, \citenamefont {Chae}, \citenamefont {Ravi}, \citenamefont
  {Anderegg}, \citenamefont {Drayna}, \citenamefont {Hutzler}, \citenamefont
  {Collopy}, \citenamefont {Ye}, \citenamefont {Ketterle},\ and\ \citenamefont
  {Doyle}}]{hemmerling2016laser}%
  \BibitemOpen
  \bibfield  {author} {\bibinfo {author} {\bibfnamefont {B.}~\bibnamefont
  {Hemmerling}}, \bibinfo {author} {\bibfnamefont {E.}~\bibnamefont {Chae}},
  \bibinfo {author} {\bibfnamefont {A.}~\bibnamefont {Ravi}}, \bibinfo {author}
  {\bibfnamefont {L.}~\bibnamefont {Anderegg}}, \bibinfo {author}
  {\bibfnamefont {G.~K.}\ \bibnamefont {Drayna}}, \bibinfo {author}
  {\bibfnamefont {N.~R.}\ \bibnamefont {Hutzler}}, \bibinfo {author}
  {\bibfnamefont {A.~L.}\ \bibnamefont {Collopy}}, \bibinfo {author}
  {\bibfnamefont {J.}~\bibnamefont {Ye}}, \bibinfo {author} {\bibfnamefont
  {W.}~\bibnamefont {Ketterle}},\ and\ \bibinfo {author} {\bibfnamefont
  {J.~M.}\ \bibnamefont {Doyle}},\ }\bibfield  {title} {\bibinfo {title} {Laser
  slowing of {CaF} molecules to near the capture velocity of a molecular
  {MOT}},\ }\href {https://doi.org/10.1088/0953-4075/49/17/174001} {\bibfield
  {journal} {\bibinfo  {journal} {J. Phys. B: At. Mol. Opt. Phys.}\ }\textbf
  {\bibinfo {volume} {49}},\ \bibinfo {pages} {174001} (\bibinfo {year}
  {2016})}\BibitemShut {NoStop}%
\bibitem [{\citenamefont {Anderegg}\ \emph {et~al.}(2017)\citenamefont
  {Anderegg}, \citenamefont {Augenbraun}, \citenamefont {Chae}, \citenamefont
  {Hemmerling}, \citenamefont {Hutzler}, \citenamefont {Ravi}, \citenamefont
  {Collopy}, \citenamefont {Ye}, \citenamefont {Ketterle},\ and\ \citenamefont
  {Doyle}}]{anderegg2017radio}%
  \BibitemOpen
  \bibfield  {author} {\bibinfo {author} {\bibfnamefont {L.}~\bibnamefont
  {Anderegg}}, \bibinfo {author} {\bibfnamefont {B.~L.}\ \bibnamefont
  {Augenbraun}}, \bibinfo {author} {\bibfnamefont {E.}~\bibnamefont {Chae}},
  \bibinfo {author} {\bibfnamefont {B.}~\bibnamefont {Hemmerling}}, \bibinfo
  {author} {\bibfnamefont {N.~R.}\ \bibnamefont {Hutzler}}, \bibinfo {author}
  {\bibfnamefont {A.}~\bibnamefont {Ravi}}, \bibinfo {author} {\bibfnamefont
  {A.}~\bibnamefont {Collopy}}, \bibinfo {author} {\bibfnamefont
  {J.}~\bibnamefont {Ye}}, \bibinfo {author} {\bibfnamefont {W.}~\bibnamefont
  {Ketterle}},\ and\ \bibinfo {author} {\bibfnamefont {J.~M.}\ \bibnamefont
  {Doyle}},\ }\bibfield  {title} {\bibinfo {title} {Radio frequency
  magneto-optical trapping of {CaF} with high density},\ }\href
  {https://doi.org/10.1103/PhysRevLett.119.103201} {\bibfield  {journal}
  {\bibinfo  {journal} {Phys. Rev. Lett.}\ }\textbf {\bibinfo {volume} {119}},\
  \bibinfo {pages} {103201} (\bibinfo {year} {2017})}\BibitemShut {NoStop}%
\bibitem [{\citenamefont {Lim}\ \emph {et~al.}(2018)\citenamefont {Lim},
  \citenamefont {Almond}, \citenamefont {Trigatzis}, \citenamefont {Devlin},
  \citenamefont {Fitch}, \citenamefont {Sauer}, \citenamefont {Tarbutt},\ and\
  \citenamefont {Hinds}}]{lim2018laser}%
  \BibitemOpen
  \bibfield  {author} {\bibinfo {author} {\bibfnamefont {J.}~\bibnamefont
  {Lim}}, \bibinfo {author} {\bibfnamefont {J.~R.}\ \bibnamefont {Almond}},
  \bibinfo {author} {\bibfnamefont {M.~A.}\ \bibnamefont {Trigatzis}}, \bibinfo
  {author} {\bibfnamefont {J.~A.}\ \bibnamefont {Devlin}}, \bibinfo {author}
  {\bibfnamefont {N.~J.}\ \bibnamefont {Fitch}}, \bibinfo {author}
  {\bibfnamefont {B.~E.}\ \bibnamefont {Sauer}}, \bibinfo {author}
  {\bibfnamefont {M.~R.}\ \bibnamefont {Tarbutt}},\ and\ \bibinfo {author}
  {\bibfnamefont {E.~A.}\ \bibnamefont {Hinds}},\ }\bibfield  {title} {\bibinfo
  {title} {Laser cooled {YbF} molecules for measuring the electron’s electric
  dipole moment},\ }\href {https://doi.org/10.1103/PhysRevLett.120.123201}
  {\bibfield  {journal} {\bibinfo  {journal} {Phys. Rev. Lett.}\ }\textbf
  {\bibinfo {volume} {120}},\ \bibinfo {pages} {123201} (\bibinfo {year}
  {2018})}\BibitemShut {NoStop}%
\bibitem [{\citenamefont {McNally}\ \emph {et~al.}(2020)\citenamefont
  {McNally}, \citenamefont {Kozyryev}, \citenamefont {Vazquez-Carson},
  \citenamefont {Wenz}, \citenamefont {Wang},\ and\ \citenamefont
  {Zelevinsky}}]{mcnally2020optical}%
  \BibitemOpen
  \bibfield  {author} {\bibinfo {author} {\bibfnamefont {R.~L.}\ \bibnamefont
  {McNally}}, \bibinfo {author} {\bibfnamefont {I.}~\bibnamefont {Kozyryev}},
  \bibinfo {author} {\bibfnamefont {S.}~\bibnamefont {Vazquez-Carson}},
  \bibinfo {author} {\bibfnamefont {K.}~\bibnamefont {Wenz}}, \bibinfo {author}
  {\bibfnamefont {T.}~\bibnamefont {Wang}},\ and\ \bibinfo {author}
  {\bibfnamefont {T.}~\bibnamefont {Zelevinsky}},\ }\bibfield  {title}
  {\bibinfo {title} {Optical cycling, radiative deflection and laser cooling of
  barium monohydride (${}^{138}${Ba}${}^1${H})},\ }\href
  {https://doi.org/10.1088/1367-2630/aba3e9} {\bibfield  {journal} {\bibinfo
  {journal} {New J. Phys.}\ }\textbf {\bibinfo {volume} {22}},\ \bibinfo
  {pages} {083047} (\bibinfo {year} {2020})}\BibitemShut {NoStop}%
\bibitem [{\citenamefont {Baum}\ \emph {et~al.}(2020)\citenamefont {Baum},
  \citenamefont {Vilas}, \citenamefont {Hallas}, \citenamefont {Augenbraun},
  \citenamefont {Raval}, \citenamefont {Mitra},\ and\ \citenamefont
  {Doyle}}]{baum20201d}%
  \BibitemOpen
  \bibfield  {author} {\bibinfo {author} {\bibfnamefont {L.}~\bibnamefont
  {Baum}}, \bibinfo {author} {\bibfnamefont {N.~B.}\ \bibnamefont {Vilas}},
  \bibinfo {author} {\bibfnamefont {C.}~\bibnamefont {Hallas}}, \bibinfo
  {author} {\bibfnamefont {B.~L.}\ \bibnamefont {Augenbraun}}, \bibinfo
  {author} {\bibfnamefont {S.}~\bibnamefont {Raval}}, \bibinfo {author}
  {\bibfnamefont {D.}~\bibnamefont {Mitra}},\ and\ \bibinfo {author}
  {\bibfnamefont {J.~M.}\ \bibnamefont {Doyle}},\ }\bibfield  {title} {\bibinfo
  {title} {{1D} magneto-optical trap of polyatomic molecules},\ }\href
  {https://doi.org/10.1103/PhysRevLett.124.133201} {\bibfield  {journal}
  {\bibinfo  {journal} {Phys. Rev. Lett.}\ }\textbf {\bibinfo {volume} {124}},\
  \bibinfo {pages} {133201} (\bibinfo {year} {2020})}\BibitemShut {NoStop}%
\bibitem [{\citenamefont {Mitra}\ \emph {et~al.}(2020)\citenamefont {Mitra},
  \citenamefont {Vilas}, \citenamefont {Hallas}, \citenamefont {Anderegg},
  \citenamefont {Augenbraun}, \citenamefont {Baum}, \citenamefont {Miller},
  \citenamefont {Raval},\ and\ \citenamefont {Doyle}}]{mitra2020direct}%
  \BibitemOpen
  \bibfield  {author} {\bibinfo {author} {\bibfnamefont {D.}~\bibnamefont
  {Mitra}}, \bibinfo {author} {\bibfnamefont {N.~B.}\ \bibnamefont {Vilas}},
  \bibinfo {author} {\bibfnamefont {C.}~\bibnamefont {Hallas}}, \bibinfo
  {author} {\bibfnamefont {L.}~\bibnamefont {Anderegg}}, \bibinfo {author}
  {\bibfnamefont {B.~L.}\ \bibnamefont {Augenbraun}}, \bibinfo {author}
  {\bibfnamefont {L.}~\bibnamefont {Baum}}, \bibinfo {author} {\bibfnamefont
  {C.}~\bibnamefont {Miller}}, \bibinfo {author} {\bibfnamefont
  {S.}~\bibnamefont {Raval}},\ and\ \bibinfo {author} {\bibfnamefont {J.~M.}\
  \bibnamefont {Doyle}},\ }\bibfield  {title} {\bibinfo {title} {Direct laser
  cooling of a symmetric top molecule},\ }\href
  {https://doi.org/10.1126/science.abc5357} {\bibfield  {journal} {\bibinfo
  {journal} {Science}\ }\textbf {\bibinfo {volume} {369}},\ \bibinfo {pages}
  {1366} (\bibinfo {year} {2020})}\BibitemShut {NoStop}%
\bibitem [{\citenamefont {Hummon}\ \emph {et~al.}(2013)\citenamefont {Hummon},
  \citenamefont {Yeo}, \citenamefont {Stuhl}, \citenamefont {Collopy},
  \citenamefont {Xia},\ and\ \citenamefont {Ye}}]{hummon2013YO}%
  \BibitemOpen
  \bibfield  {author} {\bibinfo {author} {\bibfnamefont {M.~T.}\ \bibnamefont
  {Hummon}}, \bibinfo {author} {\bibfnamefont {M.}~\bibnamefont {Yeo}},
  \bibinfo {author} {\bibfnamefont {B.~K.}\ \bibnamefont {Stuhl}}, \bibinfo
  {author} {\bibfnamefont {A.~L.}\ \bibnamefont {Collopy}}, \bibinfo {author}
  {\bibfnamefont {Y.}~\bibnamefont {Xia}},\ and\ \bibinfo {author}
  {\bibfnamefont {J.}~\bibnamefont {Ye}},\ }\bibfield  {title} {\bibinfo
  {title} {{2D} magneto-optical trapping of diatomic molecules},\ }\href
  {https://doi.org/10.1103/PhysRevLett.110.143001} {\bibfield  {journal}
  {\bibinfo  {journal} {Phys. Rev. Lett.}\ }\textbf {\bibinfo {volume} {110}},\
  \bibinfo {pages} {143001} (\bibinfo {year} {2013})}\BibitemShut {NoStop}%
\bibitem [{\citenamefont {Yeo}\ \emph {et~al.}(2015)\citenamefont {Yeo},
  \citenamefont {Hummon}, \citenamefont {Collopy}, \citenamefont {Yan},
  \citenamefont {Hemmerling}, \citenamefont {Chae}, \citenamefont {Doyle},\
  and\ \citenamefont {Ye}}]{yeo2015rotational}%
  \BibitemOpen
  \bibfield  {author} {\bibinfo {author} {\bibfnamefont {M.}~\bibnamefont
  {Yeo}}, \bibinfo {author} {\bibfnamefont {M.~T.}\ \bibnamefont {Hummon}},
  \bibinfo {author} {\bibfnamefont {A.~L.}\ \bibnamefont {Collopy}}, \bibinfo
  {author} {\bibfnamefont {B.}~\bibnamefont {Yan}}, \bibinfo {author}
  {\bibfnamefont {B.}~\bibnamefont {Hemmerling}}, \bibinfo {author}
  {\bibfnamefont {E.}~\bibnamefont {Chae}}, \bibinfo {author} {\bibfnamefont
  {J.~M.}\ \bibnamefont {Doyle}},\ and\ \bibinfo {author} {\bibfnamefont
  {J.}~\bibnamefont {Ye}},\ }\bibfield  {title} {\bibinfo {title} {Rotational
  state microwave mixing for laser cooling of complex diatomic molecules},\
  }\href {https://doi.org/10.1103/PhysRevLett.114.223003} {\bibfield  {journal}
  {\bibinfo  {journal} {Phys. Rev. Lett.}\ }\textbf {\bibinfo {volume} {114}},\
  \bibinfo {pages} {223003} (\bibinfo {year} {2015})}\BibitemShut {NoStop}%
\bibitem [{\citenamefont {Kozyryev}\ \emph {et~al.}(2017)\citenamefont
  {Kozyryev}, \citenamefont {Baum}, \citenamefont {Matsuda}, \citenamefont
  {Augenbraun}, \citenamefont {Anderegg}, \citenamefont {Sedlack},\ and\
  \citenamefont {Doyle}}]{kozyryev2017sisyphus}%
  \BibitemOpen
  \bibfield  {author} {\bibinfo {author} {\bibfnamefont {I.}~\bibnamefont
  {Kozyryev}}, \bibinfo {author} {\bibfnamefont {L.}~\bibnamefont {Baum}},
  \bibinfo {author} {\bibfnamefont {K.}~\bibnamefont {Matsuda}}, \bibinfo
  {author} {\bibfnamefont {B.~L.}\ \bibnamefont {Augenbraun}}, \bibinfo
  {author} {\bibfnamefont {L.}~\bibnamefont {Anderegg}}, \bibinfo {author}
  {\bibfnamefont {A.~P.}\ \bibnamefont {Sedlack}},\ and\ \bibinfo {author}
  {\bibfnamefont {J.~M.}\ \bibnamefont {Doyle}},\ }\bibfield  {title} {\bibinfo
  {title} {Sisyphus laser cooling of a polyatomic molecule},\ }\href
  {https://doi.org/10.1103/PhysRevLett.118.173201} {\bibfield  {journal}
  {\bibinfo  {journal} {Phys. Rev. Lett.}\ }\textbf {\bibinfo {volume} {118}},\
  \bibinfo {pages} {173201} (\bibinfo {year} {2017})}\BibitemShut {NoStop}%
\bibitem [{\citenamefont {Augenbraun}\ \emph {et~al.}(2020)\citenamefont
  {Augenbraun}, \citenamefont {Lasner}, \citenamefont {Frenett}, \citenamefont
  {Sawaoka}, \citenamefont {Miller}, \citenamefont {Steimle},\ and\
  \citenamefont {Doyle}}]{augenbraun2020laser}%
  \BibitemOpen
  \bibfield  {author} {\bibinfo {author} {\bibfnamefont {B.~L.}\ \bibnamefont
  {Augenbraun}}, \bibinfo {author} {\bibfnamefont {Z.~D.}\ \bibnamefont
  {Lasner}}, \bibinfo {author} {\bibfnamefont {A.}~\bibnamefont {Frenett}},
  \bibinfo {author} {\bibfnamefont {H.}~\bibnamefont {Sawaoka}}, \bibinfo
  {author} {\bibfnamefont {C.}~\bibnamefont {Miller}}, \bibinfo {author}
  {\bibfnamefont {T.~C.}\ \bibnamefont {Steimle}},\ and\ \bibinfo {author}
  {\bibfnamefont {J.~M.}\ \bibnamefont {Doyle}},\ }\bibfield  {title} {\bibinfo
  {title} {Laser-cooled polyatomic molecules for improved electron electric
  dipole moment searches},\ }\href {https://doi.org/10.1088/1367-2630/ab687b}
  {\bibfield  {journal} {\bibinfo  {journal} {New J. Phys.}\ }\textbf {\bibinfo
  {volume} {22}},\ \bibinfo {pages} {022003} (\bibinfo {year}
  {2020})}\BibitemShut {NoStop}%
\bibitem [{\citenamefont {Danzl}\ \emph {et~al.}(2008)\citenamefont {Danzl},
  \citenamefont {Haller}, \citenamefont {Gustavsson}, \citenamefont {Mark},
  \citenamefont {Hart}, \citenamefont {Bouloufa}, \citenamefont {Dulieu},
  \citenamefont {Ritsch},\ and\ \citenamefont {N\"agerl}}]{Danzl08Science}%
  \BibitemOpen
  \bibfield  {author} {\bibinfo {author} {\bibfnamefont {J.~G.}\ \bibnamefont
  {Danzl}}, \bibinfo {author} {\bibfnamefont {E.}~\bibnamefont {Haller}},
  \bibinfo {author} {\bibfnamefont {M.}~\bibnamefont {Gustavsson}}, \bibinfo
  {author} {\bibfnamefont {M.~J.}\ \bibnamefont {Mark}}, \bibinfo {author}
  {\bibfnamefont {R.}~\bibnamefont {Hart}}, \bibinfo {author} {\bibfnamefont
  {N.}~\bibnamefont {Bouloufa}}, \bibinfo {author} {\bibfnamefont
  {O.}~\bibnamefont {Dulieu}}, \bibinfo {author} {\bibfnamefont
  {H.}~\bibnamefont {Ritsch}},\ and\ \bibinfo {author} {\bibfnamefont {H.-C.}\
  \bibnamefont {N\"agerl}},\ }\bibfield  {title} {\bibinfo {title} {Quantum gas
  of deeply bound ground state molecules},\ }\href
  {https://doi.org/10.1126/science.1159909} {\bibfield  {journal} {\bibinfo
  {journal} {Science}\ }\textbf {\bibinfo {volume} {321}},\ \bibinfo {pages}
  {1062} (\bibinfo {year} {2008})}\BibitemShut {NoStop}%
\bibitem [{\citenamefont {Ni}\ \emph {et~al.}(2008)\citenamefont {Ni},
  \citenamefont {Ospelkaus}, \citenamefont {de~Miranda}, \citenamefont {Pe'er},
  \citenamefont {Neyenhuis}, \citenamefont {Zirbel}, \citenamefont
  {Kotochigova}, \citenamefont {Julienne}, \citenamefont {Jin},\ and\
  \citenamefont {Ye}}]{Ni08}%
  \BibitemOpen
  \bibfield  {author} {\bibinfo {author} {\bibfnamefont {K.-K.}\ \bibnamefont
  {Ni}}, \bibinfo {author} {\bibfnamefont {S.}~\bibnamefont {Ospelkaus}},
  \bibinfo {author} {\bibfnamefont {M.~H.~G.}\ \bibnamefont {de~Miranda}},
  \bibinfo {author} {\bibfnamefont {A.}~\bibnamefont {Pe'er}}, \bibinfo
  {author} {\bibfnamefont {B.}~\bibnamefont {Neyenhuis}}, \bibinfo {author}
  {\bibfnamefont {J.~J.}\ \bibnamefont {Zirbel}}, \bibinfo {author}
  {\bibfnamefont {S.}~\bibnamefont {Kotochigova}}, \bibinfo {author}
  {\bibfnamefont {P.~S.}\ \bibnamefont {Julienne}}, \bibinfo {author}
  {\bibfnamefont {D.~S.}\ \bibnamefont {Jin}},\ and\ \bibinfo {author}
  {\bibfnamefont {J.}~\bibnamefont {Ye}},\ }\bibfield  {title} {\bibinfo
  {title} {A high phase-space-density gas of polar molecules},\ }\href
  {https://doi.org/10.1126/science.1163861} {\bibfield  {journal} {\bibinfo
  {journal} {Science}\ }\textbf {\bibinfo {volume} {322}},\ \bibinfo {pages}
  {231} (\bibinfo {year} {2008})}\BibitemShut {NoStop}%
\bibitem [{\citenamefont {De~Marco}\ \emph {et~al.}(2019)\citenamefont
  {De~Marco}, \citenamefont {Valtolina}, \citenamefont {Matsuda}, \citenamefont
  {Tobias}, \citenamefont {Covey},\ and\ \citenamefont {Ye}}]{deMarco2019}%
  \BibitemOpen
  \bibfield  {author} {\bibinfo {author} {\bibfnamefont {L.}~\bibnamefont
  {De~Marco}}, \bibinfo {author} {\bibfnamefont {G.}~\bibnamefont {Valtolina}},
  \bibinfo {author} {\bibfnamefont {K.}~\bibnamefont {Matsuda}}, \bibinfo
  {author} {\bibfnamefont {W.~G.}\ \bibnamefont {Tobias}}, \bibinfo {author}
  {\bibfnamefont {J.~P.}\ \bibnamefont {Covey}},\ and\ \bibinfo {author}
  {\bibfnamefont {J.}~\bibnamefont {Ye}},\ }\bibfield  {title} {\bibinfo
  {title} {A degenerate {Fermi} gas of polar molecules},\ }\href
  {https://doi.org/10.1126/science.aau7230} {\bibfield  {journal} {\bibinfo
  {journal} {Science}\ }\textbf {\bibinfo {volume} {363}},\ \bibinfo {pages}
  {853} (\bibinfo {year} {2019})}\BibitemShut {NoStop}%
\bibitem [{\citenamefont {Takekoshi}\ \emph {et~al.}(2014)\citenamefont
  {Takekoshi}, \citenamefont {Reichs\"ollner}, \citenamefont {Schindewolf},
  \citenamefont {Hutson}, \citenamefont {Le~Sueur}, \citenamefont {Dulieu},
  \citenamefont {Ferlaino}, \citenamefont {Grimm},\ and\ \citenamefont
  {N\"agerl}}]{TakekoshiPRL14}%
  \BibitemOpen
  \bibfield  {author} {\bibinfo {author} {\bibfnamefont {T.}~\bibnamefont
  {Takekoshi}}, \bibinfo {author} {\bibfnamefont {L.}~\bibnamefont
  {Reichs\"ollner}}, \bibinfo {author} {\bibfnamefont {A.}~\bibnamefont
  {Schindewolf}}, \bibinfo {author} {\bibfnamefont {J.~M.}\ \bibnamefont
  {Hutson}}, \bibinfo {author} {\bibfnamefont {C.~R.}\ \bibnamefont
  {Le~Sueur}}, \bibinfo {author} {\bibfnamefont {O.}~\bibnamefont {Dulieu}},
  \bibinfo {author} {\bibfnamefont {F.}~\bibnamefont {Ferlaino}}, \bibinfo
  {author} {\bibfnamefont {R.}~\bibnamefont {Grimm}},\ and\ \bibinfo {author}
  {\bibfnamefont {H.-C.}\ \bibnamefont {N\"agerl}},\ }\bibfield  {title}
  {\bibinfo {title} {Ultracold dense samples of dipolar {RbCs} molecules in the
  rovibrational and hyperfine ground state},\ }\href
  {https://doi.org/10.1103/PhysRevLett.113.205301} {\bibfield  {journal}
  {\bibinfo  {journal} {Phys. Rev. Lett.}\ }\textbf {\bibinfo {volume} {113}},\
  \bibinfo {pages} {205301} (\bibinfo {year} {2014})}\BibitemShut {NoStop}%
\bibitem [{\citenamefont {Molony}\ \emph {et~al.}(2014)\citenamefont {Molony},
  \citenamefont {Gregory}, \citenamefont {Ji}, \citenamefont {Lu},
  \citenamefont {K\"oppinger}, \citenamefont {Le~Sueur}, \citenamefont
  {Blackley}, \citenamefont {Hutson},\ and\ \citenamefont
  {Cornish}}]{Molony14}%
  \BibitemOpen
  \bibfield  {author} {\bibinfo {author} {\bibfnamefont {P.~K.}\ \bibnamefont
  {Molony}}, \bibinfo {author} {\bibfnamefont {P.~D.}\ \bibnamefont {Gregory}},
  \bibinfo {author} {\bibfnamefont {Z.}~\bibnamefont {Ji}}, \bibinfo {author}
  {\bibfnamefont {B.}~\bibnamefont {Lu}}, \bibinfo {author} {\bibfnamefont
  {M.~P.}\ \bibnamefont {K\"oppinger}}, \bibinfo {author} {\bibfnamefont
  {C.~R.}\ \bibnamefont {Le~Sueur}}, \bibinfo {author} {\bibfnamefont {C.~L.}\
  \bibnamefont {Blackley}}, \bibinfo {author} {\bibfnamefont {J.~M.}\
  \bibnamefont {Hutson}},\ and\ \bibinfo {author} {\bibfnamefont {S.~L.}\
  \bibnamefont {Cornish}},\ }\bibfield  {title} {\bibinfo {title} {Creation of
  ultracold $^{87}\mathrm{Rb}^{133}\mathrm{Cs}$ molecules in the rovibrational
  ground state},\ }\href {https://doi.org/10.1103/PhysRevLett.113.255301}
  {\bibfield  {journal} {\bibinfo  {journal} {Phys. Rev. Lett.}\ }\textbf
  {\bibinfo {volume} {113}},\ \bibinfo {pages} {255301} (\bibinfo {year}
  {2014})}\BibitemShut {NoStop}%
\bibitem [{\citenamefont {Park}\ \emph {et~al.}(2015)\citenamefont {Park},
  \citenamefont {Will},\ and\ \citenamefont {Zwierlein}}]{Park15}%
  \BibitemOpen
  \bibfield  {author} {\bibinfo {author} {\bibfnamefont {J.~W.}\ \bibnamefont
  {Park}}, \bibinfo {author} {\bibfnamefont {S.~A.}\ \bibnamefont {Will}},\
  and\ \bibinfo {author} {\bibfnamefont {M.~W.}\ \bibnamefont {Zwierlein}},\
  }\bibfield  {title} {\bibinfo {title} {Ultracold dipolar gas of fermionic
  $^{23}\mathrm{Na}^{40}\mathrm{K}$ molecules in their absolute ground state},\
  }\href {https://doi.org/10.1103/PhysRevLett.114.205302} {\bibfield  {journal}
  {\bibinfo  {journal} {Phys. Rev. Lett.}\ }\textbf {\bibinfo {volume} {114}},\
  \bibinfo {pages} {205302} (\bibinfo {year} {2015})}\BibitemShut {NoStop}%
\bibitem [{\citenamefont {Guo}\ \emph {et~al.}(2016)\citenamefont {Guo},
  \citenamefont {Zhu}, \citenamefont {Lu}, \citenamefont {Ye}, \citenamefont
  {Wang}, \citenamefont {Vexiau}, \citenamefont {Bouloufa-Maafa}, \citenamefont
  {Qu\'em\'ener}, \citenamefont {Dulieu},\ and\ \citenamefont
  {Wang}}]{Guo2016}%
  \BibitemOpen
  \bibfield  {author} {\bibinfo {author} {\bibfnamefont {M.}~\bibnamefont
  {Guo}}, \bibinfo {author} {\bibfnamefont {B.}~\bibnamefont {Zhu}}, \bibinfo
  {author} {\bibfnamefont {B.}~\bibnamefont {Lu}}, \bibinfo {author}
  {\bibfnamefont {X.}~\bibnamefont {Ye}}, \bibinfo {author} {\bibfnamefont
  {F.}~\bibnamefont {Wang}}, \bibinfo {author} {\bibfnamefont {R.}~\bibnamefont
  {Vexiau}}, \bibinfo {author} {\bibfnamefont {N.}~\bibnamefont
  {Bouloufa-Maafa}}, \bibinfo {author} {\bibfnamefont {G.}~\bibnamefont
  {Qu\'em\'ener}}, \bibinfo {author} {\bibfnamefont {O.}~\bibnamefont
  {Dulieu}},\ and\ \bibinfo {author} {\bibfnamefont {D.}~\bibnamefont {Wang}},\
  }\bibfield  {title} {\bibinfo {title} {Creation of an ultracold gas of
  ground-state dipolar ${}^{23}${Na}${}^{87}${Rb} molecules},\ }\href
  {https://doi.org/10.1103/PhysRevLett.116.205303} {\bibfield  {journal}
  {\bibinfo  {journal} {Phys. Rev. Lett.}\ }\textbf {\bibinfo {volume} {116}},\
  \bibinfo {pages} {205303} (\bibinfo {year} {2016})}\BibitemShut {NoStop}%
\bibitem [{\citenamefont {Cairncross}\ \emph {et~al.}(2021)\citenamefont
  {Cairncross}, \citenamefont {Zhang}, \citenamefont {Picard}, \citenamefont
  {Yu}, \citenamefont {Wang},\ and\ \citenamefont {Ni}}]{Cairncross2021NaCs}%
  \BibitemOpen
  \bibfield  {author} {\bibinfo {author} {\bibfnamefont {W.~B.}\ \bibnamefont
  {Cairncross}}, \bibinfo {author} {\bibfnamefont {J.~T.}\ \bibnamefont
  {Zhang}}, \bibinfo {author} {\bibfnamefont {L.~R.~B.}\ \bibnamefont
  {Picard}}, \bibinfo {author} {\bibfnamefont {Y.}~\bibnamefont {Yu}}, \bibinfo
  {author} {\bibfnamefont {K.}~\bibnamefont {Wang}},\ and\ \bibinfo {author}
  {\bibfnamefont {K.-K.}\ \bibnamefont {Ni}},\ }\bibfield  {title} {\bibinfo
  {title} {Assembly of a rovibrational ground state molecule in an optical
  tweezer},\ }\href {https://doi.org/10.1103/PhysRevLett.126.123402} {\bibfield
   {journal} {\bibinfo  {journal} {Phys. Rev. Lett.}\ }\textbf {\bibinfo
  {volume} {126}},\ \bibinfo {pages} {123402} (\bibinfo {year}
  {2021})}\BibitemShut {NoStop}%
\bibitem [{\citenamefont {Stevenson}\ \emph {et~al.}(2023)\citenamefont
  {Stevenson}, \citenamefont {Lam}, \citenamefont {Bigagli}, \citenamefont
  {Warner}, \citenamefont {Yuan}, \citenamefont {Zhang},\ and\ \citenamefont
  {Will}}]{Stevenson23NaCs}%
  \BibitemOpen
  \bibfield  {author} {\bibinfo {author} {\bibfnamefont {I.}~\bibnamefont
  {Stevenson}}, \bibinfo {author} {\bibfnamefont {A.~Z.}\ \bibnamefont {Lam}},
  \bibinfo {author} {\bibfnamefont {N.}~\bibnamefont {Bigagli}}, \bibinfo
  {author} {\bibfnamefont {C.}~\bibnamefont {Warner}}, \bibinfo {author}
  {\bibfnamefont {W.}~\bibnamefont {Yuan}}, \bibinfo {author} {\bibfnamefont
  {S.}~\bibnamefont {Zhang}},\ and\ \bibinfo {author} {\bibfnamefont
  {S.}~\bibnamefont {Will}},\ }\bibfield  {title} {\bibinfo {title} {Ultracold
  gas of dipolar {NaCs} ground state molecules},\ }\href
  {https://doi.org/10.1103/PhysRevLett.130.113002} {\bibfield  {journal}
  {\bibinfo  {journal} {Phys. Rev. Lett.}\ }\textbf {\bibinfo {volume} {130}},\
  \bibinfo {pages} {113002} (\bibinfo {year} {2023})}\BibitemShut {NoStop}%
\bibitem [{\citenamefont {Bigagli}\ \emph
  {et~al.}(2023{\natexlab{a}})\citenamefont {Bigagli}, \citenamefont {Warner},
  \citenamefont {Yuan}, \citenamefont {Zhang}, \citenamefont {Stevenson},
  \citenamefont {Karman},\ and\ \citenamefont {Will}}]{Bigagli2023NaCs}%
  \BibitemOpen
  \bibfield  {author} {\bibinfo {author} {\bibfnamefont {N.}~\bibnamefont
  {Bigagli}}, \bibinfo {author} {\bibfnamefont {C.}~\bibnamefont {Warner}},
  \bibinfo {author} {\bibfnamefont {W.}~\bibnamefont {Yuan}}, \bibinfo {author}
  {\bibfnamefont {S.}~\bibnamefont {Zhang}}, \bibinfo {author} {\bibfnamefont
  {I.}~\bibnamefont {Stevenson}}, \bibinfo {author} {\bibfnamefont
  {T.}~\bibnamefont {Karman}},\ and\ \bibinfo {author} {\bibfnamefont
  {S.}~\bibnamefont {Will}},\ }\bibfield  {title} {\bibinfo {title}
  {Collisionally stable gas of bosonic dipolar ground-state molecules},\
  }\bibfield  {journal} {\bibinfo  {journal} {Nat. Phys.}\ }\href
  {https://doi.org/10.1038/s41567-023-02200-6} {10.1038/s41567-023-02200-6}
  (\bibinfo {year} {2023}{\natexlab{a}})\BibitemShut {NoStop}%
\bibitem [{\citenamefont {Softley}(2023)}]{Softley2023twenties}%
  \BibitemOpen
  \bibfield  {author} {\bibinfo {author} {\bibfnamefont {T.~P.}\ \bibnamefont
  {Softley}},\ }\bibfield  {title} {\bibinfo {title} {Cold and ultracold
  molecules in the twenties},\ }\href {https://doi.org/10.1098/rspa.2022.0806}
  {\bibfield  {journal} {\bibinfo  {journal} {Proc. R. Soc. A.}\ }\textbf
  {\bibinfo {volume} {479}},\ \bibinfo {pages} {20220806} (\bibinfo {year}
  {2023})}\BibitemShut {NoStop}%
\bibitem [{\citenamefont {Bigagli}\ \emph {et~al.}(2022)\citenamefont
  {Bigagli}, \citenamefont {Savin},\ and\ \citenamefont
  {Will}}]{Bigagli22carbon}%
  \BibitemOpen
  \bibfield  {author} {\bibinfo {author} {\bibfnamefont {N.}~\bibnamefont
  {Bigagli}}, \bibinfo {author} {\bibfnamefont {D.~W.}\ \bibnamefont {Savin}},\
  and\ \bibinfo {author} {\bibfnamefont {S.}~\bibnamefont {Will}},\ }\bibfield
  {title} {\bibinfo {title} {Laser cooling scheme for the carbon dimer
  ($^{12}\mathrm{C}{}_{2}$)},\ }\href
  {https://doi.org/10.1103/PhysRevA.105.L051301} {\bibfield  {journal}
  {\bibinfo  {journal} {Phys. Rev. A}\ }\textbf {\bibinfo {volume} {105}},\
  \bibinfo {pages} {L051301} (\bibinfo {year} {2022})}\BibitemShut {NoStop}%
\bibitem [{\citenamefont {Bigagli}\ \emph
  {et~al.}(2023{\natexlab{b}})\citenamefont {Bigagli}, \citenamefont {Savin},\
  and\ \citenamefont {Will}}]{Bigagli2023OH}%
  \BibitemOpen
  \bibfield  {author} {\bibinfo {author} {\bibfnamefont {N.}~\bibnamefont
  {Bigagli}}, \bibinfo {author} {\bibfnamefont {D.~W.}\ \bibnamefont {Savin}},\
  and\ \bibinfo {author} {\bibfnamefont {S.}~\bibnamefont {Will}},\ }\bibfield
  {title} {\bibinfo {title} {Laser scheme for {Doppler} cooling of the hydroxyl
  cation ({OH}$^+$)},\ }\href {https://doi.org/10.1021/acs.jpca.3c03248}
  {\bibfield  {journal} {\bibinfo  {journal} {J. Phys. Chem. A}\ }\textbf
  {\bibinfo {volume} {127}},\ \bibinfo {pages} {8194} (\bibinfo {year}
  {2023}{\natexlab{b}})}\BibitemShut {NoStop}%
\bibitem [{\citenamefont {Tennyson}\ \emph {et~al.}(2016)\citenamefont
  {Tennyson}, \citenamefont {Yurchenko}, \citenamefont {Al-Refaie},
  \citenamefont {Barton}, \citenamefont {Chubb}, \citenamefont {Coles},
  \citenamefont {Diamantopoulou}, \citenamefont {Gorman}, \citenamefont {Hill},
  \citenamefont {Lam}, \citenamefont {Lodi}, \citenamefont {McKemmish},
  \citenamefont {Na}, \citenamefont {Owens}, \citenamefont {Polyansky},
  \citenamefont {Rivlin}, \citenamefont {Sousa-Silva}, \citenamefont
  {Underwood}, \citenamefont {Yachmenev},\ and\ \citenamefont
  {Zak}}]{Tennyson2016Exomol}%
  \BibitemOpen
  \bibfield  {author} {\bibinfo {author} {\bibfnamefont {J.}~\bibnamefont
  {Tennyson}}, \bibinfo {author} {\bibfnamefont {S.~N.}\ \bibnamefont
  {Yurchenko}}, \bibinfo {author} {\bibfnamefont {A.~F.}\ \bibnamefont
  {Al-Refaie}}, \bibinfo {author} {\bibfnamefont {E.~J.}\ \bibnamefont
  {Barton}}, \bibinfo {author} {\bibfnamefont {K.~L.}\ \bibnamefont {Chubb}},
  \bibinfo {author} {\bibfnamefont {P.~A.}\ \bibnamefont {Coles}}, \bibinfo
  {author} {\bibfnamefont {S.}~\bibnamefont {Diamantopoulou}}, \bibinfo
  {author} {\bibfnamefont {M.~N.}\ \bibnamefont {Gorman}}, \bibinfo {author}
  {\bibfnamefont {C.}~\bibnamefont {Hill}}, \bibinfo {author} {\bibfnamefont
  {A.~Z.}\ \bibnamefont {Lam}}, \bibinfo {author} {\bibfnamefont
  {L.}~\bibnamefont {Lodi}}, \bibinfo {author} {\bibfnamefont {L.~K.}\
  \bibnamefont {McKemmish}}, \bibinfo {author} {\bibfnamefont {Y.}~\bibnamefont
  {Na}}, \bibinfo {author} {\bibfnamefont {A.}~\bibnamefont {Owens}}, \bibinfo
  {author} {\bibfnamefont {O.~L.}\ \bibnamefont {Polyansky}}, \bibinfo {author}
  {\bibfnamefont {T.}~\bibnamefont {Rivlin}}, \bibinfo {author} {\bibfnamefont
  {C.}~\bibnamefont {Sousa-Silva}}, \bibinfo {author} {\bibfnamefont {D.~S.}\
  \bibnamefont {Underwood}}, \bibinfo {author} {\bibfnamefont {A.}~\bibnamefont
  {Yachmenev}},\ and\ \bibinfo {author} {\bibfnamefont {E.}~\bibnamefont
  {Zak}},\ }\bibfield  {title} {\bibinfo {title} {The {ExoMol} database:
  Molecular line lists for exoplanet and other hot atmospheres},\ }\href
  {https://doi.org/10.1016/j.jms.2016.05.002} {\bibfield  {journal} {\bibinfo
  {journal} {J. Mol. Spectrosc.}\ }\textbf {\bibinfo {volume} {327}},\ \bibinfo
  {pages} {73} (\bibinfo {year} {2016})}\BibitemShut {NoStop}%
\bibitem [{\citenamefont {Gordon}\ \emph {et~al.}(2022)\citenamefont {Gordon},
  \citenamefont {Rothman}, \citenamefont {Hargreaves}, \citenamefont {Hashemi},
  \citenamefont {Karlovets}, \citenamefont {Skinner}, \citenamefont {Conway},
  \citenamefont {Hill}, \citenamefont {Kochanov}, \citenamefont {Tan},\ and\
  \citenamefont {{et al}}}]{Gordon2022HITRAN}%
  \BibitemOpen
  \bibfield  {author} {\bibinfo {author} {\bibfnamefont {I.}~\bibnamefont
  {Gordon}}, \bibinfo {author} {\bibfnamefont {L.}~\bibnamefont {Rothman}},
  \bibinfo {author} {\bibfnamefont {R.}~\bibnamefont {Hargreaves}}, \bibinfo
  {author} {\bibfnamefont {R.}~\bibnamefont {Hashemi}}, \bibinfo {author}
  {\bibfnamefont {E.}~\bibnamefont {Karlovets}}, \bibinfo {author}
  {\bibfnamefont {F.}~\bibnamefont {Skinner}}, \bibinfo {author} {\bibfnamefont
  {E.}~\bibnamefont {Conway}}, \bibinfo {author} {\bibfnamefont
  {C.}~\bibnamefont {Hill}}, \bibinfo {author} {\bibfnamefont {R.}~\bibnamefont
  {Kochanov}}, \bibinfo {author} {\bibfnamefont {Y.}~\bibnamefont {Tan}},\ and\
  \bibinfo {author} {\bibnamefont {{et al}}},\ }\bibfield  {title} {\bibinfo
  {title} {The {HITRAN}2020 molecular spectroscopic database},\ }\href
  {https://doi.org/10.1016/j.jqsrt.2021.107949} {\bibfield  {journal} {\bibinfo
   {journal} {J. Quant. Spectrosc. Radiat. Transf.}\ }\textbf {\bibinfo
  {volume} {277}},\ \bibinfo {pages} {107949} (\bibinfo {year}
  {2022})}\BibitemShut {NoStop}%
\bibitem [{Spl(2023)}]{Splatalogue23}%
  \BibitemOpen
  \href@noop {} {\bibinfo {title} {Splatalogue: Database for astronomical
  spectroscopy}},\ \bibinfo {howpublished} {\url{https://splatalogue.online/}}
  (\bibinfo {year} {2023}),\ \bibinfo {note} {accessed: 2023-07-13}\BibitemShut
  {NoStop}%
\bibitem [{\citenamefont {Lovas}(2002)}]{NISTdatabase}%
  \BibitemOpen
  \bibfield  {author} {\bibinfo {author} {\bibfnamefont {F.}~\bibnamefont
  {Lovas}},\ }\href {https://doi.org/10.18434/T4T59X} {\bibinfo {title}
  {{Diatomic Spectral Database, NIST Standard Reference Database 114}}}
  (\bibinfo {year} {2002})\BibitemShut {NoStop}%
\bibitem [{CDM(2023)}]{CDMS23}%
  \BibitemOpen
  \href@noop {} {\bibinfo {title} {{CDMS: The Cologne Database for Molecular
  Spectroscopy}}},\ \bibinfo {howpublished}
  {\url{https://cdms.astro.uni-koeln.de/}} (\bibinfo {year} {2023}),\ \bibinfo
  {note} {accessed: 2023-07-13}\BibitemShut {NoStop}%
\bibitem [{\citenamefont {Liu}\ \emph {et~al.}(2020)\citenamefont {Liu},
  \citenamefont {Truppe}, \citenamefont {Meijer},\ and\ \citenamefont
  {P{\'{e}}rez-R{\'{\i}}os}}]{Liu2020database}%
  \BibitemOpen
  \bibfield  {author} {\bibinfo {author} {\bibfnamefont {X.}~\bibnamefont
  {Liu}}, \bibinfo {author} {\bibfnamefont {S.}~\bibnamefont {Truppe}},
  \bibinfo {author} {\bibfnamefont {G.}~\bibnamefont {Meijer}},\ and\ \bibinfo
  {author} {\bibfnamefont {J.}~\bibnamefont {P{\'{e}}rez-R{\'{\i}}os}},\
  }\bibfield  {title} {\bibinfo {title} {The diatomic molecular spectroscopy
  database},\ }\href {https://doi.org/10.1186/s13321-020-00433-8} {\bibfield
  {journal} {\bibinfo  {journal} {J. Cheminform.}\ }\textbf {\bibinfo {volume}
  {12}},\ \bibinfo {pages} {31} (\bibinfo {year} {2020})}\BibitemShut {NoStop}%
\bibitem [{\citenamefont {Bernath}(2020)}]{Bernath2020MoLLIST}%
  \BibitemOpen
  \bibfield  {author} {\bibinfo {author} {\bibfnamefont {P.~F.}\ \bibnamefont
  {Bernath}},\ }\bibfield  {title} {\bibinfo {title} {{MoLLIST}: Molecular line
  lists, intensities and spectra},\ }\href
  {https://doi.org/10.1016/j.jqsrt.2019.106687} {\bibfield  {journal} {\bibinfo
   {journal} {J. Quant. Spectrosc. Radiat. Transf.}\ }\textbf {\bibinfo
  {volume} {240}},\ \bibinfo {pages} {106687} (\bibinfo {year}
  {2020})}\BibitemShut {NoStop}%
\bibitem [{\citenamefont {Trudeau}(1994)}]{Trudeau1994book}%
  \BibitemOpen
  \bibfield  {author} {\bibinfo {author} {\bibfnamefont {R.~J.}\ \bibnamefont
  {Trudeau}},\ }\href@noop {} {\emph {\bibinfo {title} {Introduction to Graph
  Theory}}}\ (\bibinfo  {publisher} {Dover Publications},\ \bibinfo {address}
  {Mineola, New York},\ \bibinfo {year} {1994})\BibitemShut {NoStop}%
\bibitem [{\citenamefont {Barab{\'a}si}\ and\ \citenamefont
  {Frangos}(2002)}]{barabasi2002linked}%
  \BibitemOpen
  \bibfield  {author} {\bibinfo {author} {\bibfnamefont {A.-L.}\ \bibnamefont
  {Barab{\'a}si}}\ and\ \bibinfo {author} {\bibfnamefont {J.}~\bibnamefont
  {Frangos}},\ }\href@noop {} {\emph {\bibinfo {title} {Linked: the new science
  of networks}}}\ (\bibinfo  {publisher} {Basic Books},\ \bibinfo {year}
  {2002})\BibitemShut {NoStop}%
\bibitem [{Neo(2023{\natexlab{a}})}]{Neo4j}%
  \BibitemOpen
  \href@noop {} {\bibinfo {title} {{Neo4j Graph Data Platform}}},\ \bibinfo
  {howpublished} {\url{https://neo4j.com/}} (\bibinfo {year}
  {2023}{\natexlab{a}}),\ \bibinfo {note} {accessed: 2023-04-28}\BibitemShut
  {NoStop}%
\bibitem [{\citenamefont {Needham}\ and\ \citenamefont
  {Hodler}(2019)}]{Needham2019book}%
  \BibitemOpen
  \bibfield  {author} {\bibinfo {author} {\bibfnamefont {M.}~\bibnamefont
  {Needham}}\ and\ \bibinfo {author} {\bibfnamefont {A.~E.}\ \bibnamefont
  {Hodler}},\ }\href@noop {} {\emph {\bibinfo {title} {Graph algorithms:
  Practical examples in Apache Spark and Neo4j}}}\ (\bibinfo  {publisher}
  {O'Reilly Media},\ \bibinfo {address} {Sebastopol, California},\ \bibinfo
  {year} {2019})\BibitemShut {NoStop}%
\bibitem [{\citenamefont {Miao}\ \emph {et~al.}(2023)\citenamefont {Miao},
  \citenamefont {Ma},\ and\ \citenamefont {Huang}}]{Miao2023toxicity}%
  \BibitemOpen
  \bibfield  {author} {\bibinfo {author} {\bibfnamefont {Y.}~\bibnamefont
  {Miao}}, \bibinfo {author} {\bibfnamefont {H.}~\bibnamefont {Ma}},\ and\
  \bibinfo {author} {\bibfnamefont {J.}~\bibnamefont {Huang}},\ }\bibfield
  {title} {\bibinfo {title} {Recent advances in toxicity prediction:
  {A}pplications of deep graph learning},\ }\href
  {https://doi.org/10.1021/acs.chemrestox.2c00384} {\bibfield  {journal}
  {\bibinfo  {journal} {Chem. Res. Toxicol.}\ }\textbf {\bibinfo {volume}
  {36}},\ \bibinfo {pages} {1206} (\bibinfo {year} {2023})}\BibitemShut
  {NoStop}%
\bibitem [{\citenamefont {Wellnitz}\ \emph {et~al.}(2023)\citenamefont
  {Wellnitz}, \citenamefont {Kekić}, \citenamefont {Heiss}, \citenamefont
  {Gertz}, \citenamefont {Weidemüller},\ and\ \citenamefont
  {Spitz}}]{Wellnitz2023network}%
  \BibitemOpen
  \bibfield  {author} {\bibinfo {author} {\bibfnamefont {D.}~\bibnamefont
  {Wellnitz}}, \bibinfo {author} {\bibfnamefont {A.}~\bibnamefont {Kekić}},
  \bibinfo {author} {\bibfnamefont {J.}~\bibnamefont {Heiss}}, \bibinfo
  {author} {\bibfnamefont {M.}~\bibnamefont {Gertz}}, \bibinfo {author}
  {\bibfnamefont {M.}~\bibnamefont {Weidemüller}},\ and\ \bibinfo {author}
  {\bibfnamefont {A.}~\bibnamefont {Spitz}},\ }\bibfield  {title} {\bibinfo
  {title} {A network approach to atomic spectra},\ }\href
  {https://doi.org/10.1088/2632-072X/ace1c3} {\bibfield  {journal} {\bibinfo
  {journal} {J. Phys. Complex.}\ }\textbf {\bibinfo {volume} {4}},\ \bibinfo
  {pages} {03LT01} (\bibinfo {year} {2023})}\BibitemShut {NoStop}%
\bibitem [{\citenamefont {Metcalf}\ and\ \citenamefont {van~der
  Straten}(1999)}]{Metcalf1999book}%
  \BibitemOpen
  \bibfield  {author} {\bibinfo {author} {\bibfnamefont {H.~J.}\ \bibnamefont
  {Metcalf}}\ and\ \bibinfo {author} {\bibfnamefont {P.}~\bibnamefont {van~der
  Straten}},\ }\href@noop {} {\emph {\bibinfo {title} {Laser cooling and
  trapping}}}\ (\bibinfo  {publisher} {Springer},\ \bibinfo {address} {Berlin,
  Germany},\ \bibinfo {year} {1999})\BibitemShut {NoStop}%
\bibitem [{\citenamefont {Castin}\ \emph {et~al.}(1989)\citenamefont {Castin},
  \citenamefont {Wallis},\ and\ \citenamefont {Dalibard}}]{castin1989limit}%
  \BibitemOpen
  \bibfield  {author} {\bibinfo {author} {\bibfnamefont {Y.}~\bibnamefont
  {Castin}}, \bibinfo {author} {\bibfnamefont {H.}~\bibnamefont {Wallis}},\
  and\ \bibinfo {author} {\bibfnamefont {J.}~\bibnamefont {Dalibard}},\
  }\bibfield  {title} {\bibinfo {title} {Limit of {Doppler} cooling},\ }\href
  {https://pro.college-de-france.fr/jean.dalibard/publi3/josaB_89.pdf}
  {\bibfield  {journal} {\bibinfo  {journal} {J. Opt. Soc. Am. B}\ }\textbf
  {\bibinfo {volume} {6}},\ \bibinfo {pages} {2046} (\bibinfo {year}
  {1989})}\BibitemShut {NoStop}%
\bibitem [{\citenamefont {Fitch}\ and\ \citenamefont
  {Tarbutt}(2021{\natexlab{b}})}]{FITCH2021157}%
  \BibitemOpen
  \bibfield  {author} {\bibinfo {author} {\bibfnamefont {N.}~\bibnamefont
  {Fitch}}\ and\ \bibinfo {author} {\bibfnamefont {M.}~\bibnamefont
  {Tarbutt}},\ }\bibfield  {title} {\bibinfo {title} {Chapter 3 -
  {Laser}-cooled molecules}\ }(\bibinfo  {publisher} {Academic Press},\
  \bibinfo {year} {2021})\ pp.\ \bibinfo {pages} {157--262}\BibitemShut
  {NoStop}%
\bibitem [{\citenamefont {Vilas}\ \emph {et~al.}(2022)\citenamefont {Vilas},
  \citenamefont {Hallas}, \citenamefont {Anderegg}, \citenamefont {Robichaud},
  \citenamefont {Winnicki}, \citenamefont {Mitra},\ and\ \citenamefont
  {Doyle}}]{Vilas2022CaOH}%
  \BibitemOpen
  \bibfield  {author} {\bibinfo {author} {\bibfnamefont {N.~B.}\ \bibnamefont
  {Vilas}}, \bibinfo {author} {\bibfnamefont {C.}~\bibnamefont {Hallas}},
  \bibinfo {author} {\bibfnamefont {L.}~\bibnamefont {Anderegg}}, \bibinfo
  {author} {\bibfnamefont {P.}~\bibnamefont {Robichaud}}, \bibinfo {author}
  {\bibfnamefont {A.}~\bibnamefont {Winnicki}}, \bibinfo {author}
  {\bibfnamefont {D.}~\bibnamefont {Mitra}},\ and\ \bibinfo {author}
  {\bibfnamefont {J.~M.}\ \bibnamefont {Doyle}},\ }\bibfield  {title} {\bibinfo
  {title} {Magneto-optical trapping and sub-doppler cooling of a polyatomic
  molecule},\ }\href {https://doi.org/10.1038/s41586-022-04620-5} {\bibfield
  {journal} {\bibinfo  {journal} {Nature}\ }\textbf {\bibinfo {volume} {606}},\
  \bibinfo {pages} {70} (\bibinfo {year} {2022})}\BibitemShut {NoStop}%
\bibitem [{\citenamefont {Rosa}(2004)}]{Rosa2004EPJD}%
  \BibitemOpen
  \bibfield  {author} {\bibinfo {author} {\bibfnamefont {M.~D.}\ \bibnamefont
  {Rosa}},\ }\bibfield  {title} {\bibinfo {title} {Laser-cooling molecules},\
  }\href {https://doi.org/10.1140/epjd/e2004-00167-2} {\bibfield  {journal}
  {\bibinfo  {journal} {Eur. Phys. J. D}\ }\textbf {\bibinfo {volume} {31}},\
  \bibinfo {pages} {395} (\bibinfo {year} {2004})}\BibitemShut {NoStop}%
\bibitem [{\citenamefont {Tarbutt}\ \emph {et~al.}(2013)\citenamefont
  {Tarbutt}, \citenamefont {Sauer}, \citenamefont {Hudson},\ and\ \citenamefont
  {Hinds}}]{Tarbutt2013NJP}%
  \BibitemOpen
  \bibfield  {author} {\bibinfo {author} {\bibfnamefont {M.~R.}\ \bibnamefont
  {Tarbutt}}, \bibinfo {author} {\bibfnamefont {B.~E.}\ \bibnamefont {Sauer}},
  \bibinfo {author} {\bibfnamefont {J.~J.}\ \bibnamefont {Hudson}},\ and\
  \bibinfo {author} {\bibfnamefont {E.~A.}\ \bibnamefont {Hinds}},\ }\bibfield
  {title} {\bibinfo {title} {Design for a fountain of {YbF} molecules to
  measure the electron{\textquotesingle}s electric dipole moment},\ }\href
  {https://doi.org/10.1088/1367-2630/15/5/053034} {\bibfield  {journal}
  {\bibinfo  {journal} {New J. Phys.}\ }\textbf {\bibinfo {volume} {15}},\
  \bibinfo {pages} {053034} (\bibinfo {year} {2013})}\BibitemShut {NoStop}%
\bibitem [{\citenamefont {Yurchenko}\ \emph {et~al.}(2020)\citenamefont
  {Yurchenko}, \citenamefont {Mellor}, \citenamefont {Freedman},\ and\
  \citenamefont {Tennyson}}]{Yurchenko2020CO2}%
  \BibitemOpen
  \bibfield  {author} {\bibinfo {author} {\bibfnamefont {S.~N.}\ \bibnamefont
  {Yurchenko}}, \bibinfo {author} {\bibfnamefont {T.~M.}\ \bibnamefont
  {Mellor}}, \bibinfo {author} {\bibfnamefont {R.~S.}\ \bibnamefont
  {Freedman}},\ and\ \bibinfo {author} {\bibfnamefont {J.}~\bibnamefont
  {Tennyson}},\ }\bibfield  {title} {\bibinfo {title} {{ExoMol} line lists
  {\textendash} {XXXIX}. {Ro-vibrational} molecular line list for {CO}$_2$},\
  }\href {https://doi.org/10.1093/mnras/staa1874} {\bibfield  {journal}
  {\bibinfo  {journal} {Mon. Notices Royal Astron. Soc.}\ }\textbf {\bibinfo
  {volume} {496}},\ \bibinfo {pages} {5282} (\bibinfo {year}
  {2020})}\BibitemShut {NoStop}%
\bibitem [{Note1()}]{Note1}%
  \BibitemOpen
  \bibinfo {note} {To keep the notation simple, we chose to use $S_0$, $S_1$,
  and $S_2$ interchangeably to symbolize individual states belonging to the set
  of starting states, reachable excited states, and reachable states and to
  symbolize the sets themselves.}\BibitemShut {Stop}%
\bibitem [{\citenamefont {Hutzler}\ \emph {et~al.}(2012)\citenamefont
  {Hutzler}, \citenamefont {Lu},\ and\ \citenamefont
  {Doyle}}]{hutzler2012buffer}%
  \BibitemOpen
  \bibfield  {author} {\bibinfo {author} {\bibfnamefont {N.~R.}\ \bibnamefont
  {Hutzler}}, \bibinfo {author} {\bibfnamefont {H.-I.}\ \bibnamefont {Lu}},\
  and\ \bibinfo {author} {\bibfnamefont {J.~M.}\ \bibnamefont {Doyle}},\
  }\bibfield  {title} {\bibinfo {title} {The buffer gas beam: An intense, cold,
  and slow source for atoms and molecules},\ }\href
  {https://doi.org/10.1021/cr200362u} {\bibfield  {journal} {\bibinfo
  {journal} {Chem. Rev.}\ }\textbf {\bibinfo {volume} {112}},\ \bibinfo {pages}
  {4803} (\bibinfo {year} {2012})}\BibitemShut {NoStop}%
\bibitem [{Exo(2023)}]{Exomol23}%
  \BibitemOpen
  \href@noop {} {\bibinfo {title} {Exomol: High temperature molecular line
  lists for modelling exoplanet atmospheres database}},\ \bibinfo
  {howpublished} {\url{https://www.exomol.com/data/molecules/}} (\bibinfo
  {year} {2023}),\ \bibinfo {note} {accessed: 2023-04-28}\BibitemShut {NoStop}%
\bibitem [{\citenamefont {Yurchenko}\ \emph {et~al.}(2023)\citenamefont
  {Yurchenko}, \citenamefont {Brady}, \citenamefont {Tennyson}, \citenamefont
  {Smirnov}, \citenamefont {Vasilyev},\ and\ \citenamefont
  {Solomonik}}]{Yurchenko2023YO}%
  \BibitemOpen
  \bibfield  {author} {\bibinfo {author} {\bibfnamefont {S.}~\bibnamefont
  {Yurchenko}}, \bibinfo {author} {\bibfnamefont {R.}~\bibnamefont {Brady}},
  \bibinfo {author} {\bibfnamefont {J.}~\bibnamefont {Tennyson}}, \bibinfo
  {author} {\bibfnamefont {S.}~\bibnamefont {Smirnov}}, \bibinfo {author}
  {\bibfnamefont {O.}~\bibnamefont {Vasilyev}},\ and\ \bibinfo {author}
  {\bibfnamefont {V.}~\bibnamefont {Solomonik}},\ }\bibfield  {title} {\bibinfo
  {title} {{ExoMol} line lists - {LIII: E}mpirical rovibronic spectra yttrium
  oxide ({YO})},\ }\href
  {https://doi.org/https://doi.org/10.1093/mnras/stad3225} {\bibfield
  {journal} {\bibinfo  {journal} {Mon. Notices Royal Astron. Soc.}\ }\textbf
  {\bibinfo {volume} {527}},\ \bibinfo {pages} {4899–4912} (\bibinfo {year}
  {2023})}\BibitemShut {NoStop}%
\bibitem [{\citenamefont {Yurchenko}\ \emph {et~al.}(2018)\citenamefont
  {Yurchenko}, \citenamefont {Szab{\'{o}}}, \citenamefont {Pyatenko},\ and\
  \citenamefont {Tennyson}}]{Yurchenko2018C2}%
  \BibitemOpen
  \bibfield  {author} {\bibinfo {author} {\bibfnamefont {S.~N.}\ \bibnamefont
  {Yurchenko}}, \bibinfo {author} {\bibfnamefont {I.}~\bibnamefont
  {Szab{\'{o}}}}, \bibinfo {author} {\bibfnamefont {E.}~\bibnamefont
  {Pyatenko}},\ and\ \bibinfo {author} {\bibfnamefont {J.}~\bibnamefont
  {Tennyson}},\ }\bibfield  {title} {\bibinfo {title} {{ExoMol} line lists
  {XXXI}: {S}pectroscopy of lowest eights electronic states of {C}$_2$},\
  }\href {https://doi.org/10.1093/mnras/sty2050} {\bibfield  {journal}
  {\bibinfo  {journal} {Mon. Notices Royal Astron. Soc.}\ }\textbf {\bibinfo
  {volume} {480}},\ \bibinfo {pages} {3397} (\bibinfo {year}
  {2018})}\BibitemShut {NoStop}%
\bibitem [{\citenamefont {McKemmish}\ \emph {et~al.}(2020)\citenamefont
  {McKemmish}, \citenamefont {Syme}, \citenamefont {Borsovszky}, \citenamefont
  {Yurchenko}, \citenamefont {Tennyson}, \citenamefont {Furtenbacher},\ and\
  \citenamefont {Cs{\'a}sz{\'a}r}}]{mckemmish2020update}%
  \BibitemOpen
  \bibfield  {author} {\bibinfo {author} {\bibfnamefont {L.~K.}\ \bibnamefont
  {McKemmish}}, \bibinfo {author} {\bibfnamefont {A.-M.}\ \bibnamefont {Syme}},
  \bibinfo {author} {\bibfnamefont {J.}~\bibnamefont {Borsovszky}}, \bibinfo
  {author} {\bibfnamefont {S.~N.}\ \bibnamefont {Yurchenko}}, \bibinfo {author}
  {\bibfnamefont {J.}~\bibnamefont {Tennyson}}, \bibinfo {author}
  {\bibfnamefont {T.}~\bibnamefont {Furtenbacher}},\ and\ \bibinfo {author}
  {\bibfnamefont {A.~G.}\ \bibnamefont {Cs{\'a}sz{\'a}r}},\ }\bibfield  {title}
  {\bibinfo {title} {An update to the \texttt{Marvel} data set and {ExoMol}
  line list for $^{12}${C}$_2$},\ }\href
  {https://doi.org/10.1093/mnras/staa1954} {\bibfield  {journal} {\bibinfo
  {journal} {Mon. Notices Royal Astron. Soc.}\ }\textbf {\bibinfo {volume}
  {497}},\ \bibinfo {pages} {1081} (\bibinfo {year} {2020})}\BibitemShut
  {NoStop}%
\bibitem [{\citenamefont {Hodges}\ and\ \citenamefont
  {Bernath}(2017)}]{Hodges2017OH}%
  \BibitemOpen
  \bibfield  {author} {\bibinfo {author} {\bibfnamefont {J.~N.}\ \bibnamefont
  {Hodges}}\ and\ \bibinfo {author} {\bibfnamefont {P.~F.}\ \bibnamefont
  {Bernath}},\ }\bibfield  {title} {\bibinfo {title} {Fourier transform
  spectroscopy of the {$A^3 \Pi - X^3 \Sigma^-$} transition of {OH}$^+$},\
  }\href {https://doi.org/10.3847/1538-4357/aa6bf5} {\bibfield  {journal}
  {\bibinfo  {journal} {Astrophys. J.}\ }\textbf {\bibinfo {volume} {840}},\
  \bibinfo {pages} {81} (\bibinfo {year} {2017})}\BibitemShut {NoStop}%
\bibitem [{\citenamefont {Brooke}\ \emph {et~al.}(2014)\citenamefont {Brooke},
  \citenamefont {Ram}, \citenamefont {Western}, \citenamefont {Li},
  \citenamefont {Schwenke},\ and\ \citenamefont {Bernath}}]{Brooke2014CN}%
  \BibitemOpen
  \bibfield  {author} {\bibinfo {author} {\bibfnamefont {J.~S.~A.}\
  \bibnamefont {Brooke}}, \bibinfo {author} {\bibfnamefont {R.~S.}\
  \bibnamefont {Ram}}, \bibinfo {author} {\bibfnamefont {C.~M.}\ \bibnamefont
  {Western}}, \bibinfo {author} {\bibfnamefont {G.}~\bibnamefont {Li}},
  \bibinfo {author} {\bibfnamefont {D.~W.}\ \bibnamefont {Schwenke}},\ and\
  \bibinfo {author} {\bibfnamefont {P.~F.}\ \bibnamefont {Bernath}},\
  }\bibfield  {title} {\bibinfo {title} {Einstein {$A$} coefficients and
  oscillator strengths for the {$A^2 \Pi - X^2 \Sigma^+$} (red) and {$B^2
  \Sigma^+ - X^2 \Sigma^+$} (violet) rovibrational transitions in the {$X^2
  \Sigma^+$} state of {CN}},\ }\href
  {https://doi.org/10.1088/0067-0049/210/2/23} {\bibfield  {journal} {\bibinfo
  {journal} {Astrophys. J.}\ }\textbf {\bibinfo {volume} {210}},\ \bibinfo
  {pages} {23} (\bibinfo {year} {2014})}\BibitemShut {NoStop}%
\bibitem [{\citenamefont {Syme}\ and\ \citenamefont
  {McKemmish}(2020)}]{Syme2020CN}%
  \BibitemOpen
  \bibfield  {author} {\bibinfo {author} {\bibfnamefont {A.-M.}\ \bibnamefont
  {Syme}}\ and\ \bibinfo {author} {\bibfnamefont {L.~K.}\ \bibnamefont
  {McKemmish}},\ }\bibfield  {title} {\bibinfo {title} {Experimental energy
  levels of {${}^{12}$C${}^{14}$N} through \texttt{MARVEL} analysis},\ }\href
  {https://doi.org/10.1093/mnras/staa2791} {\bibfield  {journal} {\bibinfo
  {journal} {Mon. Notices Royal Astron. Soc.}\ }\textbf {\bibinfo {volume}
  {499}},\ \bibinfo {pages} {25} (\bibinfo {year} {2020})}\BibitemShut
  {NoStop}%
\bibitem [{\citenamefont {Syme}\ and\ \citenamefont
  {McKemmish}(2021)}]{Syme2021CN}%
  \BibitemOpen
  \bibfield  {author} {\bibinfo {author} {\bibfnamefont {A.-M.}\ \bibnamefont
  {Syme}}\ and\ \bibinfo {author} {\bibfnamefont {L.~K.}\ \bibnamefont
  {McKemmish}},\ }\bibfield  {title} {\bibinfo {title} {Full spectroscopic
  model and trihybrid experimental-perturbative-variational line list for
  {CN}},\ }\href {https://doi.org/10.1093/mnras/stab1551} {\bibfield  {journal}
  {\bibinfo  {journal} {Mon. Notices Royal Astron. Soc.}\ }\textbf {\bibinfo
  {volume} {505}},\ \bibinfo {pages} {4383} (\bibinfo {year}
  {2021})}\BibitemShut {NoStop}%
\bibitem [{Neo(2023{\natexlab{b}})}]{Neo4jGDSLibrary}%
  \BibitemOpen
  \href@noop {} {\bibinfo {title} {{The Neo4j Graph Data Science Library Manual
  v2.3}}},\ \bibinfo {howpublished}
  {\url{https://neo4j.com/docs/graph-data-science/current/}} (\bibinfo {year}
  {2023}{\natexlab{b}}),\ \bibinfo {note} {accessed: 2023-04-28}\BibitemShut
  {NoStop}%
\bibitem [{\citenamefont {Gioran}(2021)}]{Neo4jtrillion}%
  \BibitemOpen
  \bibfield  {author} {\bibinfo {author} {\bibfnamefont {C.}~\bibnamefont
  {Gioran}},\ }\href@noop {} {\bibinfo {title} {Behind the scenes of creating
  the world’s biggest graph database}},\ \bibinfo {howpublished}
  {\href{https://medium.com/neo4j/behind-the-scenes-of-creating-the-worlds-biggest-graph-database-cd22f477c843}{Medium}}
  (\bibinfo {year} {2021}),\ \bibinfo {note} {accessed: 2023-04-28}\BibitemShut
  {NoStop}%
\bibitem [{Cyp(2023)}]{Cypher}%
  \BibitemOpen
  \href@noop {} {\bibinfo {title} {{Introduction to Cypher}}},\ \bibinfo
  {howpublished} {\url{https://neo4j.com/docs/getting-started/cypher-intro/}}
  (\bibinfo {year} {2023}),\ \bibinfo {note} {accessed: 2023-04-28}\BibitemShut
  {NoStop}%
\bibitem [{\citenamefont {Dawid}\ \emph {et~al.}(2024)\citenamefont {Dawid},
  \citenamefont {Bigagli}, \citenamefont {Savin},\ and\ \citenamefont
  {Will}}]{OurMolRepo}%
  \BibitemOpen
  \bibfield  {author} {\bibinfo {author} {\bibfnamefont {A.}~\bibnamefont
  {Dawid}}, \bibinfo {author} {\bibfnamefont {N.}~\bibnamefont {Bigagli}},
  \bibinfo {author} {\bibfnamefont {D.~W.}\ \bibnamefont {Savin}},\ and\
  \bibinfo {author} {\bibfnamefont {S.}~\bibnamefont {Will}},\ }\href
  {https://doi.org/10.5281/zenodo.13983497} {}\bibinfo {howpublished}
  {\url{https://doi.org/10.5281/zenodo.13983497}} (\bibinfo {year} {2024}),\
  \bibinfo {note} {{GitHub repository: Detection-Of-Laser-Cooled-Molecules
  (Version arXiv2.0).}}\BibitemShut {Stop}%
\bibitem [{\citenamefont {Collopy}\ \emph {et~al.}(2015)\citenamefont
  {Collopy}, \citenamefont {Hummon}, \citenamefont {Yeo}, \citenamefont {Yan},\
  and\ \citenamefont {Ye}}]{Collopy2015YO}%
  \BibitemOpen
  \bibfield  {author} {\bibinfo {author} {\bibfnamefont {A.~L.}\ \bibnamefont
  {Collopy}}, \bibinfo {author} {\bibfnamefont {M.~T.}\ \bibnamefont {Hummon}},
  \bibinfo {author} {\bibfnamefont {M.}~\bibnamefont {Yeo}}, \bibinfo {author}
  {\bibfnamefont {B.}~\bibnamefont {Yan}},\ and\ \bibinfo {author}
  {\bibfnamefont {J.}~\bibnamefont {Ye}},\ }\bibfield  {title} {\bibinfo
  {title} {Prospects for a narrow line {MOT} in {YO}},\ }\href
  {https://doi.org/10.1088/1367-2630/17/5/055008} {\bibfield  {journal}
  {\bibinfo  {journal} {New J. Phys.}\ }\textbf {\bibinfo {volume} {17}},\
  \bibinfo {pages} {055008} (\bibinfo {year} {2015})}\BibitemShut {NoStop}%
\bibitem [{\citenamefont {Collopy}\ \emph {et~al.}(2018)\citenamefont
  {Collopy}, \citenamefont {Ding}, \citenamefont {Wu}, \citenamefont
  {Finneran}, \citenamefont {Anderegg}, \citenamefont {Augenbraun},
  \citenamefont {Doyle},\ and\ \citenamefont {Ye}}]{collopy20183d}%
  \BibitemOpen
  \bibfield  {author} {\bibinfo {author} {\bibfnamefont {A.~L.}\ \bibnamefont
  {Collopy}}, \bibinfo {author} {\bibfnamefont {S.}~\bibnamefont {Ding}},
  \bibinfo {author} {\bibfnamefont {Y.}~\bibnamefont {Wu}}, \bibinfo {author}
  {\bibfnamefont {I.~A.}\ \bibnamefont {Finneran}}, \bibinfo {author}
  {\bibfnamefont {L.}~\bibnamefont {Anderegg}}, \bibinfo {author}
  {\bibfnamefont {B.~L.}\ \bibnamefont {Augenbraun}}, \bibinfo {author}
  {\bibfnamefont {J.~M.}\ \bibnamefont {Doyle}},\ and\ \bibinfo {author}
  {\bibfnamefont {J.}~\bibnamefont {Ye}},\ }\bibfield  {title} {\bibinfo
  {title} {{3D} magneto-optical trap of yttrium monoxide},\ }\href
  {https://doi.org/10.1103/PhysRevLett.121.213201} {\bibfield  {journal}
  {\bibinfo  {journal} {Phys. Rev. Lett.}\ }\textbf {\bibinfo {volume} {121}},\
  \bibinfo {pages} {213201} (\bibinfo {year} {2018})}\BibitemShut {NoStop}%
\bibitem [{\citenamefont {Ding}\ \emph {et~al.}(2020)\citenamefont {Ding},
  \citenamefont {Wu}, \citenamefont {Finneran}, \citenamefont {Burau},\ and\
  \citenamefont {Ye}}]{ding2020sub}%
  \BibitemOpen
  \bibfield  {author} {\bibinfo {author} {\bibfnamefont {S.}~\bibnamefont
  {Ding}}, \bibinfo {author} {\bibfnamefont {Y.}~\bibnamefont {Wu}}, \bibinfo
  {author} {\bibfnamefont {I.~A.}\ \bibnamefont {Finneran}}, \bibinfo {author}
  {\bibfnamefont {J.~J.}\ \bibnamefont {Burau}},\ and\ \bibinfo {author}
  {\bibfnamefont {J.}~\bibnamefont {Ye}},\ }\bibfield  {title} {\bibinfo
  {title} {Sub-doppler cooling and compressed trapping of {YO} molecules at
  $\mu$ k temperatures},\ }\href {https://doi.org/10.1103/PhysRevX.10.021049}
  {\bibfield  {journal} {\bibinfo  {journal} {Phys. Rev. X}\ }\textbf {\bibinfo
  {volume} {10}},\ \bibinfo {pages} {021049} (\bibinfo {year}
  {2020})}\BibitemShut {NoStop}%
\bibitem [{\citenamefont {Bernard}\ \emph {et~al.}(1979)\citenamefont
  {Bernard}, \citenamefont {Bacis},\ and\ \citenamefont
  {Luc}}]{bernard1979fourier}%
  \BibitemOpen
  \bibfield  {author} {\bibinfo {author} {\bibfnamefont {A.}~\bibnamefont
  {Bernard}}, \bibinfo {author} {\bibfnamefont {R.}~\bibnamefont {Bacis}},\
  and\ \bibinfo {author} {\bibfnamefont {P.}~\bibnamefont {Luc}},\ }\bibfield
  {title} {\bibinfo {title} {Fourier transform spectroscopy: {E}xtensive
  analysis of the ${A} {}^{2} {\Pi} \rightarrow {X} {}^{2} {\Sigma}^+$ and ${B}
  {}^{2} {\Sigma}^+ \rightarrow {X} {}^{2} {\Sigma}^+$ systems of yttrium
  oxide},\ }\href {https://adsabs.harvard.edu/full/1979ApJ...227..338B}
  {\bibfield  {journal} {\bibinfo  {journal} {Astrophys. J.}\ }\textbf
  {\bibinfo {volume} {227}},\ \bibinfo {pages} {338} (\bibinfo {year}
  {1979})}\BibitemShut {NoStop}%
\bibitem [{\citenamefont {Merer}\ \emph {et~al.}(1975)\citenamefont {Merer},
  \citenamefont {Malm}, \citenamefont {Martin}, \citenamefont {Horani},\ and\
  \citenamefont {Rostas}}]{merer1975ultraviolet}%
  \BibitemOpen
  \bibfield  {author} {\bibinfo {author} {\bibfnamefont {A.}~\bibnamefont
  {Merer}}, \bibinfo {author} {\bibfnamefont {D.}~\bibnamefont {Malm}},
  \bibinfo {author} {\bibfnamefont {R.}~\bibnamefont {Martin}}, \bibinfo
  {author} {\bibfnamefont {M.}~\bibnamefont {Horani}},\ and\ \bibinfo {author}
  {\bibfnamefont {J.}~\bibnamefont {Rostas}},\ }\bibfield  {title} {\bibinfo
  {title} {The ultraviolet emission spectra of $\mathrm{OH}^+$ and
  $\mathrm{OD}^+$. {R}otational structure and perturbations in the a$^3
  \pi$--x$^3 \sigma^-$ transition},\ }\href {https://doi.org/10.1139/p75-037}
  {\bibfield  {journal} {\bibinfo  {journal} {Can. J. Phys.}\ }\textbf
  {\bibinfo {volume} {53}},\ \bibinfo {pages} {251} (\bibinfo {year}
  {1975})}\BibitemShut {NoStop}%
\bibitem [{\citenamefont {Zhang}\ \emph {et~al.}(2018)\citenamefont {Zhang},
  \citenamefont {Yang}, \citenamefont {Wang}, \citenamefont {Ma},\ and\
  \citenamefont {Liu}}]{zhang2018theoretical}%
  \BibitemOpen
  \bibfield  {author} {\bibinfo {author} {\bibfnamefont {Q.-Q.}\ \bibnamefont
  {Zhang}}, \bibinfo {author} {\bibfnamefont {C.-L.}\ \bibnamefont {Yang}},
  \bibinfo {author} {\bibfnamefont {M.-S.}\ \bibnamefont {Wang}}, \bibinfo
  {author} {\bibfnamefont {X.-G.}\ \bibnamefont {Ma}},\ and\ \bibinfo {author}
  {\bibfnamefont {W.-W.}\ \bibnamefont {Liu}},\ }\bibfield  {title} {\bibinfo
  {title} {A theoretical study on the laser cooling scheme for the
  three-energy-level system of the {CN} molecule},\ }\href
  {https://doi.org/10.1088/1361-6455/aad032} {\bibfield  {journal} {\bibinfo
  {journal} {J. Phys. B: At. Mol. Opt. Phys}\ }\textbf {\bibinfo {volume}
  {51}},\ \bibinfo {pages} {155102} (\bibinfo {year} {2018})}\BibitemShut
  {NoStop}%
\bibitem [{\citenamefont {Corder}\ \emph {et~al.}(2015)\citenamefont {Corder},
  \citenamefont {Arnold}, \citenamefont {Hua},\ and\ \citenamefont
  {Metcalf}}]{corder2015laser}%
  \BibitemOpen
  \bibfield  {author} {\bibinfo {author} {\bibfnamefont {C.}~\bibnamefont
  {Corder}}, \bibinfo {author} {\bibfnamefont {B.}~\bibnamefont {Arnold}},
  \bibinfo {author} {\bibfnamefont {X.}~\bibnamefont {Hua}},\ and\ \bibinfo
  {author} {\bibfnamefont {H.}~\bibnamefont {Metcalf}},\ }\bibfield  {title}
  {\bibinfo {title} {Laser cooling without spontaneous emission using the
  bichromatic force},\ }\href@noop {} {\bibfield  {journal} {\bibinfo
  {journal} {JOSA B}\ }\textbf {\bibinfo {volume} {32}},\ \bibinfo {pages}
  {B75} (\bibinfo {year} {2015})}\BibitemShut {NoStop}%
\bibitem [{\citenamefont {Wenz}\ \emph {et~al.}(2020)\citenamefont {Wenz},
  \citenamefont {Kozyryev}, \citenamefont {McNally}, \citenamefont {Aldridge},\
  and\ \citenamefont {Zelevinsky}}]{wenz2020large}%
  \BibitemOpen
  \bibfield  {author} {\bibinfo {author} {\bibfnamefont {K.}~\bibnamefont
  {Wenz}}, \bibinfo {author} {\bibfnamefont {I.}~\bibnamefont {Kozyryev}},
  \bibinfo {author} {\bibfnamefont {R.~L.}\ \bibnamefont {McNally}}, \bibinfo
  {author} {\bibfnamefont {L.}~\bibnamefont {Aldridge}},\ and\ \bibinfo
  {author} {\bibfnamefont {T.}~\bibnamefont {Zelevinsky}},\ }\bibfield  {title}
  {\bibinfo {title} {Large molasses-like cooling forces for molecules using
  polychromatic optical fields: A theoretical description},\ }\href@noop {}
  {\bibfield  {journal} {\bibinfo  {journal} {Phys. Rev. Res.}\ }\textbf
  {\bibinfo {volume} {2}},\ \bibinfo {pages} {043377} (\bibinfo {year}
  {2020})}\BibitemShut {NoStop}%
\bibitem [{\citenamefont {Yelin}\ \emph {et~al.}(2006)\citenamefont {Yelin},
  \citenamefont {Kirby},\ and\ \citenamefont {C\^ot\'e}}]{Yelin2006molqubit}%
  \BibitemOpen
  \bibfield  {author} {\bibinfo {author} {\bibfnamefont {S.~F.}\ \bibnamefont
  {Yelin}}, \bibinfo {author} {\bibfnamefont {K.}~\bibnamefont {Kirby}},\ and\
  \bibinfo {author} {\bibfnamefont {R.}~\bibnamefont {C\^ot\'e}},\ }\bibfield
  {title} {\bibinfo {title} {Schemes for robust quantum computation with polar
  molecules},\ }\href {https://doi.org/10.1103/PhysRevA.74.050301} {\bibfield
  {journal} {\bibinfo  {journal} {Phys. Rev. A}\ }\textbf {\bibinfo {volume}
  {74}},\ \bibinfo {pages} {050301(R)} (\bibinfo {year} {2006})}\BibitemShut
  {NoStop}%
\bibitem [{\citenamefont {Zelevinsky}\ \emph {et~al.}(2008)\citenamefont
  {Zelevinsky}, \citenamefont {Kotochigova},\ and\ \citenamefont
  {Ye}}]{Zelevinsky2008massratio}%
  \BibitemOpen
  \bibfield  {author} {\bibinfo {author} {\bibfnamefont {T.}~\bibnamefont
  {Zelevinsky}}, \bibinfo {author} {\bibfnamefont {S.}~\bibnamefont
  {Kotochigova}},\ and\ \bibinfo {author} {\bibfnamefont {J.}~\bibnamefont
  {Ye}},\ }\bibfield  {title} {\bibinfo {title} {Precision test of mass-ratio
  variations with lattice-confined ultracold molecules},\ }\href
  {https://doi.org/10.1103/PhysRevLett.100.043201} {\bibfield  {journal}
  {\bibinfo  {journal} {Phys. Rev. Lett.}\ }\textbf {\bibinfo {volume} {100}},\
  \bibinfo {pages} {043201} (\bibinfo {year} {2008})}\BibitemShut {NoStop}%
\bibitem [{\citenamefont {Vitanov}\ \emph {et~al.}(2017)\citenamefont
  {Vitanov}, \citenamefont {Rangelov}, \citenamefont {Shore},\ and\
  \citenamefont {Bergmann}}]{Vitanov2017stirap}%
  \BibitemOpen
  \bibfield  {author} {\bibinfo {author} {\bibfnamefont {N.~V.}\ \bibnamefont
  {Vitanov}}, \bibinfo {author} {\bibfnamefont {A.~A.}\ \bibnamefont
  {Rangelov}}, \bibinfo {author} {\bibfnamefont {B.~W.}\ \bibnamefont
  {Shore}},\ and\ \bibinfo {author} {\bibfnamefont {K.}~\bibnamefont
  {Bergmann}},\ }\bibfield  {title} {\bibinfo {title} {Stimulated {Raman}
  adiabatic passage in physics, chemistry, and beyond},\ }\href
  {https://doi.org/10.1103/RevModPhys.89.015006} {\bibfield  {journal}
  {\bibinfo  {journal} {Rev. Mod. Phys.}\ }\textbf {\bibinfo {volume} {89}},\
  \bibinfo {pages} {015006} (\bibinfo {year} {2017})}\BibitemShut {NoStop}%
\bibitem [{\citenamefont {Panda}\ \emph {et~al.}(2016)\citenamefont {Panda},
  \citenamefont {O'Leary}, \citenamefont {West}, \citenamefont {Baron},
  \citenamefont {Hess}, \citenamefont {Hoffman}, \citenamefont {Kirilov},
  \citenamefont {Overstreet}, \citenamefont {West}, \citenamefont {DeMille},
  \citenamefont {Doyle},\ and\ \citenamefont {Gabrielse}}]{Panda2016STIRAP}%
  \BibitemOpen
  \bibfield  {author} {\bibinfo {author} {\bibfnamefont {C.~D.}\ \bibnamefont
  {Panda}}, \bibinfo {author} {\bibfnamefont {B.~R.}\ \bibnamefont {O'Leary}},
  \bibinfo {author} {\bibfnamefont {A.~D.}\ \bibnamefont {West}}, \bibinfo
  {author} {\bibfnamefont {J.}~\bibnamefont {Baron}}, \bibinfo {author}
  {\bibfnamefont {P.~W.}\ \bibnamefont {Hess}}, \bibinfo {author}
  {\bibfnamefont {C.}~\bibnamefont {Hoffman}}, \bibinfo {author} {\bibfnamefont
  {E.}~\bibnamefont {Kirilov}}, \bibinfo {author} {\bibfnamefont {C.~B.}\
  \bibnamefont {Overstreet}}, \bibinfo {author} {\bibfnamefont {E.~P.}\
  \bibnamefont {West}}, \bibinfo {author} {\bibfnamefont {D.}~\bibnamefont
  {DeMille}}, \bibinfo {author} {\bibfnamefont {J.~M.}\ \bibnamefont {Doyle}},\
  and\ \bibinfo {author} {\bibfnamefont {G.}~\bibnamefont {Gabrielse}},\
  }\bibfield  {title} {\bibinfo {title} {Stimulated {Raman} adiabatic passage
  preparation of a coherent superposition of tho
  ${H}^{3}{\mathrm{\ensuremath{\Delta}}}_{1}$ states for an improved electron
  electric-dipole-moment measurement},\ }\href
  {https://doi.org/10.1103/PhysRevA.93.052110} {\bibfield  {journal} {\bibinfo
  {journal} {Phys. Rev. A}\ }\textbf {\bibinfo {volume} {93}},\ \bibinfo
  {pages} {052110} (\bibinfo {year} {2016})}\BibitemShut {NoStop}%
\bibitem [{NIS(2018)}]{NIST}%
  \BibitemOpen
  \href@noop {} {\bibinfo {title} {{The NIST Reference on Constants, Units, and
  Uncertainty. CODATA recommended values}}} (\bibinfo {year} {2018}),\ \bibinfo
  {note} {used values:
  \href{https://physics.nist.gov/cgi-bin/cuu/Value?kshcminv|search_for=Boltzmann}{Boltzmann
  constant}, \href{https://physics.nist.gov/cgi-bin/cuu/Value?h\#mid}{Planck
  constant},
  \href{https://physics.nist.gov/cgi-bin/cuu/Convert?exp=0&num=&From=k&To=j&Action=Only+show+factor}{K
  to J conversion factor},
  \href{https://physics.nist.gov/cgi-bin/cuu/Convert?exp=0&num=&From=u&To=kg&Action=Only+show+factor}{u
  to kg conversion factor},
  \href{https://physics.nist.gov/cgi-bin/cuu/Value?c\#mid}{speed of light in
  vacuum}}\BibitemShut {NoStop}%
\end{thebibliography}%

\appendix
\onecolumngrid

\newpage
\clearpage
\section{Extending graph search to laser cooling schemes with two $S_1$ states}\label{app:algo_extension}

Expanding the graph search to include an extra excited state $S_1$ is of interest, as it usually leads to a shorter cooling time $t_{\rm cool}$. We focus here on cooling schemes of the type as in Fig.~\ref{fig:schemes}(b). This involves searching for more complex subgraphs, so we need to modify the search conditions. We also need to use formulas for cooling time, closure, and scattering rate adjusted for the presence of an additional subscheme. We follow here the rate model by Fitch \& Tarbutt, described in Sec.~3.1 in Ref.~\cite{Fitch2021chapter}.

The graph search for double-$S_1$ laser cooling schemes of the type as in Fig.~\ref{fig:schemes}(b) requires first running the search for single-$S_1$ schemes as in Fig.~\ref{fig:schemes}(a), but with a relaxed condition on their viability. Recall from Sec.~\ref{ssec:graph_algo} that we consider a cooling scheme to be viable if the cooling to a desired temperature is completed before 90\% of molecules are lost from the sample, i.e., $t_{\rm cool} < t_{10\%}$ (or equivalently $n_{\rm cool} < n_{10\%}$). Anticipating at best a double speedup from including a second excited state, we consider all single-$S_1$ cooling schemes that meet the condition of $\frac{t_{\rm cool}}{2} < t_{10\%}$ (or equivalently $\frac{n_{\rm cool}}{2} < n_{10\%}$) and label participating $S_1$ and $S_2$ states as $S_1$ and $S_2$ candidates to participate in double-$S_1$ schemes. Thanks to this preprocessing step, we greatly reduce the number of states that the more complex search needs to comb through, and the resulting runtime of this more complex search query is shorter than the one of the search for single-$S_1$ schemes.

Next, among candidates for $S_1$ states, we search for two excited states, $S_1^A$ and $S_1^B$, connected with the same starting state. Moreover, note that the rate model for multi-$S_1$ states that we follow here assumes the same lifetime of the excited states (see Eq.~(34) in Ref.~\cite{Fitch2021chapter}). This is an approximation. To minimize any potential errors, we limit the search to $S_1$ pairs with similar lifetimes. We assumed that they can differ at most by 50\%, but this condition can be modified by a user. Then we assume that the $S_1$ lifetime is an average lifetime of both states:
\begin{equation}\label{eq:lifetime_2S}
    \tau = \frac{1}{2} (\tau_A + \tau_B)\,.
\end{equation}
Finally, we require that both excited states $S_1^A$ and $S_1^B$ decay to the same reachable states $S_2$, and we identify them among candidates for $S_2$ states. This requirement also comes from the assumptions built into the rate model by Fitch \& Tarbutt.

To estimate the properties of the identified double-$S_1$ schemes, we need to employ modified formulas for cooling time, closure, and scattering rate, taking into account an existence of two subschemes within a cooling scheme. The closure of a double-$S_1$ cooling scheme is simply given by an average of closures of two composing subschemes:
\begin{equation}\label{eq:closure_doubleS}
    p = \frac{1}{2} ( \sum_i \mathrm{BR^A}_{i} + \sum_j \mathrm{BR^B}_{j} ) = \frac{1}{2} ( p_A + p_B ) \,,
\end{equation}
where $i$ and $j$ run only over driven transitions of subschemes A and B, respectively. The scattering rate for a double-$S_1$ cooling scheme \cite{Fitch2021chapter} is
\begin{equation}\label{eq:R_doubleS}
    R^{-1} = \tau\left[ \frac{(N_{\rm g}+N_{\rm e})}{N_{\rm e}} + \frac{2}{3} \frac{\Gamma \pi h c}{N_{\rm e}} \sum_l \frac{\sum_k \mathrm{BR}_{lk}}{ \sum_k I_{lk} \lambda_{lk}^3 \mathrm{BR}_{lk}}\right] = \tau\left[ ( 1 + \frac{N_{\rm g}}{2} ) + \frac{1}{3} \Gamma \pi h c \sum_l \frac{\mathrm{BR}_{l,\mathrm{A}} + \mathrm{BR}_{l,\mathrm{B}}}{ I_{l,\mathrm{A}} \lambda_{l,\mathrm{A}}^3 \mathrm{BR}_{l,\mathrm{A}} + I_{l,\mathrm{B}} \lambda_{l,\mathrm{B}}^3 \mathrm{BR}_{l,\mathrm{B}}}\right]\,.
\end{equation}
Here, $N_{\rm e} = 2$ is the number of excited states $S_1$, $N_{\rm g}$ is the number of $S_2$ states (from which molecules are driven to $S_1$ states), $\Gamma \equiv 1/\tau$, $h$ is Planck’s constant, $c$ is the speed of light, $I_{kl}$ is the intensity of the laser addressing the transition between $k$-th excited $S_1$ state and $l$-th lower-energy $S_2$ state, and $\lambda_{kl}$ is the wavelength of the $kl$-th addressed transition.
Finally, there is a difference in an average momentum of a photon that participates in a cooling scheme, $p_{\rm mean}$. This impacts the formula for the number of scatterings necessary to reach rest, $n_{\rm cool}$. The average photon momentum is an average of mean photon momenta of two subschemes weighted by the respective closures:
\begin{equation}\label{eq:pmean_doubleS}
   p_{\rm mean} = \frac{h}{2} (\frac{1}{p_A} \sum_i \frac{\mathrm{BR^A}_i}{\lambda^A_i} + \frac{1}{p_B} \sum_j \frac{\mathrm{BR^B}_j}{\lambda^{\rm B}_j} )\,,
\end{equation}
where $i$ and $j$ run over driven transitions of subschemes A and B, respectively. The number of scattering processes is therefore
\begin{equation}\label{eq:ncool_doubleS}
    n_{\rm cool} = \frac{p_{\rm init}}{p_{\rm mean}} = \frac{2}{h} \sqrt{3 k_{\rm B} T_{\rm init} m} \left( \frac{1}{p_A} \sum_i \frac{\mathrm{BR^A}_i}{\lambda^A_i} + \frac{1}{p_B} \sum_j \frac{\mathrm{BR^B}_j}{\lambda^{\rm B}_j} \right)^{-1} \,.
\end{equation}

\section{Selected identified laser cooling schemes}\label{app:schemes}
In this appendix, we present the most promising laser cooling schemes identified for ${}^{89}$Y${}^{14}$O~\cite{Yurchenko2023YO}, ${}^{12}$C$_2$~\cite{Yurchenko2018C2, mckemmish2020update}, ${}^{16}$O${}^{1}$H$^+$~\cite{Hodges2017OH, Bernath2020MoLLIST}, ${}^{12}$C${}^{14}$N~\cite{Brooke2014CN, Syme2020CN, Syme2021CN}, and ${}^{12}$C${}^{16}$O$_2$~\cite{Yurchenko2020CO2}, based on data from ExoMol. It would be unfeasible to present every scheme identified here since, even just for the simplest studied molecule, ${}^{12}$C$_2$, we detected in total 388 laser cooling schemes. We count schemes separately if they are composed of a unique set of $S_0$, $S_1$, and $S_2$ states. Therefore, we refer the interested reader to the GitHub repository accompanying this paper~\cite{OurMolRepo}, where we store the full list of schemes. Note that the schemes identified meet the conditions described in Sec.~\ref{ssec:graph_algo}. In particular, their $t_{\rm cool} (4\,\mathrm{K}) < t_{10\%}$; but we put no other constraint on the total cooling time. So depending on the molecule, the detected schemes may have very long cooling times of up to seconds (if closure permits it) and therefore may not yet be infeasible, given current technological capabilities for ultracold experiments.

\subsection{YO}\label{app-ss:YO}

\begin{figure}[t]
    \centering
    \includegraphics[width=0.95\columnwidth]{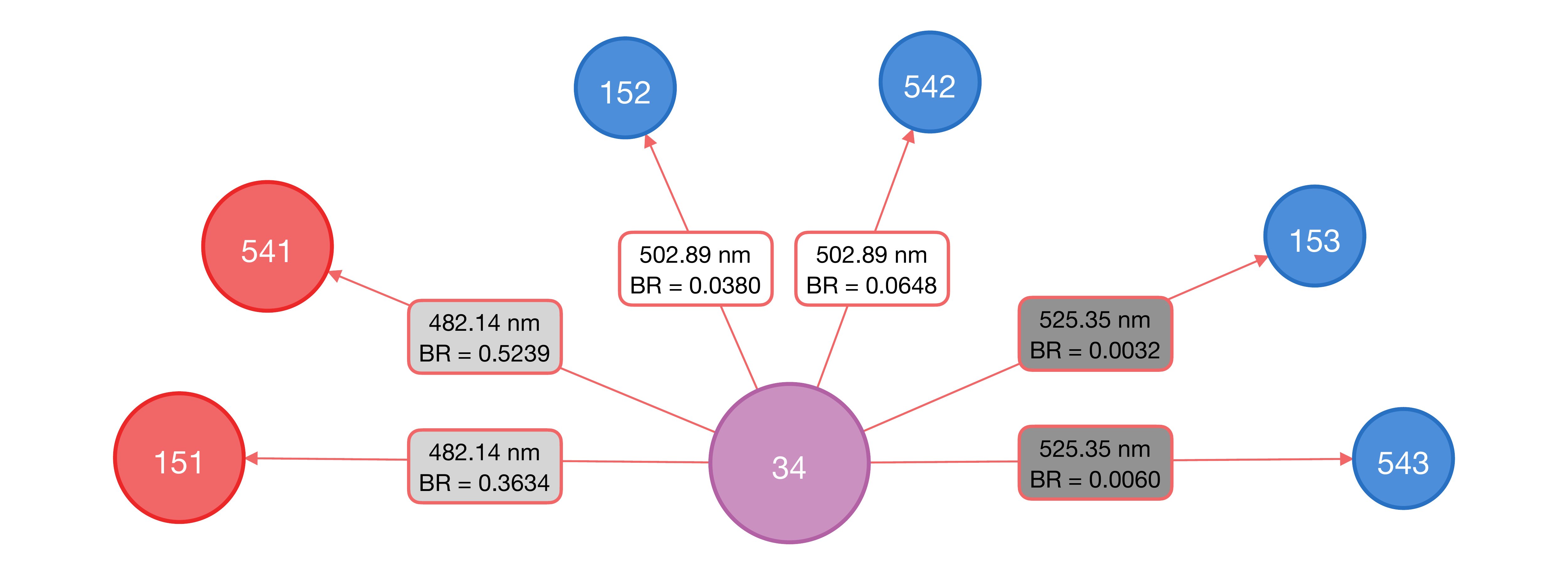}
    \vspace{-0.3cm}
    \caption{The fastest identified cooling scheme for YO with three lasers (can be much improved by adding two more lasers). The numbers in each circle give the state ID (each state is described in detail in Table~\ref{tab:YO_scheme}). The box on each arrow gives the laser wavelength and the BR of the transition from the higher energy state to the lower energy state. $S_0$ states are red and $S_1$ states are purple. States of $S_2 \notin \{S_0, S_1\}$ are shown in blue. Note that each pair of transitions here can be addressed with one laser. The resulting three-laser cooling scheme is characterized by $T_{\rm init} = 4\,$K, $R^{-1} = 0.216 \, \mu \mathrm{s}$, $n_{\rm cool} = 2845$, $t_{\rm cool} = 615\,\mu$s, $\frac{n_{\rm cool}}{n_{10\%}} = 0.966$, and $p = 0.9992$.  The closure can be improved up to by adding two lasers of $\lambda = 549.73$ (achieving $\frac{n_{\rm cool}}{n_{10\%}} = 0.232$, $t_{\rm cool} = 790\,\mu$s, and $p = 0.9998$) and $1602.16\,$nm (achieving $\frac{n_{\rm cool}}{n_{10\%}} = 0.078$, $t_{\rm cool} = 878\,\mu$s, and $p = 0.99994$), as described in the text.}
    \vspace{-0.3cm}
    \label{fig:YO-scheme}
\end{figure}

${}^{89}$Y${}^{14}$O has already been laser cooled experimentally \cite{Collopy2015YO,yeo2015rotational,collopy20183d,ding2020sub}. Here, besides recovering these laser cooling schemes, we have identified over 1700 new schemes, based on about 140 excited states, from \href{https://www.exomol.com/data/molecules/YO/89Y-16O/BRYTS/}{ExoMol} data~\cite{Yurchenko2023YO}. Interestingly, the fastest schemes are not the known ones based on $A^2 \Pi_{1/2}$ as the excited $S_1$ state, but the newly identified ones based on $B^2 \Sigma$. Given that schemes employing states from the $A^2 \Pi_{1/2}$ manifold have been presented extensively in the context of their experimental realization, we focus here on the $B^2 \Sigma$ manifold. In particular, we present the fastest of such schemes in Fig.~\ref{fig:YO-scheme}. It also has a twin counterpart with states of opposite parity.

The scheme in Fig.~\ref{fig:YO-scheme} is based on the main transition $X^2 \Sigma^+ \rightarrow B^2 \Sigma$. With just three lasers (each addressing a pair of transitions that are at the same frequency but lead to states of different $e/f$ parity) it achieves a closure of $p = 0.9992$ and $t_{\rm cool} = 615\,\mu$s with $R^{-1} = 0.216 \, \mu \mathrm{s}$ and $n_{\rm cool} = 2845$. One more laser of $\lambda = 549.73\,$nm can be added to address two additional transitions $34 \rightarrow 154$ and $34 \rightarrow 544$, resulting in $R^{-1} = 0.278 \, \mu \mathrm{s}$, $t_{\rm cool} = 790\,\mu$s, $\frac{n_{\rm cool}}{n_{10\%}} = 0.232$, and $p = 0.9998$. One can continue to address leakage to pairs of higher vibrational states, but it is more beneficial to use the next laser of $\lambda = 1602.16\,$nm to address the transition $34 \rightarrow 559$ ($A^{\prime 2} \Delta$), which produces $R^{-1} = 0.309 \, \mu \mathrm{s}$, $t_{\rm cool} = 878\,\mu$s, $\frac{n_{\rm cool}}{n_{10\%}} = 0.078$, and $p = 0.99994$, yielding a very efficient cycling scheme using only five lasers.

\begin{table}[b]
\vspace{-0.3cm}
\caption{Molecular states involved in the laser cooling scheme presented in Figs.~\ref{fig:YO-scheme}-\ref{fig:YO-double-scheme}. Abbreviations follow Table 1 in Ref.~\cite{Yurchenko2023YO}}
\label{tab:YO_scheme}
\begin{tabular}{|c|c|l|l}
\cline{1-3}
state id  & role   & state properties \footnotesize{$^*$Energies are in cm$^{-1}$, $\tau$ in s, $\tau = 10000\,$s is the ground or metastable state}                    &  \\ \cline{1-3}
\multirow{2}{*}{22} & \multirow{2}{*}{\textcolor{Orchid}{$S_1'$}} & \{ef: "e", gi: 4, parity: "+", g: 0.007737, tau: 3.5473e-8, J: 0.5, Omega: 0.5, v: 0, &  \\
                 &     & Epsilon: -0.5, id: "22", state: "A2Pi", Lambda: 1, energy: 16295.142495\}                          &  \\ \cline{1-3}
\multirow{2}{*}{34} & \multirow{2}{*}{\textcolor{Orchid}{$S_1$}} & \{ef: "e", gi: 4, parity: "+", g: 1.995791, tau: 3.073e-8, J: 0.5, Omega: 0.5, v: 0, &  \\
                   &   & Epsilon: 0.5, id: "34", state: "B2Sigma", Lambda: 0, energy: 20741.6877\}                          &  \\ \cline{1-3}
\multirow{2}{*}{151} & \multirow{2}{*}{\textcolor{RedOrange}{$S_0$}} & \{ef: "f", gi: 4, parity: "-", g: -0.66741, tau: 942950.0, J: 0.5, Omega: -0.5, v: 0,  &  \\
                  &    & Epsilon: -0.5, id: "151", state: "X2Sigma+", Lambda: 0, energy: 0.7761\}                               &  \\ \cline{1-3}
\multirow{2}{*}{152} & \multirow{2}{*}{\textcolor{NavyBlue}{$S_2$}} & \{ef: "f", gi: 4, parity: "-", g: -0.66741, tau: 0.05211, J: 0.5, Omega: -0.5, v: 1, &  \\
                      & & Epsilon: -0.5, id: "152", state: "X2Sigma+", Lambda: 0, energy: 856.519\}                              &  \\ \cline{1-3}
\multirow{2}{*}{153} & \multirow{2}{*}{\textcolor{NavyBlue}{$S_2$}} & \{ef: "f", gi: 4, parity: "-", g: -0.66741, tau: 0.026311, J: 0.5, Omega: -0.5, v: 2, &  \\
                     & & Epsilon: -0.5, id: "153", state: "X2Sigma+", Lambda: 0, energy: 1706.6031\}                             &  \\ \cline{1-3}
\multirow{2}{*}{154} & \multirow{2}{*}{\textcolor{NavyBlue}{$S_2$}} & \{ef: "f", gi: 4, parity: "-", g: -0.667411, tau: 0.017713, J: 0.5, Omega: -0.5, v: 3, &  \\
                     & & Epsilon: -0.5, id: "154", state: "X2Sigma+", Lambda: 0, energy: 2551.0341\}                             &  \\ \cline{1-3}
\multirow{2}{*}{541} & \multirow{2}{*}{\textcolor{RedOrange}{$S_0$}} & \{ef: "e", gi: 8, parity: "-", g: 0.667402, tau: 997510.0, J: 1.5, Omega: -0.5, v: 0, &  \\
                     & & Epsilon: -0.5, id: "541", state: "X2Sigma+", Lambda: 0, energy: 0.7761\}                          &  \\ \cline{1-3}
\multirow{2}{*}{542} & \multirow{2}{*}{\textcolor{NavyBlue}{$S_2$}} & \{ef: "e", gi: 8, parity: "-", g: 0.667402, tau: 0.052102, J: 1.5, Omega: -0.5, v: 1, &  \\
                     & & Epsilon: -0.5, id: "542", state: "X2Sigma+", Lambda: 0, energy: 856.519\}                          &  \\ \cline{1-3}
\multirow{2}{*}{543} & \multirow{2}{*}{\textcolor{NavyBlue}{$S_2$}} & \{ef: "e", gi: 8, parity: "-", g: 0.667403, tau: 0.026307, J: 1.5, Omega: -0.5, v: 2, &  \\
                    & & Epsilon: -0.5, id: "543", state: "X2Sigma+", Lambda: 0, energy: 1706.6031\}                          &  \\ \cline{1-3}
\multirow{2}{*}{544} & \multirow{2}{*}{\textcolor{NavyBlue}{$S_2$}} & \{ef: "e", gi: 8, parity: "-", g: 0.667403, tau: 0.01771, J: 1.5, Omega: -0.5, v: 3, &  \\
                    &  & Epsilon: -0.5, id: "544", state: "X2Sigma+", Lambda: 0, energy: 2551.0341\}                          &  \\ \cline{1-3}
\multirow{2}{*}{559} & \multirow{2}{*}{\textcolor{NavyBlue}{$S_2$}} & \{ef: "e", gi: 8, parity: "-", g: 0.39954, tau: 0.0009435, J: 1.5, Omega: -1.5, v: 0, &  \\
                    &  & Epsilon: 0.5, id: "559", state: "Ap2Delta", Lambda: -2, energy: 14500.12397\}                          &  \\ \cline{1-3}                      
\end{tabular}
\end{table}

\begin{figure}[t]
    \centering
    \includegraphics[width=0.95\columnwidth]{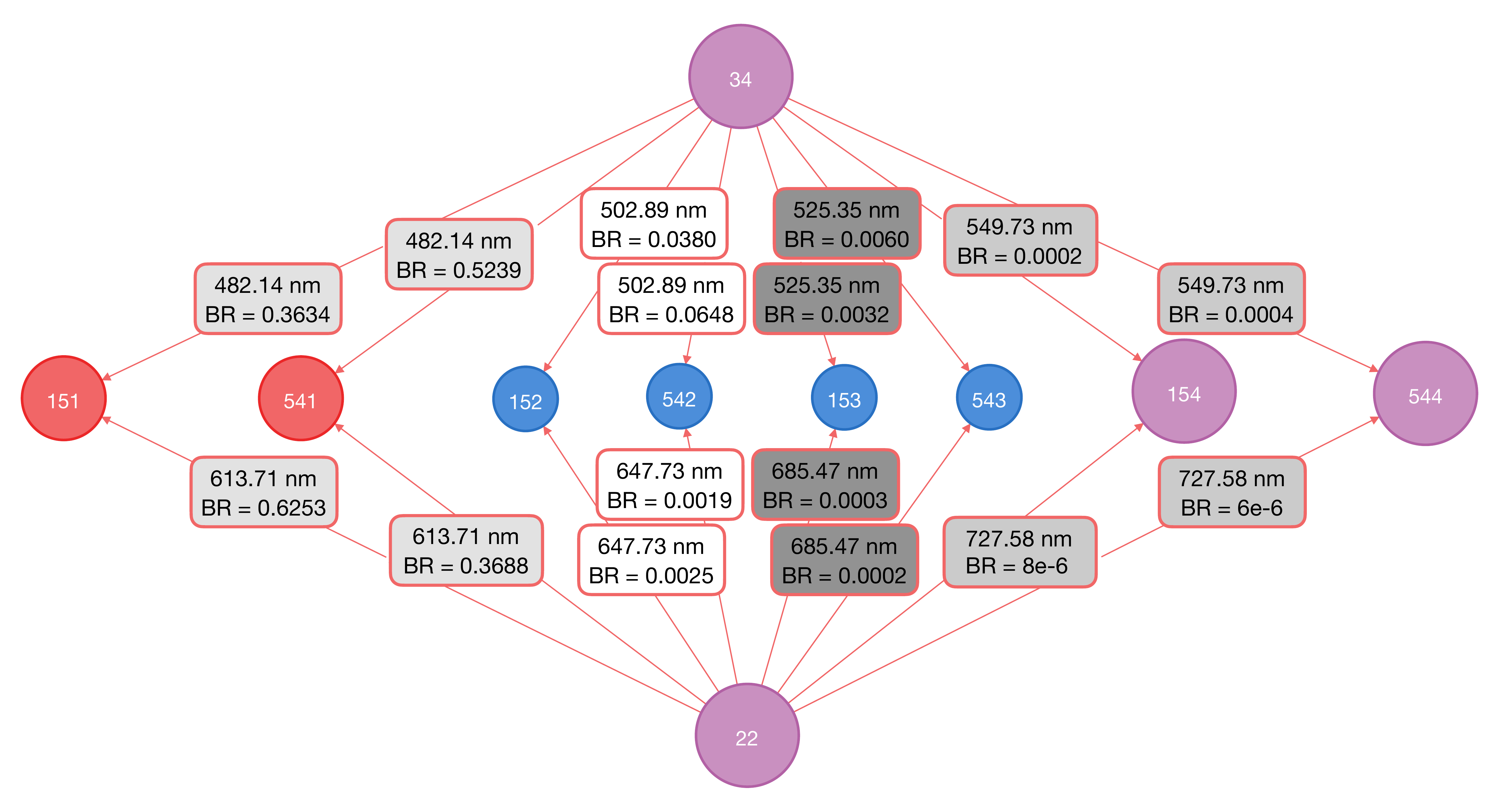}
    \vspace{-0.3cm}
    \caption{The fastest identified cooling scheme for YO based on two $S_1$ states, requiring eight lasers. Note that each pair of transitions here can be addressed with one laser. The resulting eight-laser cooling scheme is characterized by $T_{\rm init} = 4\,$K, $R^{-1} = 0.167 \, \mu \mathrm{s}$, $n_{\rm cool} = 3180$, $t_{\rm cool} = 530\,\mu$s, $\frac{n_{\rm cool}}{n_{10\%}} = 0.823$, and $p = 0.9994$. The color convention is described in the caption of Fig.~\ref{fig:YO-scheme}.}
    \vspace{-0.3cm}
    \label{fig:YO-double-scheme}
\end{figure}

We also identified several viable laser cooling schemes based on two $S_1$ states, of the type as in Fig.~\ref{fig:schemes}(b), so using the algorithm extension described in App.~\ref{app:algo_extension}. We present the most interesting double-$S_1$ scheme in Fig.~\ref{fig:YO-double-scheme}. It is an extension of the scheme discussed above and presented in Fig.~\ref{fig:YO-scheme}, constructed by adding a lowest-energy state ($v=0$, $J=0.5$) from another electronic manifold, $A^2 \Pi$, that decays to the same $S_2$ states. Most excitingly, it offers an additional speedup ($R^{-1} = 0.167 \, \mu \mathrm{s}$, $t_{\rm cool} = 530\,\mu$s) to the single-$S_1$ cooling scheme at the expense of a larger number of lasers (eight) and worse closure ($\frac{n_{\rm cool}}{n_{10\%}} = 0.824$, and $p = 0.9994$). Interestingly, another identified viable double-$S_1$ cooling scheme is also an extension of the scheme in Fig.~\ref{fig:YO-scheme}, but this time the second $S_1$ state is a higher vibrational state from $A^2 \Pi$ electronic manifold. Its properties are: $T_{\rm init} = 4\,$K, $R^{-1} = 0.185 \, \mu \mathrm{s}$, $n_{\rm cool} = 3184$, $t_{\rm cool} = 588\,\mu$s, $\frac{n_{\rm cool}}{n_{10\%}} = 0.705$, and $p = 0.9994$.

For YO, we identified in total 1781 schemes based on 140 different $S_1$ states: 4 cooling schemes based on 6 transitions, 10 with 7 transitions, 33 with 8 transitions, 97 with 9 transitions, 155 with 10 transitions, 230 with 11 transitions, 256 with 12 transitions, 286 with 13 transitions, 344 with 14 transitions, and 366 with 15 transitions. Note that some pairs of transitions require only one laser because they lead to degenerate states. Almost all schemes have very short cooling times below 2 ms.

\newpage
\clearpage
\subsection{\texorpdfstring{C\textsubscript{2}}{C2}}\label{app-ss:C2}

\begin{figure}[b]
    \centering
    \includegraphics[width=0.99\columnwidth]{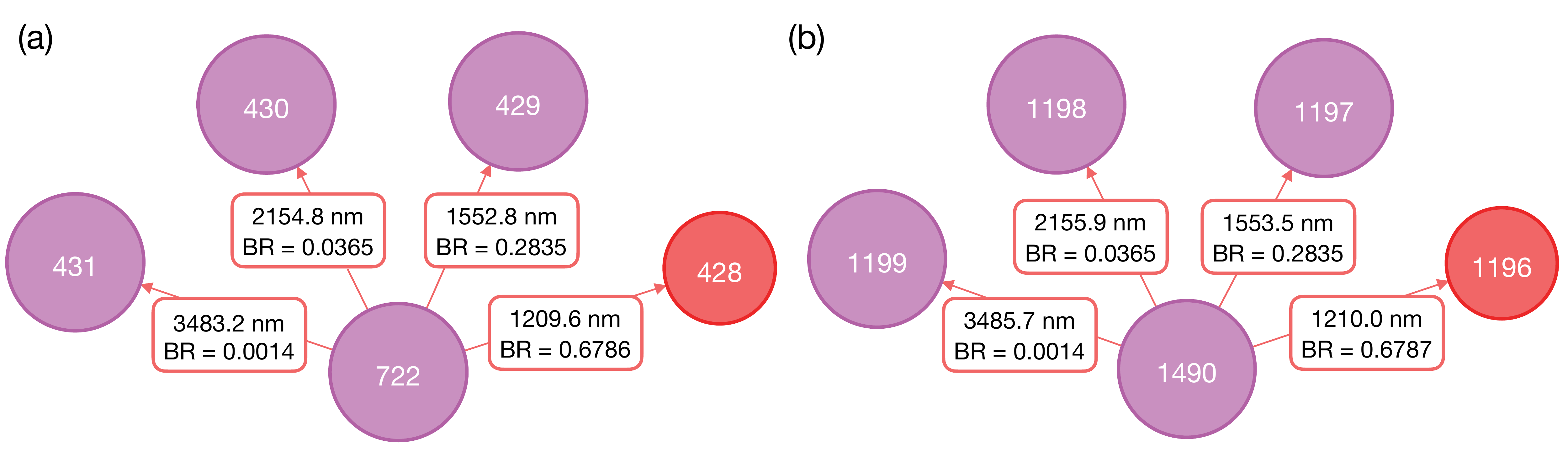}
    \caption{Laser cooling schemes based on higher rotational branches for C$_2$. We present here the cooling time for the lowest energy $S_0$. (a) The cooling scheme with four lasers based on $J=2$ states with $T_{\rm init} = 6.8\,$K, $R^{-1} = 65.7 \, \mu \mathrm{s}$, $n_{\rm cool} = 6643$, $t_{\rm cool} = 436\,$ms, $\frac{n_{\rm cool}}{n_{10\%}} = 0.016$, and closure $p = 0.999994$. (b) Same as (a) but for $J=4$, with $T_{\rm init} = 22.6\,$K, $R^{-1} = 65.8 \, \mu \mathrm{s}$, $n_{\rm cool} = 12126$, $t_{\rm cool} = 797\,$s, $\frac{n_{\rm cool}}{n_{10\%}} = 0.030$, and $p = 0.999994$. The color convention is described in the caption of Fig.~\ref{fig:YO-scheme}.}
    \label{fig:C2-J-schemes}
\end{figure}

We present here three of the most interesting cooling schemes identified for ${}^{12}$C$_2$ based on \href{https://www.exomol.com/data/molecules/C2/12C2/8states/}{ExoMol} data~\cite{Yurchenko2018C2, mckemmish2020update}. The first interesting observation is that we identified a series of laser cooling schemes relying on states with increasing even rotational quantum numbers, $J=2,4,6,\ldots$. The schemes for $J=2$ and $J=4$ are presented in Fig.~\ref{fig:C2-J-schemes}(a) and (b), respectively. The figure caption contains more information such as the cooling time and closure. The states taking part in those schemes are listed in Tab.~\ref{tab:C2_J2_scheme} and~\ref{tab:C2_J4_scheme}, respectively. Each of these four-laser schemes for $J=2,4,6,\ldots$ involve an $S_1$ with $v=0$.

\begin{table}[h]
\caption{Molecular states involved in the laser cooling schemes presented in Fig.~\ref{fig:C2-J-schemes}(a) and Fig.~\ref{fig:C2-fastest-scheme}. Abbreviations follow Table 4 in Ref.~\cite{Yurchenko2018C2}.}
\label{tab:C2_J2_scheme}
\begin{tabular}{|l|l|}
\hline
state id             & state properties \footnotesize{$^*$Energies are in cm$^{-1}$, $\tau$ in s, $\tau = 10000\,$s is the ground or metastable state}                                                              \\ \hline
\multirow{2}{*}{428} & \{gtot: 5, el\_state: "X1Sigmag+", Sigma: 0, tau: 10000, parity\_rot: "e", J: 2, parity\_tot: "+",   \\
                     & gLande: 0.0, Omega: 0, v: 0, id: 428, Lambda: 0, energy: 10.874961\}                                 \\ \hline
\multirow{2}{*}{429} & \{gtot: 5, el\_state: "X1Sigmag+", Sigma: 0, tau: 1702.1, parity\_rot: "e", J: 2, parity\_tot: "+",  \\
                     & gLande: 0.0, Omega: 0, v: 1, id: 429, Lambda: 0, energy: 1838.238745\}                               \\ \hline
\multirow{2}{*}{430} & \{gtot: 5, el\_state: "X1Sigmag+", Sigma: 0, tau: 1079.8, parity\_rot: "e", J: 2, parity\_tot: "+",  \\
                     & gLande: 0.0, Omega: 0, v: 2, id: 430, Lambda: 0, energy: 3637.327786\}                               \\ \hline
\multirow{2}{*}{431} & \{gtot: 5, el\_state: "X1Sigmag+", Sigma: 0, tau: 13.319, parity\_rot: "e", J: 2, parity\_tot: "+",  \\
                     & gLande: 0.0, Omega: 0, v: 3, id: 431, Lambda: 0, energy: 5407.217738\}                               \\ \hline
\multirow{2}{*}{434} & \{gtot: 5, el\_state: "X1Sigmag+", Sigma: 0, tau: 0.64954, parity\_rot: "e", J: 2, parity\_tot: "+",  \\
                     & gLande: 0.0, Omega: 0, v: 4, id: 434, Lambda: 0, energy: 7146.764074\}                               \\ \hline
\multirow{2}{*}{722} & \{gtot: 5, el\_state: "A1Piu", Sigma: 0, tau: 0.000013134, parity\_rot: "f", J: 2, parity\_tot: "-", \\
                     & gLande: 0.1667, Omega: -1, v: 0, id: 722, Lambda: -1, energy: 8278.035245\}                          \\ \hline
\multirow{2}{*}{728} & \{gtot: 5, el\_state: "A1Piu", Sigma: 0, tau: 0.000010619, parity\_rot: "f", J: 2, parity\_tot: "-", \\
                     & gLande: 0.1667, Omega: -1, v: 1, id: 728, Lambda: -1, energy: 9861.977838\}                          \\ \hline
\end{tabular}
\end{table}

\begin{figure}[t]
    \centering
    \includegraphics[width=0.99\columnwidth]{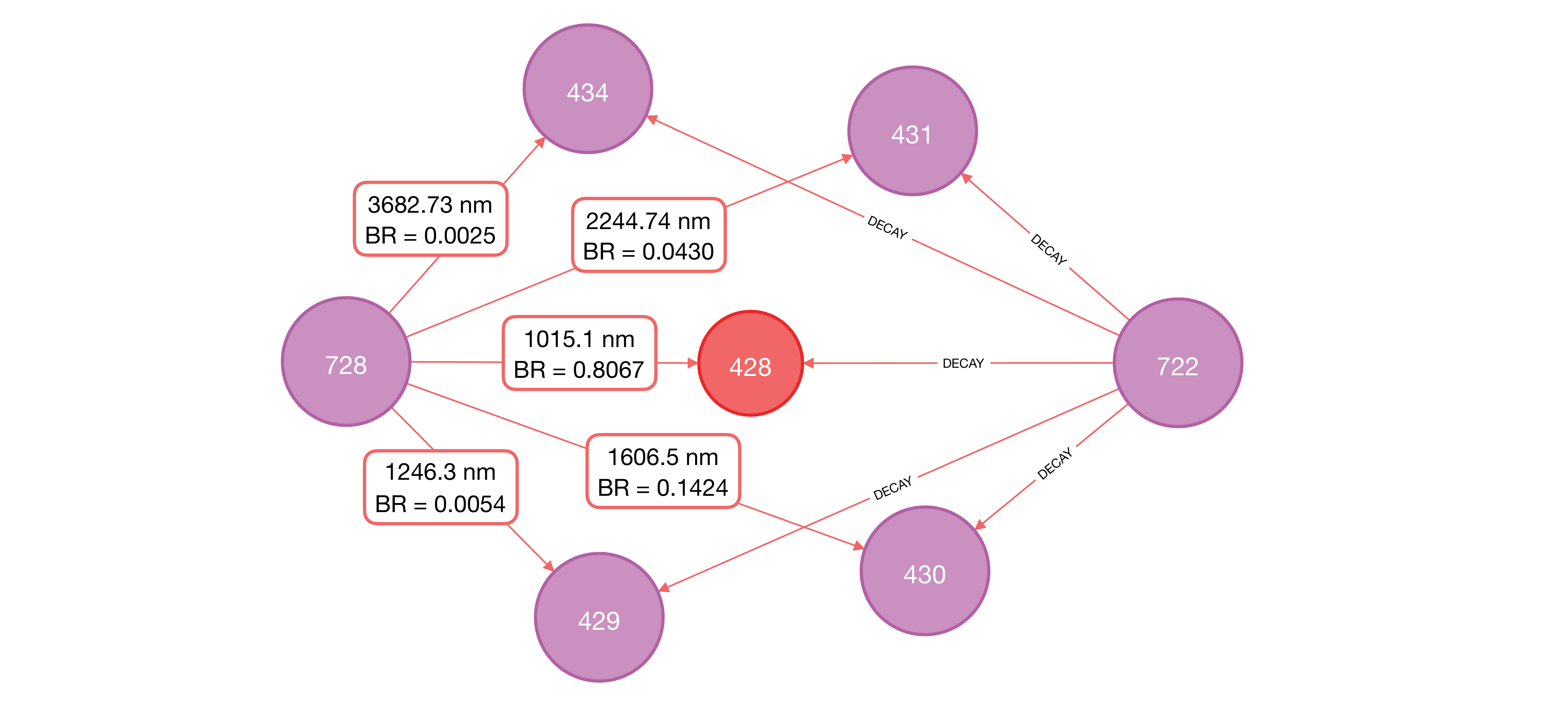}
    \caption{The cooling scheme with five lasers, based on states with $J=2$ and $S_1$ (id: 728) with $v=1$. Here, $T_{\rm init} = 6.8\,$K, $R^{-1} = 63.7 \, \mu \mathrm{s}$, $n_{\rm cool} = 5570$, $t_{\rm cool} = 355\,$ms, $\frac{n_{\rm cool}}{n_{10\%}} = 0.026$, $p = 0.999989$. It is based on largely the same $S_2$ states as the scheme in Fig.~\ref{fig:C2-J-schemes}(a) but uses a higher vibrational level as the $S_1$ state, therefore gains on the cooling time as more excited state has a shorter lifetime. The cooling time is given for the lowest-energy $S_0$. The color convention is described in the caption of Fig.~\ref{fig:YO-scheme}.}
    \label{fig:C2-fastest-scheme}
\end{figure}

\newpage
Interestingly, each of the four-laser schemes has a five-laser counterpart but involving an $S_1$ with $v=1$. We present an example for $J=2$ in Fig.~\ref{fig:C2-fastest-scheme}. The five-laser scheme has a reduced $t_{\rm cool}$ compared to the four-laser scheme.  The higher vibrational excitation of $S_1$ in the five-laser scheme, compared to that of the four-laser scheme, results in an $S_1$ with a shorter lifetime.  This leads to a shorter $t_{\rm cool}$ (e.g., Eqs.~\eqref{eq:R} and~\eqref{eq:tcool}). One can excite to even higher energy and shorter-lived vibrational levels and thereby further reduce $t_{\rm cool}$; but this will require more and more lasers due to the increased leakage. The cooling time also slowly increases with the number of lasers (see again Eqs.~\eqref{eq:R} and~\eqref{eq:tcool}, which shows that in the limit of perfect laser saturation, $t_{\rm cool} \propto \tau (G+1)$). Therefore, the sweet spot is 6 lasers and resulting cooling time is $t_{\rm cool} = 312\,$ms. This example also showcases why one should run the search for an increasing number of lasers, instead of immediately setting the maximum allowed number of lasers.

For C$_2$ we identified in total 388 cooling schemes, including: 9 laser cooling schemes requiring 4 lasers, another 9 with 5 lasers, 15 with 6 lasers, 16 with 7 lasers, 43 with 8 lasers, 29 with 9 lasers, 47 with 10 lasers, 33 with 11 lasers, 49 with 12 lasers, 39 with 13 lasers, 48 with 14 lasers, and 51 with 15 lasers. As one may expect, we observe a general increase in the number of viable laser cooling schemes as the number of lasers increases.

\begin{table}
\caption{Molecular states involved in the laser cooling scheme presented in Fig.~\ref{fig:C2-J-schemes}(b). Abbreviations follow Table 4 in Ref.~\cite{Yurchenko2018C2}}
\label{tab:C2_J4_scheme}
\begin{tabular}{|l|l|l}
\cline{1-2}
state id     & state properties \footnotesize{$^*$Energies are in cm$^{-1}$, $\tau$ in s, $\tau = 10000\,$s is the ground or metastable state}           &  \\ \cline{1-2}
\multirow{2}{*}{1196} & \{gtot: 9, el\_state: "X1Sigmag+", Sigma: 0, tau: 10000, parity\_rot: "e", J: 4, parity\_tot: "+",  &  \\
                      & gLande: 0.0, Omega: 0, v: 0, id: 1196, Lambda: 0, energy: 36.212883\}                               &  \\ \cline{1-2}
\multirow{2}{*}{1197} & \{gtot: 9, el\_state: "X1Sigmag+", Sigma: 0, tau: 1693.4, parity\_rot: "e", J: 4, parity\_tot: "+", &  \\
                      & gLande: 0.0, Omega: 0, v: 1, id: 1197, Lambda: 0, energy: 1863.33762\}                              &  \\ \cline{1-2}
\multirow{2}{*}{1198} & \{gtot: 9, el\_state: "X1Sigmag+", Sigma: 0, tau: 1078.6, parity\_rot: "e", J: 4, parity\_tot: "+", &  \\
                      & gLande: 0.0, Omega: 0, v: 2, id: 1198, Lambda: 0, energy: 3662.164972\}                             &  \\ \cline{1-2}
\multirow{2}{*}{1199} & \{gtot: 9, el\_state: "X1Sigmag+", Sigma: 0, tau: 13.113, parity\_rot: "e", J: 4, parity\_tot: "+", &  \\
                      & gLande: 0.0, Omega: 0, v: 3, id: 1199, Lambda: 0, energy: 5431.792252\}                             &  \\ \cline{1-2}
\multirow{2}{*}{1490} & \{gtot: 9, el\_state: "A1Piu", Sigma: 0, tau: 0.00001315, parity\_rot: "f", J: 4, parity\_tot: "-", &  \\
                      & gLande: 0.05, Omega: -1, v: 0, id: 1490, Lambda: -1, energy: 8300.549519\}                          &  \\ \cline{1-2}
\end{tabular}
\end{table}

\newpage
\clearpage
\subsection{\texorpdfstring{OH$^+$}{OH+}}\label{app-ss:OH}

\begin{figure}[t]
    \centering
    \includegraphics[width=0.99\columnwidth]{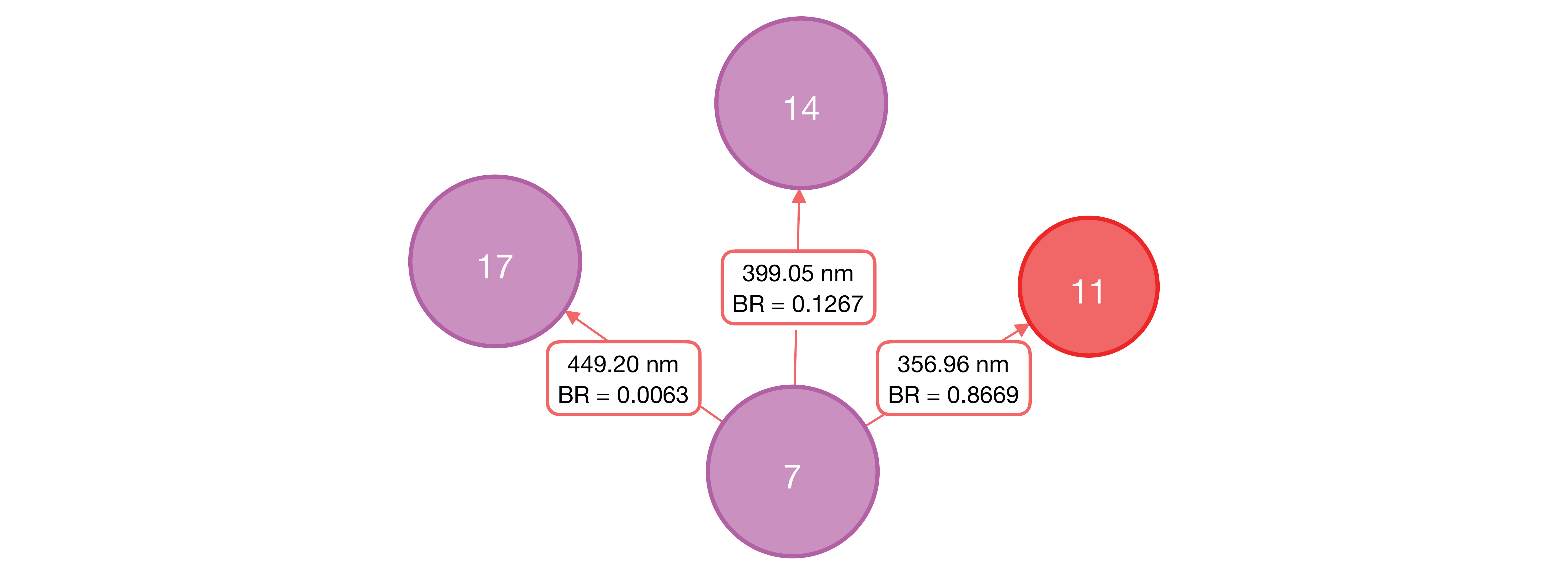}
    \caption{Three-laser cooling scheme for OH$^+$. Here $T_{\rm init} = 21.5\,$K, $R^{-1} = 11.2 \, \mu \mathrm{s}$, $n_{\rm cool} = 3260$, $t_{\rm cool} = 37\,$ms, $\frac{n_{\rm cool}}{n_{10\%}} = 0.081$, and $p = 0.99994$. Adding one more laser to address the $7 \rightarrow 29$ transition results in $T_{\rm init} = 21.5\,$K,  $R^{-1} = 14.0 \, \mu \mathrm{s}$, $n_{\rm cool} = 3260$, $t_{\rm cool} = 46\,$ms, $\frac{n_{\rm cool}}{n_{10\%}} = 4 \times 10^{-5}$, and $p = 0.99999997$. The color convention is described in the caption of Fig.~\ref{fig:YO-scheme}.}
    \label{fig:OH-scheme}
\end{figure}

We present here a cooling schemes identified for ${}^{16}$O${}^{1}$H$^+$ based on \href{https://www.exomol.com/data/molecules/OH_p/16O-1H_p/MoLLIST/}{ExoMol} data~\cite{Hodges2017OH, Bernath2020MoLLIST} that outperforms both the scheme that we found manually in Ref.~\cite{Bigagli2023OH} as well as all of the others identified in our automated search. In specific, we have found a scheme that outperforms the others in terms of the number of lasers, accessibility of laser wavelengths, and cooling time. The cooling scheme is shown in Fig.~\ref{fig:OH-scheme}. It requires only three lasers, achieves $t_{\rm cool} = 37\,$ms and $p = 0.99994$. The states participating in the scheme are listed in Tab.~\ref{tab:OH_scheme}. A probable reason for overlooking this scheme during the manual search is that the starting state $S_0$ has $N=1$, whereas the absolute ground state has $N=0$. In addition, one could add one more laser (with $\lambda = 509.82\,$nm) to this scheme, covering the transition between states 7 and 29. That would result in a bit slower cooling ($t_{\rm cool} = 46\,$ms), but in almost perfect closure ($p = 0.99999997$).

\begin{table}[h]
\caption{Molecular states involved in the laser cooling scheme presented in Fig.~\ref{fig:OH-scheme}. Abbreviations follow Ref.~\cite{Hodges2017OH}, more details in the footnote}
\label{tab:OH_scheme}
\begin{tabular}{|l|l|l}
\cline{1-2}
state id     & state properties \footnotesize{$^*$Energies are in cm$^{-1}$, $\tau$ in s, $\tau = 10000\,$s is the ground or metastable state}           &  \\ \cline{1-2}
\multirow{2}{*}{7} & \{degeneracy: 2, Fef: "F3f", v: 0, el\_state: "A", tau: 0.0000028007634990774732,  &  \\
                      & J: 0, id: 7, N: 1, energy: 28049.05116\}                               &  \\ \cline{1-2}
\multirow{2}{*}{11} & \{degeneracy: 6, Fef: "F2f", v: 0, el\_state: "X", tau: 10000, &  \\
                      & J: 1, id: 11, N: 1, energy: 34.4599\}                              &  \\ \cline{1-2}
\multirow{2}{*}{14} & \{degeneracy: 6, Fef: "F2f", v: 1, el\_state: "X", tau: 0.003614466540883231, &  \\
                      & J: 1, id: 14, N: 1, energy: 2989.351896\}                             &  \\ \cline{1-2}
\multirow{2}{*}{17} & \{degeneracy: 6, Fef: "F2f", v: 2, el\_state: "X", tau: 0.0019477898817769452, &  \\
                      & J: 1, id: 17, N: 1, energy: 5787.256617\}                             &  \\ \cline{1-2}
\multirow{2}{*}{29} & \{degeneracy: 6, Fef: "F2f", v: 3, el\_state: "X", tau: 0.0014024511171105165, &  \\
                      & J: 1, id: 20, N: 1, energy: 8434.347592\}                          &  \\ \cline{1-2}
\end{tabular}

\footnotesize{$^*$Fef - F/(e/f) - F1/F2/F3 spin component and rotationless parity; v - state vibrational quantum number; \\J - total angular momentum (without the nuclear spin); N - nuclear and electronic rotation.}
\end{table}

For OH$^+$ we identified in total 509 laser cooling schemes, including: 2 schemes requiring 2 lasers, 28 with 3 lasers, 79 with 4 lasers, 51 with 5 lasers, 2 with 6 lasers, 2 with 7 lasers, 8 with 8 lasers, 10 with 9 lasers, 39 with 10 lasers, 61 with 11 lasers, 72 with 12 lasers, 60 with 13 lasers, 56 with 14 lasers, and 39 with 15 lasers.

\newpage
\clearpage
\subsection{CN}\label{app-ss:CN}

\begin{figure}[t]
    \centering
    \includegraphics[width=0.89\columnwidth]{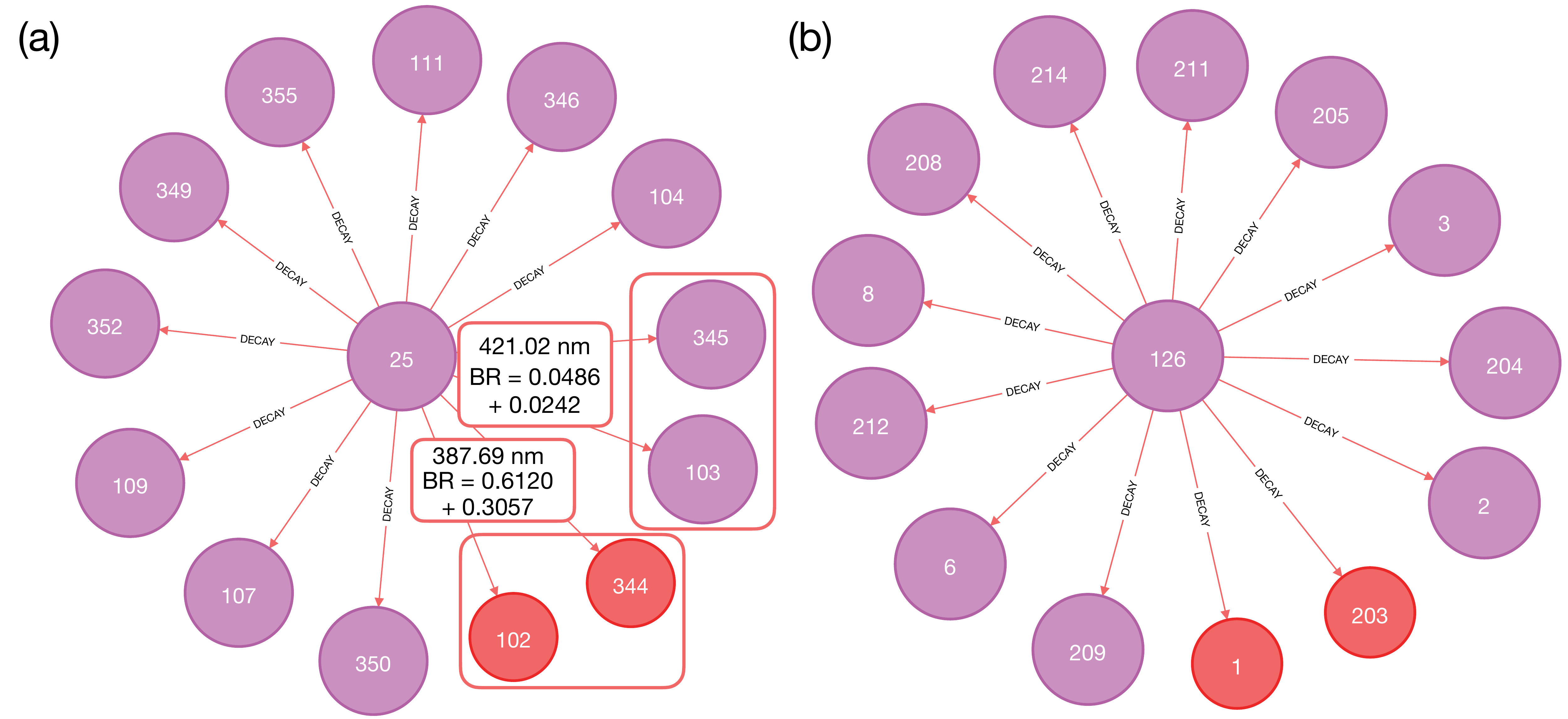}
    \caption{Two of the fastest laser cooling schemes identified for CN. Cooling times are given for the lowest-energy $S_0$ in each. (a) The cooling scheme using 11 lasers and $T_{\rm init} = 4\,$K, $R^{-1} = 0.9 \, \mu \mathrm{s}$, $n_{\rm cool} = 1516$, $t_{\rm cool} = 1.34\,$ms, $\frac{n_{\rm cool}}{n_{10\%}} = 0.650$, and $p = 0.99901$. Only the dominant transitions are described in detail in the figure. The remaining transitions have $\lambda \in [387.69, 763.2]\,$nm and $\mathrm{BR} \in [0.0003, 0.0029]$. (b) This scheme has the opposite parity of (a) and the same properties, except for a smaller $p = 0.998904$ and therefore larger $\frac{n_{\rm cool}}{n_{10\%}} = 0.721$. The color convention is described in the caption of Fig.~\ref{fig:YO-scheme}.}
    \label{fig:CN-fastest-schemes}
\end{figure}

We identified a few notable laser cooling schemes for ${}^{12}$C${}^{14}$N based on \href{https://www.exomol.com/data/molecules/CN/12C-14N/Trihybrid/}{ExoMol} data~\cite{Brooke2014CN, Syme2020CN, Syme2021CN}. Interestingly, each has a twin of the opposite parity, allowing for the state selection at the stage of cooling.
One of the fastest laser cooling schemes is presented in Fig.~\ref{fig:CN-fastest-schemes}(a). The states that take part are listed in Tab.~\ref{tab:CN-fastest-schemes}. It requires 11 lasers addressing 13 transitions (there are two pairs of states that differ in energy by less than 1 GHz). Its closure can be improved by adding further lasers. This scheme also has a twin of the opposite parity, shown in Fig.~\ref{fig:CN-fastest-schemes}(b). Its closure is a bit worse ($p = 0.998904$), but it is equally fast.

\begin{figure}[b]
    \centering
    \includegraphics[width=0.89\columnwidth]{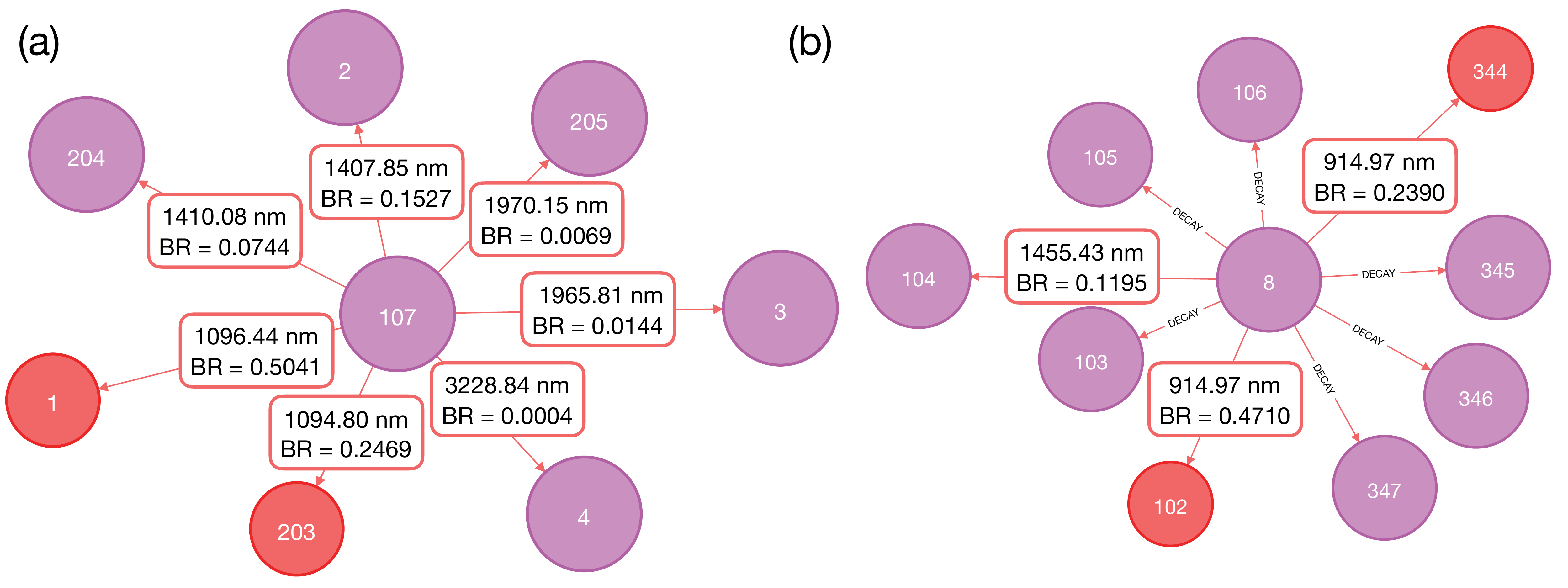}
    \caption{(a) A cooling scheme for CN with 7 lasers, $T_{\rm init} = 4\,$K, $R^{-1} = 85.9 \, \mu \mathrm{s}$, $n_{\rm cool} = 4515$, $t_{\rm cool} = 388\,$ms, $\frac{n_{\rm cool}}{n_{10\%}} = 0.401$, and $p = 0.9998$. (b) A cooling scheme with 9 lasers, $T_{\rm init} = 4\,$K, $R^{-1} = 94.5 \, \mu \mathrm{s}$, $n_{\rm cool} = 3944$, $t_{\rm cool} = 373\,$ms, $\frac{n_{\rm cool}}{n_{10\%}} = 0.99$, and $p = 0.9994$. Only the dominant transitions are described in detail in the figure. The remaining transitions have $\lambda \in [914.65, 2051.82]\,$nm and $\mathrm{BR} \in [0.001, 0.012]$. The color convention is described in the caption of Fig.~\ref{fig:YO-scheme}.}
    \label{fig:CN-less-laser-schemes}
\end{figure}

In addition, we identified schemes that require fewer lasers (Fig.~\ref{fig:CN-less-laser-schemes}), though they are slower (on the order of 400 ms). The scheme in Fig.~\ref{fig:CN-less-laser-schemes}(a) has a closure of $p=0.9998$.  But adding a laser at $\lambda = 3240.37\,$nm, to address the transition between states 107 and 206, reaches $p=0.999998$ and $\frac{n_{\rm cool}}{n_{10\%}} = 0.004$, though it slows down to $t_{\rm cool} = 436\,$ms. States taking part in the scheme in Fig.~\ref{fig:CN-less-laser-schemes}(a) are listed in Tab.~\ref{tab:CN-less-laser-schemes-a}. In Fig.~\ref{fig:CN-less-laser-schemes}(b), we show a counterpart of (a) that relies on $S_1$ in $v=1$ state instead of $v=0$ state. Both schemes have twins of opposite parity and both also use states of $J=1/2$ and $J=3/2$ allowing for further state selection.

For CN, we identified in total 843 cooling schemes: 2 requiring 2 lasers, 4 with 7 lasers, 4 with 8 lasers, 4 with 9 lasers, 220 with 10 lasers, 190 with 11 lasers, 169 with 12 lasers, 4 with 13 lasers, 121 with 14 lasers, and 125 with 15 lasers.

\begin{table}[h]
\caption{Molecular states involved in the laser cooling scheme presented in Fig.~\ref{fig:CN-fastest-schemes}(a). States in Fig.~\ref{fig:CN-fastest-schemes}(b) have opposite parity, as an example we put in this table state 126. Abbreviations follow Table 5 in Ref.~\cite{Syme2021CN}}
\label{tab:CN-fastest-schemes}
\begin{tabular}{|l|l|l}
\cline{1-2}
state id     & state properties \footnotesize{$^*$Energies are in cm$^{-1}$, $\tau$ in s, $\tau = 10000\,$s is the ground or metastable state}              &  \\ \cline{1-2}
\multirow{2}{*}{25} & \{ef: "e", gtot: 6, parity: "+", g: 2.002312, Sigma: 0.5, tau: 6.276881037662201e-8,  &  \\
                      & J: 0.5, Omega: 0.5, v: 0, id: 25, state: "B(2SIGMA+)", Lambda: 0, energy: 25797.87218\}                               &  \\ \cline{1-2}
\multirow{2}{*}{126} & \{ef: "f", gtot: 6, parity: "-", g: -0.667445, Sigma: -0.5, tau: 6.276783543771163e-8,  &  \\
                      & J: 0.5, Omega: -0.5, v: 0, id: 126, state: "B(2SIGMA+)", Lambda: 0, energy: 25801.76964\}                               &  \\ \cline{1-2}
\multirow{2}{*}{102} & \{ef: "f", gtot: 6, parity: "-", g: -0.667444, Sigma: -0.5, tau: 99900.0999000999, &  \\
                      & J: 0.5, Omega: -0.5, v: 0, id: 102, state: "X(2SIGMA+)", Lambda: 0, energy: 3.777245\}                              &  \\ \cline{1-2}
\multirow{2}{*}{103} & \{ef: "f", gtot: 6, parity: "-", g: -0.667444, Sigma: -0.5, tau: 0.11327196802158737, &  \\
                      & J: 0.5, Omega: -0.5, v: 1, id: 103, state: "X(2SIGMA+)", Lambda: 0, energy: 2046.162428\}                             &  \\ \cline{1-2}
\multirow{2}{*}{104} & \{ef: "f", gtot: 6, parity: "-", g: -0.667443, Sigma: -0.5, tau: 0.058569375608610474, &  \\
                      & J: 0.5, Omega: -0.5, v: 2, id: 104, state: "X(2SIGMA+)", Lambda: 0, energy: 4062.252487\}                             &  \\ \cline{1-2}
\multirow{2}{*}{107} & \{ef: "f", gtot: 6, parity: "-", g: -0.000771, Sigma: 0.5, tau: 0.000010735097750447735, &  \\
                      & J: 0.5, Omega: -0.5, v: 0, id: 107, state: "A(2PI)", Lambda: -1, energy: 9145.438068\}                          &  \\ \cline{1-2}
\multirow{2}{*}{109} & \{ef: "f", gtot: 6, parity: "-", g: -0.000773, Sigma: 0.5, tau: 0.000009456513821230949, &  \\
                      & J: 0.5, Omega: -0.5, v: 1, id: 109, state: "A(2PI)", Lambda: -1, energy: 10933.09055\}                          &  \\ \cline{1-2}
\multirow{2}{*}{111} & \{ef: "f", gtot: 6, parity: "-", g: -0.000777, Sigma: 0.5, tau: 0.00000851467606252028, &  \\
                      & J: 0.5, Omega: -0.5, v: 2, id: 111, state: "A(2PI)", Lambda: -1, energy: 12695.1779\}                          &  \\ \cline{1-2}
\multirow{2}{*}{344} & \{ef: "e", gtot: 12, parity: "-", g: 0.667436, Sigma: -0.5, tau: 98183.60333824251, &  \\
                      & J: 1.5, Omega: -0.5, v: 0, id: 344, state: "X(2SIGMA+)", Lambda: 0, energy: 3.781184\}                          &  \\ \cline{1-2}
\multirow{2}{*}{345} & \{ef: "e", gtot: 12, parity: "-", g: 0.667436, Sigma: -0.5, tau: 0.11329429533686929, &  \\
                      & J: 1.5, Omega: -0.5, v: 1, id: 345, state: "X(2SIGMA+)", Lambda: 0, energy: 2046.173377\}                          &  \\ \cline{1-2}
\multirow{2}{*}{346} & \{ef: "e", gtot: 12, parity: "-", g: 0.667435, Sigma: -0.5, tau: 0.058579863542777465, &  \\
                      & J: 1.5, Omega: -0.5, v: 2, id: 346, state: "X(2SIGMA+)", Lambda: 0, energy: 4062.274444\}                          &  \\ \cline{1-2}    
\multirow{2}{*}{349} & \{ef: "e", gtot: 12, parity: "-", g: 0.846607, Sigma: -0.5, tau: 0.000010936572155162485, &  \\
                      & J: 1.5, Omega: -1.5, v: 0, id: 349, state: "A(2PI)", Lambda: -1, energy: 9094.326452\}                          &  \\ \cline{1-2}   
\multirow{2}{*}{350} & \{ef: "e", gtot: 12, parity: "-", g: -0.046296, Sigma: 0.5, tau: 0.000010738508018653686, &  \\
                      & J: 1.5, Omega: -0.5, v: 0, id: 350, state: "A(2PI)", Lambda: -1, energy: 9150.698066\}                          &  \\ \cline{1-2}    
\multirow{2}{*}{352} & \{ef: "e", gtot: 12, parity: "-", g: 0.846298, Sigma: -0.5, tau: 0.000009617387193930866, &  \\
                      & J: 1.5, Omega: -1.5, v: 1, id: 352, state: "A(2PI)", Lambda: -1, energy: 10882.03968\}                          &  \\ \cline{1-2}    
\multirow{2}{*}{355} & \{ef: "e", gtot: 12, parity: "-", g: 0.845985, Sigma: -0.5, tau: 0.000008648673715972027, &  \\
                      & J: 1.5, Omega: -1.5, v: 2, id: 355, state: "A(2PI)", Lambda: -1, energy: 12644.18487\}                          &  \\ \cline{1-2}    
\end{tabular}
\end{table}

\begin{table}[h]
\caption{Molecular states involved in the laser cooling scheme presented in Fig.~\ref{fig:CN-less-laser-schemes}(a). Abbreviations follow Table 5 in Ref.~\cite{Syme2021CN}}
\label{tab:CN-less-laser-schemes-a}
\begin{tabular}{|l|l|l}
\cline{1-2}
state id     & state properties \footnotesize{$^*$Energies are in cm$^{-1}$, $\tau$ in s, $\tau = 10000\,$s is the ground or metastable state}                 &  \\ \cline{1-2}
\multirow{2}{*}{1} & \{ef: "e", gtot: 6, parity: "+", g: 2.002305, Sigma: 0.5, tau: 10000,  &  \\
                      & J: 0.5, Omega: 0.5, v: 0, id: 1, state: "X(2SIGMA+)", Lambda: 0, energy: 0.0\}                               &  \\ \cline{1-2}
\multirow{2}{*}{2} & \{ef: "e", gtot: 6, parity: "+", g: 2.002304, Sigma: 0.5, tau: 0.11329005653173821,  &  \\
                      & J: 0.5, Omega: 0.5, v: 1, id: 2, state: "X(2SIGMA+)", Lambda: 0, energy: 2042.387527\}                               &  \\ \cline{1-2}
\multirow{2}{*}{3} & \{ef: "e", gtot: 6, parity: "+", g: 2.002302, Sigma: 0.5, tau: 0.05857887645714955, &  \\
                      & J: 0.5, Omega: 0.5, v: 2, id: 3, state: "X(2SIGMA+)", Lambda: 0, energy: 4058.490117\}                              &  \\ \cline{1-2}
\multirow{2}{*}{4} & \{ef: "e", gtot: 6, parity: "+", g: 2.002299, Sigma: 0.5, tau: 0.039867860368878974, &  \\
                      & J: 0.5, Omega: 0.5, v: 3, id: 4, state: "X(2SIGMA+)", Lambda: 0, energy: 6048.345767\}                             &  \\ \cline{1-2}
\multirow{2}{*}{107} & \{ef: "f", gtot: 6, parity: "-", g: -0.000771, Sigma: 0.5, tau: 0.000010735097750447735, &  \\
                      & J: 0.5, Omega: -0.5, v: 0, id: 107, state: "A(2PI)", Lambda: -1, energy: 9145.438068\}                             &  \\ \cline{1-2}
\multirow{2}{*}{203} & \{ef: "f", gtot: 12, parity: "+", g: -0.400464, Sigma: 0.5, tau: 10352.717070595178, &  \\
                      & J: 1.5, Omega: 0.5, v: 0, id: 203, state: "X(2SIGMA+)", Lambda: 0, energy: 11.339227\}                          &  \\ \cline{1-2}
\multirow{2}{*}{204} & \{ef: "f", gtot: 12, parity: "+", g: -0.400463, Sigma: 0.5, tau: 0.11325474129663118, &  \\
                      & J: 1.5, Omega: 0.5, v: 1, id: 204, state: "X(2SIGMA+)", Lambda: 0, energy: 2053.656583\}                          &  \\ \cline{1-2}
\multirow{2}{*}{205} & \{ef: "f", gtot: 12, parity: "+", g: -0.400463, Sigma: 0.5, tau: 0.05856064868254569, &  \\
                      & J: 1.5, Omega: 0.5, v: 2, id: 205, state: "X(2SIGMA+)", Lambda: 0, energy: 4069.679601\}                          &  \\ \cline{1-2}
\end{tabular}
\end{table}

\newpage
\clearpage
\subsection{\texorpdfstring{CO$_2$}{CO2}}\label{app-ss:CO2}

\begin{figure}[t]
    \centering
    \includegraphics[width=\columnwidth]{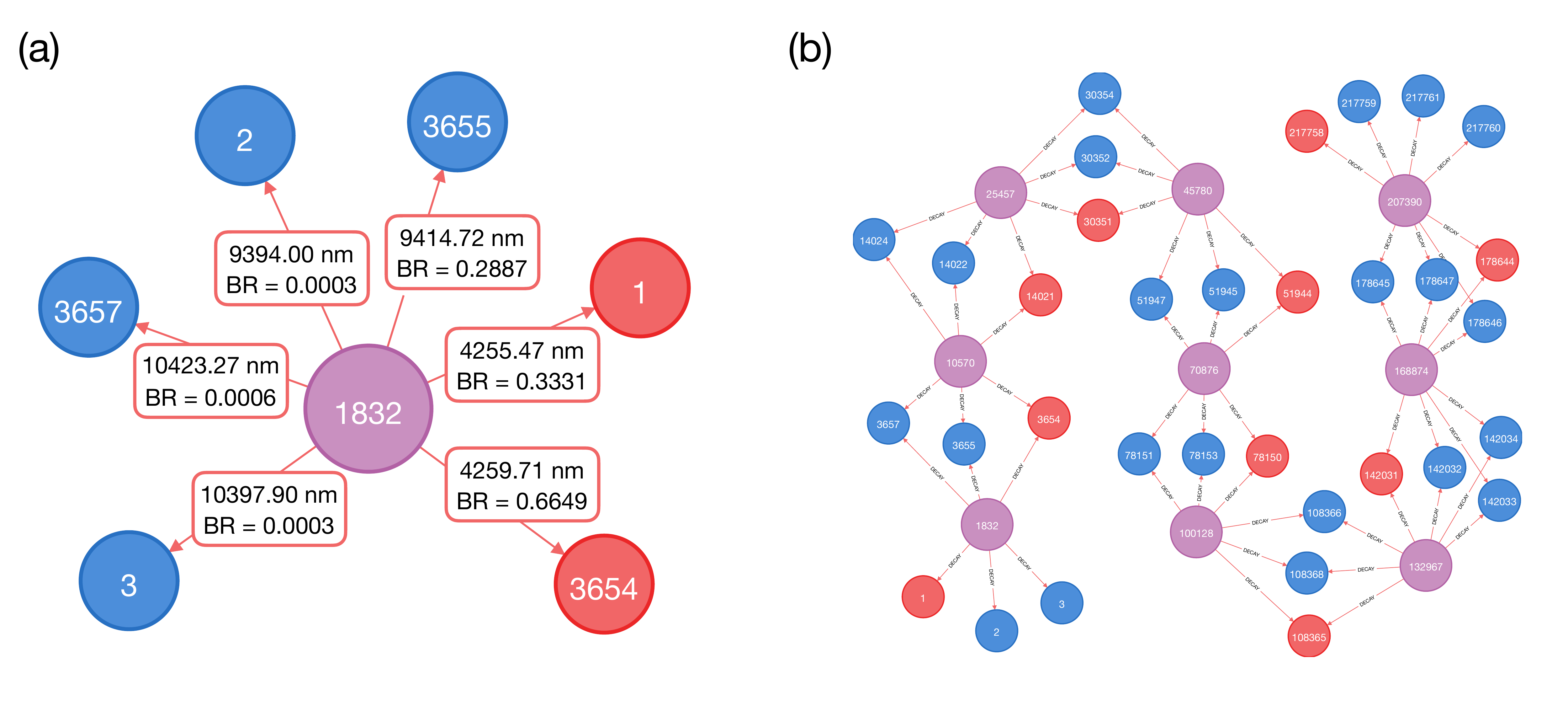}
    \caption{Laser cooling schemes for CO$_2$. (a) The best detected laser cooling scheme with $R^{-1} = 16576 \, \mu \mathrm{s}$, $n_{\rm cool} = 22384$, $t_{\rm cool} = 371\,$s, $\frac{n_{\rm cool}}{n_{10\%}} = 10^{-6}$, $p = 1$. (b) Interestingly, all detected laser cooling schemes are interconnected. Additional details are given in Figs.~\ref{fig:C2-J-schemes}, \ref{fig:CN-fastest-schemes}, and \ref{fig:YO-scheme}. The cooling time is given for the lowest-energy $S_0$ (state 1).}
    \label{fig:CO2-scheme}
\end{figure}

In Fig.~\ref{fig:CO2-scheme}(a), we present the quickest cooling scheme identified for ${}^{12}$C${}^{16}$O$_2$ based on \href{https://www.exomol.com/data/molecules/CO2/12C-16O2/UCL-4000/}{ExoMol} data~\cite{Yurchenko2020CO2}. The states participating in the scheme are listed in Tab.~\ref{tab:CO2_scheme}. Its long cooling time of 371 s probably does not allow for Doppler cooling but may be a basis for indirect cooling methods. This scheme is the first of a series of interconnected cooling schemes, as shown in Fig.~\ref{fig:CO2-scheme}(b). Each of these require six or seven lasers to achieve perfect closure. Interestingly, each pair of adjacent cooling schemes shares three or four states as their $S_2$ states, as illustrated in Fig.~\ref{fig:CO2-scheme}(b). Together, these pairs form a ladder of cooling schemes requiring higher and higher initial temperature and therefore having longer and longer cooling times.

\begin{table}[h]
\caption{States participating in the cooling scheme of CO$_2$. Abbreviations follow Table 2 in Ref.~\cite{Yurchenko2020CO2}}
\label{tab:CO2_scheme}
\begin{tabular}{|l|l|}
\hline
State id              & state properties \footnotesize{$^*$Energies are in cm$^{-1}$, $\tau$ in s, $\tau = 10000\,$s is the ground or metastable state} \\ \hline
\multirow{3}{*}{1}    & \{ef: "e", gtot: 1, n1: 0, m1: 0, n3: 0, m2: 0, l2: 0, Ci: 1.0, m3: 0, \\
                      & m4: 0, tau: 10000, J: 0, n2lin: 0, r: 1, id: "1", v1: 0, v2: 0, v3: 0,   \\
                      & energy: 0.0, tot\_symmetry: "A1"\}                                      \\ \hline
\multirow{3}{*}{1832} & \{ef: "e", gtot: 3, n1: 0, m1: 0, n3: 1, m2: 0, l2: 0, Ci: 1.0, m3: 0,  \\
                      & m4: 1, tau: 0.0023679621911542378, J: 1, n2lin: 0, r: 1, id: "1832",    \\
                      & v1: 0, v2: 1, v3: 0, energy: 2349.917065, tot\_symmetry: "A2"\}         \\ \hline
\multirow{3}{*}{2}    & \{ef: "e", gtot: 1, n1: 0, m1: 1, n3: 0, m2: 0, l2: 0, Ci: 1.0, m3: 0,  \\
                      & m4: 0, tau: 0.7935248373274083, J: 0, n2lin: 2, r: 2, id: "2", v1: 0,   \\
                      & v2: 0, v3: 1, energy: 1285.408201, tot\_symmetry: "A1"\}                \\ \hline
\multirow{3}{*}{3}    & \{ef: "e", gtot: 1, n1: 1, m1: 1, n3: 0, m2: 0, l2: 0, Ci: 1.0, m3: 0,  \\
                      & m4: 0, tau: 0.565866908103214, J: 0, n2lin: 0, r: 1, id: "3", v1: 1,    \\
                      & v2: 0, v3: 0, energy: 1388.1842, tot\_symmetry: "A1"\}                  \\ \hline
\multirow{3}{*}{3654} & \{ef: "e", gtot: 5, n1: 0, m1: 0, n3: 0, m2: 0, l2: 0, Ci: 1.0, m3: 0,  \\
                      & m4: 0, tau: 10000, J: 2, n2lin: 0, r: 1, id: "3654", v1: 0, v2: 0,       \\
                      & v3: 0, energy: 2.3413, tot\_symmetry: "A1"\}                            \\ \hline
\multirow{3}{*}{3655} & \{ef: "e", gtot: 5, n1: 0, m1: 1, n3: 0, m2: 0, l2: 0, Ci: 1.0, m3: 0,  \\
                      & m4: 0, tau: 0.7936130025554339, J: 2, n2lin: 2, r: 2, id: "3655",       \\
                      & v1:0, v2: 0, v3: 1, energy: 1287.7511, tot\_symmetry: "A1"\}            \\ \hline
\multirow{3}{*}{3657} & \{ef: "e", gtot: 5, n1: 1, m1: 1, n3: 0, m2: 0, l2: 0, Ci: 1.0, m3: 0,  \\
                      & m4: 0, tau: 0.5657644609396215, J: 2, n2lin: 0, r: 1, id: "3657",       \\
                      & v1: 1, v2: 0, v3: 0, energy: 1390.525301, tot\_symmetry: "A1"\}         \\ \hline
\end{tabular}
\end{table}

\newpage
\clearpage
\section{Build your own NH database from scratch}\label{app:NH-tutorial}

To avoid the installation process, we will build this small database fully in the cloud, taking advantage of the \href{https://neo4j.com/cloud/platform/aura-graph-database/}{Neo4j AuraDB cloud service}. Note it requires setting up an account. The free instance of the database has limits (up to 1 GB of the database instance, and up to 200k nodes and 400k relationships, as of Jul 7, 2023), so we will tackle the NH molecule available in Exomol \cite{Exomol23}. For molecules with more energy levels and decay channels, a local installation of (free) Neo4j Desktop will be necessary (or a paid cloud instance).

\begin{enumerate}
    \item Set up an account and create a free database instance of \href{https://neo4j.com/cloud/platform/aura-graph-database/}{Neo4j Aura}. This will generate the password that needs to be saved to access the database again. Open the database. This will also lead to additional tutorials available to users. 
    
    \item Download the data on ${}^{14}$NH energy levels and decay channels from the \href{https://www.exomol.com/data/molecules/NH/14N-1H/MoLLIST/}{Exomol website} \cite{Exomol23}. ``MoLLIST: line list'' contains three files: \texttt{14N-1H\_\_MoLLIST.trans.bz2}, \texttt{14N-1H\_\_MoLLIST.states.bz2}, and \texttt{14N-1H\_\_MoLLIST.readme} \cite{Bernath2020MoLLIST}.
    \texttt{14N-1H\_\_MoLLIST.states.bz2} contains a compressed list of energy levels. In particular, every energy level is characterized by its identification number, energy in cm$^{-1}$, and various quantum numbers.
    \texttt{14N-1H\_\_MoLLIST.states.bz2} contains a compressed list of transitions with the identification numbers $f$ and $i$ for upper (final) and lower (initial) levels, respectively, Einstein-A coefficients denoted by $A$ (s$^{-1}$) and transition frequencies $\nu$ (cm$^{-1}$).
    The content detailed description is in \texttt{14N-1H\_\_MoLLIST.readme}.

    \item Unpack the files and convert them to csv (e.g., with the Linux commands \texttt{sed 's/ $\backslash$+/,/g' 14N-1H\_\_MoLLIST.trans > NH\_transitions.csv} and \texttt{sed 's/ $\backslash$+/,/g' 14N-1H\_\_MoLLIST.states > NH\_states.csv}) and remove the first empty column (e.g., with \texttt{cut -c 2- file.csv > file\_without\_the\_first\_character\_in\_each\_line.csv})
    
    \item Add a header to each file that tells the database the name of each column.
    Below are headers for \texttt{NH\_states.csv} and \texttt{NH\_transitions.csv}, respectively. 

\begin{minted}{vim}
id,energy,gtot,J,N,F,parity,v,state
\end{minted}

\begin{minted}{vim}
:START_ID,:END_ID,Einstein_coeff,frequency
\end{minted}

    \item Go to ``Import'' tab in the Neo4j AuraDB instance and transfer there two csv files. 
    
    \item Build your graph data model. Add node label. Name it ``State'', select \texttt{NH\_states.csv} as the file describing it, and map its properties from the file (``Map from file'' and ``Select all''). Hover over the edge of the node till you see a plus appearing, click and hold for a relationship to appear that starts and ends in the same node label. Name it ``DECAYS'', select \texttt{NH\_transitions.csv} as the file that describes it, select ``From'' as \texttt{:START\_ID} and ``To'' as \texttt{:END\_ID}, and map its properties from the file (this time ignore \texttt{:START\_ID} and \texttt{:END\_ID}).
    
    \item Click ``Run import''. It will create 1,285 nodes and 22,545 relationships. Click ``Explore results''. It will visualize results for you.

    \item As the database is small compared to other molecules, we need only a few steps to run the graph search. We start by computing the branching ratios as in Eq.~\eqref{eq:br}.

    \begin{minted}{cypher}
// 1 Computing branching ratios
// find every node that has an outgoing decay channel and name the node `s1`
// find also all decay channels exiting the `s1' node
match (s1)-[r:DECAYS]->() 
// for a fixed node `s1', sum up the Einstein coefficients of outgoing decay channels 
// and save it as `decays_sum' (in parallel for every node!)
with s1, sum(r.Einstein_coeff) as decays_sum
// now again find the same nodes but this time match the pattern `energy level s1 decays to
// energy level s2' with every decay channel considered separately. Name the relationship `r1'
match(s1)-[r1:DECAYS]->(s2)
// set a new property of the relationship, namely `branching_ratio'
set r1.branching_ratio = r1.Einstein_coeff/decays_sum
// return a number of relationships to which we added the new property
return count(r1)
// output: 22 545
\end{minted}

    \item Change all transition frequencies (in cm$^{-1}$) to laser wavelengths (in nm).
    \begin{minted}{cypher}
// 2 Computing energy differences between S1 and S2 connected with a decay channel
match (s1)-[r:DECAYS]->(s2)
set r.energy_diff = 10000000/(s1.energy-s2.energy)
return count(r)
// output: 22 545
\end{minted}

    \item Compute the lifetime $\tau$ of every energy level as in Eq.~\eqref{eq:lifetime}.
    \begin{minted}{cypher}
// 3 Computing tau
match (s1)-[r:DECAYS]->()
with s1, sum(r.Einstein_coeff) as decays_sum
set s1.tau = 1/decays_sum
return count(s1)
// output: 1 284 (we have 1 285 nodes, but one of them is the ground state with no outgoing 
// decay channels)
    \end{minted}

    \item Set also lifetimes of ground and metastable states to a large number.
    \begin{minted}{cypher}
// 4 Tau for special nodes - ground and metastable states
match (s1) WHERE NOT (s1)-[:DECAYS]->()
set s1.tau=1000
return count(s1)
// output: 1 (the ground state)
\end{minted}

    \item Search the graph for subgraphs representing laser cooling schemes meeting preset conditions. Your output should look like the one presented in Tab.~\ref{tab:output_query_5}. To understand where various constants in the queries come from, see App.~\ref{app:math}.

\begin{minted}{cypher}
// 5 Find schemes for specific number of lasers (G = num_decays) with preset parameters 
with 4 as num_decays, 500 as max_initial_temp_K, 320 as min_laser_nm, 1500 as max_laser_nm, 
1e-8 as br_cutoff, 1e-6 as min_starting_tau

/// 5.1 Find reachable starting stable states S0 (depending on max_initial_temp_K) 
/// and excited states S1 connected with S0 via a decay channel with energy differences 
/// addressable with lasers of wavelengths between min_laser_nm and max_laser_nm
/// E0 = kB * T0, 1 K = 0.6950348004 cm-1 
match (s0)<-[r1:DECAYS]-(s1) where s0.energy < -log(0.1)*max_initial_temp_K*0.6950348004
and s0.tau > min_starting_tau and r1.energy_diff > min_laser_nm and r1.energy_diff < max_laser_nm

/// 5.2 Remember all found excited states S1 and compute the initial temperature 
/// of the potential laser cooling scheme from the S0 energy
/// T_init = max(4 K, E(S0)/(-log(0.1)*kB)), 1 cm-1 = 1.438776877 K
with s1, apoc.coll.max([4, s0.energy*0.62485285866738]) as initial_temp, s0.id as starting_id, 
num_decays as num_decays, min_laser_nm as min_laser_nm, max_laser_nm as max_laser_nm, br_cutoff 
as br_cutoff, min_starting_tau as min_starting_tau

/// 5.3 For each remembered S1, find the G outgoing decay channels with the largest BR
/// 5.3.1 Start by finding all outgoing decay channels for each remembered S1
match (s1)-[r2:DECAYS]->(s2) where r2.energy_diff > min_laser_nm 
and r2.energy_diff < max_laser_nm and r2.branching_ratio > br_cutoff

/// 5.3.2 Store BR of all outgoing decay channels in br_list for each remembered S1
with s1, collect(r2.branching_ratio) as br_list, num_decays as num_decays, initial_temp as 
initial_temp, starting_id as starting_id, min_laser_nm as min_laser_nm, max_laser_nm as 
max_laser_nm, br_cutoff as br_cutoff, min_starting_tau as min_starting_tau

/// 5.3.3 Find the G'th largest BR (= minimal_considered_BR_per_scheme) for each remembered S1 
with s1, apoc.coll.min(reverse(apoc.coll.sort(br_list))[0..num_decays]) as 
minimal_considered_BR_per_scheme, initial_temp as initial_temp, starting_id as starting_id, 
min_laser_nm as min_laser_nm, max_laser_nm as max_laser_nm, br_cutoff as br_cutoff, 
min_starting_tau as min_starting_tau

/// 5.3.4 For each remembered S1, find outgoing decay channels with BR larger and equal
/// to the corresponding G'th largest BR computed in 5.3.3.
/// All found patterns are potential laser cooling schemes.
match (s1)-[r2:DECAYS]->(s2) where r2.energy_diff > min_laser_nm 
and r2.energy_diff < max_laser_nm and s2.tau > min_starting_tau and 
r2.branching_ratio > br_cutoff and r2.branching_ratio >= minimal_considered_BR_per_scheme

/// 5.4 Compute n_cool, n_10%, and R^-1 of each found potential cooling scheme
/// Also prepare for comparing lifetime(S2) > t_cool * BR(S1->S2) =>
/// => min( lifetime(S2) / BR(S1->S2) ) > t_cool
/// 5.4.1 Compute closure, sum_inv_lambdas_3, and min_tau_br_ratio
with s1.id as id, count(DISTINCT r2) as num_decays, sum(DISTINCT r2.branching_ratio) as closure, 
sum(DISTINCT 1 / (r2.energy_diff*r2.energy_diff*r2.energy_diff)) as sum_inv_lambdas_3,
s1.tau as lifetime, sum(DISTINCT r2.branching_ratio / r2.energy_diff) as sum_of_BR_lambda_ratios, 
min(DISTINCT s2.tau/r2.branching_ratio) as min_tau_br_ratio, initial_temp as initial_temp, 
starting_id as starting_id, collect(round(r2.energy_diff,2)) as lambda_list

/// 5.4.2 Compute n10, R^-1, n_cool
/// Molecular mass of 14NH is 15 u
with id as id, num_decays as num_decays, closure as closure, initial_temp as initial_temp, 
starting_id as starting_id, min_tau_br_ratio as min_tau_br_ratio, log(0.1)/log(closure) as n10, 
lifetime * (num_decays + 1) + 0.04160402474381969 * sum_inv_lambdas_3 as inv_R_s, 
sqrt(initial_temp * 15) * closure * 0.39579544150466855 / sum_of_BR_lambda_ratios as n_cool,
lambda_list as lambda_list

/// 5.5 Return only the schemes that meet n_cool < n10 and whose shortest-lived S2's live longer
/// than the average time that the molecule spends in these S2's during the cooling procedure
/// (lifetime_S2 > t_cool * BR(S1->S2))
where n_cool < n10 and n_cool*inv_R_s < min_tau_br_ratio
return id as S1_id, starting_id as S0_id, round(initial_temp,1) as T_init, num_decays, 
round(n_cool,0) as n_cool, round(n_cool*inv_R_s*1000,1) as t_cool_ms, 
round(n_cool/n10,3) as n_cool_n10_ratio, round(closure,8) as closure, lambda_list as lambda_list_nm
order by t_cool_ms, num_decays, n_cool_n10_ratio

/// As discussed in sec. IID, it can be useful to output the results for T_init = 4K
/// use instead: where sqrt(4)/sqrt(initial_temp)*n_cool < n10 and 
/// sqrt(4)/sqrt(initial_temp)*n_cool*inv_R_s < min_tau_br_ratio
/// return id as S1_id, starting_id as S0_id, round(initial_temp,1) as approx_T_init, num_decays, 
/// round(sqrt(4)/sqrt(initial_temp)*n_cool,0) as n_cool_3K, 
/// round(sqrt(4)/sqrt(initial_temp)*n_cool*inv_R_s*1000,1) as t_cool_ms_3K, 
/// round(sqrt(4)/sqrt(initial_temp)*n_cool/n10,3) as n_cool_n10_ratio_3K, round(closure,8) as 
/// closure, lambda_list as lambda_list_nm
/// order by t_cool_ms_4K, num_decays, n_cool_n10_ratio_4K
\end{minted}

\begin{table}[t]
\centering
\caption{Output of the query no. 5 to find schemes for specific number of lasers (G = num\_decays) with preset parameters}
\label{tab:output_query_5}
\resizebox{\columnwidth}{!}{%
\begin{tabular}{lllllllll}
\hline
S1\_id & S0\_id & T\_init & num\_decays & n\_cool & t\_cool\_ms & n\_cool\_n10\_ratio & closure & lambda\_list\_nm \\ \hline
744 & 1 & 4.0   & 4 & 1030  & 2  & 0.02  & 0.99994935 & {[}335.27, 336.37, 374.51, 375.83{]} \\ \hline
745 & 3 & 20.8    & 4 & 2352 & 5  & 0.00  & 0.99999997 & {[}335.61, 374.93, 421.87, 478.64{]} \\ \hline
744 & 7 & 61.0 & 4 & 4023 & 8 & 0.09 & 0.99994935 & {[}335.27, 336.37, 374.51, 375.83{]} \\ \hline
\end{tabular}%
}
\end{table}

    \item Visualize selected laser cooling scheme for further investigation.

\begin{minted}{cypher}
// 6 Show the scheme with a set number of decays, starting from a specific excited state
with 4 as num_decays, 744 as s1_id
match (s1)-[r2:DECAYS]->(s2) where s1.id = s1_id

with s1, collect(r2.branching_ratio) as list, num_decays as num_decays, s1_id as s1_id
with s1, s1_id as s1_id, apoc.coll.min(reverse(apoc.coll.sort(list))[0..num_decays]) as 
minimal_considered_BR_per_scheme 

match p = (s1)-[r2:DECAYS]->(s2) where s1.id = s1_id and 
r2.branching_ratio >= minimal_considered_BR_per_scheme
return p
// output: one of the schemes in Fig. 2
\end{minted}

\end{enumerate}

\section{A bit of math to explain constants in the code}\label{app:math}
To identify viable laser cooling schemes, the graph algorithm needs to compute $t_{\rm cool}$ from Eq.~\eqref{eq:tcool}, $t_{\rm 10\%}$ from Eq.~\eqref{eq:t10}, and check the lifetimes of participating reachable states $S_2$ so that the average time that a molecule spends in this state during the cooling process is shorter than the state lifetime. 

The simple math below is just to convince the reader that the constants that appear out of nowhere in the algorithm presented in App.~\ref{app:NH-tutorial} are valid. In the code, we compute separately $n_{\rm cool}$ from Eq.~\eqref{eq:ncool}, $n_{\rm 10\%}$ from Eq.~\eqref{eq:n10}, and $R^{-1}$ from Eq.~\eqref{eq:R}, and then compute $t_{\rm cool}$ and $t_{\rm 10\%}$ as $n_{\rm cool} \times R^{-1}$ and $n_{\rm 10\%} \times R^{-1}$, respectively. 
\begin{itemize}
    \item The calculation of $n_{\rm 10\%}$ appears in the code simply as \texttt{log(0.1)/log(closure)}, where \texttt{closure} is $p$ from Eq.~\eqref{eq:closure} computed in the code as \texttt{sum(DISTINCT r2.branching\_ratio)}. Everything here is unitless.

    \item The calculation of $n_{\rm cool}$ takes a few steps. 
    \begin{itemize}
        \item Firstly, we compute the $T_{\rm init}$ in K as \texttt{apoc.coll.max([4, s0.energy*0.62485285866738])}, following $T_{\rm init} = \max{(4\,\mathrm{K}, \frac{E_{S_0}}{- \ln{0.1} k_{\rm B}})}$, where $E_{S_0}$ is \texttt{s0.energy} in cm$^{-1}$, and $k_{\rm B} = 0.6950348004\,\,\frac{\rm{cm}^{-1}}{\rm K}$ \cite{NIST}.

        \item Then, we compute $\sum_i \frac{\mathrm{BR}_i}{\lambda_i}$ present in Eq.~\eqref{eq:ncool} as \texttt{sum(DISTINCT r2.branching\_ratio / r2.energy\_diff) as sum\_of\_BR\_lambda\_ratios}.

        \item Finally, we compute $n_{\rm cool}$ from Eq.~\eqref{eq:ncool} equal to $\frac{p}{h} \sqrt{3 k_{\rm B} T_{\rm init} m} \left( \sum_i \frac{\mathrm{BR}_i}{\lambda_i} \right)^{-1}$ with \texttt{sqrt(initial\_temp * 15) * closure * 0.39579544150466855 / sum\_of\_BR\_lambda\_ratios as n\_cool}, where $T_{\rm init}$ is in K, $15$ is the molecular mass of ${}^{14}$NH in u, $\lambda_i$ in nm, and 
    \begin{gather*}
        0.39579544150466855 = h^{-1} \sqrt{3 \times \mathrm{KtoJ} \times \mathrm{utokg}} \times \mathrm{nmtom} =\\
        = (6.62607015 \times 10^{-34} \,\, \mathrm{J} \times \mathrm{s})^{-1} \sqrt{3 \times 1.380649 \times 10^{-23} \,\, \mathrm{J} \times 1.66053906660 \times 10^{-27} \,\, \mathrm{kg}} \times 10^{-9} \,.
    \end{gather*}
    The value of the Planck constant $h$ and conversion factors between K and J and between u (Da) and kg are taken from Ref.~\cite{NIST}.
    \end{itemize}
      
    \item Now we need to compute $R^{-1}$ from Eq.~\eqref{eq:R}, simplified as 
    $R^{-1} = \tau (G+1) + \frac{2}{3} \pi h c I^{-1} \sum_i \frac{1}{\lambda_i^3}$, where \[I = 1\,\, \mathrm{W} \times \mathrm{cm}^{-2} = 1 \,\, \mathrm{J} \times \mathrm{s}^{-1} \times (10^{-2} \,\, \mathrm{m})^{-2}\,.\]
    \begin{itemize}
        \item We first compute $\sum_i \frac{1}{\lambda_i^3}$ with \texttt{sum(DISTINCT 1 / (r2.energy\_diff*r2.energy\_diff*r2.energy\_diff)) as sum\_inv\_lambdas\_3}, where $\lambda$ is in nm. 

        \item Then we compute $R^{-1}$ as \texttt{lifetime * (num\_decays + 1) + 0.04160402474381969 * sum\_inv\_lambdas\_3 as inv\_R\_s}, where $\tau$ is in s, resulting $R^{-1}$ is also in s, and
        \begin{gather*}
            0.04160402474381969 = \frac{2}{3} \pi h c I^{-1} \mathrm{nmtom}^{-3} = \\
            = \frac{2}{3} \pi \times 6.62607015 \times 10^{-34} \,\, \mathrm{J} \times \mathrm{s} \times 299792458 \,\, \mathrm{m} \mathrm{s}^{-1} \times 1\,\, \mathrm{J}^{-1} \times \mathrm{s} \times 10^{-4} \,\, \mathrm{m}^2 \times (10^{-9} \,\, \mathrm{m})^{-3}\,.
        \end{gather*}
    \end{itemize}

\end{itemize}

\end{document}